\documentclass[a4paper]{article}
\usepackage[margin=2.5cm]{geometry}
\usepackage{amsmath,amsfonts,amssymb,amsthm}
\usepackage[hyperfootnotes=false]{hyperref}
\usepackage{cleveref}
\usepackage{etoolbox}
\usepackage[style=authoryear,backend=biber]{biblatex}
\usepackage{setspace}
\usepackage{parskip}
\usepackage{caption}
\usepackage{booktabs}
\usepackage{threeparttable}
\usepackage{graphicx}
\usepackage{pgf}
\usepackage{adjustbox}
\usepackage{tabularray}
\usepackage{float}
\usepackage{graphicx}
\usepackage[normalem]{ulem}
\usepackage{tikz}
\usepackage{placeins}
\usepackage[svgnames]{xcolor}
\usepackage{xfp}

\hypersetup{colorlinks=true,linkcolor=Crimson,citecolor=Crimson,urlcolor=Crimson}
\newtheorem{proposition}{Proposition}

\newtheorem{assumption}{Assumption}
\crefname{assumption}{assumption}{assumptions}
\Crefname{assumption}{Assumption}{Assumptions}
\UseTblrLibrary{booktabs}
\UseTblrLibrary{siunitx}

\NewTableCommand{\tinytableDefineColor}[3]{\definecolor{#1}{#2}{#3}}
\AtBeginEnvironment{table}{\footnotesize}
\AtBeginEnvironment{talltblr}{\footnotesize}
\AtBeginEnvironment{longtblr}{\footnotesize}

\newcommand{\DescAgeDiffYears}{6}
\newcommand{\DescWageRatioPct}{96}
\newcommand{\DescGovtSharePct}{6}
\newcommand{\DescPrivateFormalSharePct}{33}
\newcommand{\DescSelfEmployedSharePct}{39}
\newcommand{\DescUnpaidSharePct}{13}
\newcommand{\DescCasualSharePct}{3}
\newcommand{\DescGovtSocialServicesPct}{80}
\newcommand{\DescGovtAgriculturePct}{7}
\newcommand{\DescPrivateSocialServicesPct}{22}
\newcommand{\DescPrivateManufacturingPct}{27}
\newcommand{\DescGovtTeachersPct}{25}
\newcommand{\DescGovtOfficialsPct}{8}
\newcommand{\DescGovtOtherServicePct}{9}
\newcommand{\DescLaborerRatio}{21}

\newcommand{\DescAgricRatio}{13}
\newcommand{\DescMoverSharePct}{10}
\newcommand{\DescInMoverSharePct}{52.8}
\newcommand{\DescOutMoverSharePct}{47.2}
\newcommand{\DescBothWayMoverPct}{40}

\newcommand{\LassoOofRsqPct}{10}
\newcommand{\RfOofRsqPct}{10}
\newcommand{\GbmOofRsqPct}{11}
\newcommand{\BasisOofRsqPct}{-6}
\newcommand{\RfInRsqPct}{60}
\newcommand{\BasisInRsqPct}{34}
\newcommand{\RsqGainMoreVarsPct}{64}

\newcommand{\MlVsOlsGainPct}{24}
\newcommand{\GbmOofRsqComplementPct}{89}
\newcommand{\CorrLassoGbmPct}{84}
\newcommand{\TrainingSampleSize}{13,311}
\newcommand{\CorrMaxOlsPct}{77}
\newcommand{\SensMinYears}{7}
\newcommand{\SensGbmOofRsqPct}{24}
\newcommand{\SensLassoOofRsqPct}{27}

\newcommand{\SkillFirstYear}{1989}
\newcommand{\SkillLastYear}{2014}
\newcommand{\RelGovtControlFirstPct}{12}
\newcommand{\RelGovtAllFirstPct}{58}
\newcommand{\GrowthGovtPct}{21}
\newcommand{\GrowthPrivatePct}{29}

\newcommand{\HiringShareCohortChangePct}{7}
\newcommand{\HiringShareCohortChangeDiffPct}{16}

\newcommand{\RsqGovtSkillPct}{7}

\newcommand{\RsqGovtExperiencePct}{18}

\newcommand{\RsqCtrlSkillPct}{15}

\newcommand{\RsqCtrlExperiencePct}{19}
\newcommand{\RsqCtrlVsGovtRatio}{2.11}
\newcommand{\WpUncondLogPts}{0.50}
\newcommand{\WpUncondPct}{65}
\newcommand{\WpLcJobLogPts}{0.36}
\newcommand{\WpLcJobPct}{43}
\newcommand{\GgWithinJobLogPts}{0.49}
\newcommand{\GgWithinJobPct}{39}
\newcommand{\GgGovtVsPrivateRatio}{1/6}
\newcommand{\GgExplainsWpPct}{29}
\newcommand{\WpCondToUncondFrac}{66}

\newcommand{\HeckmanFirstWald}{21.8}
\newcommand{\HeckmanFirstP}{< 0.001}
\newcommand{\HeckmanExclWald}{1.065}
\newcommand{\HeckmanExclP}{0.30}

\newcommand{\HeckmanImrCoef}{0.29}

\newcommand{\StabPeriodBeforePct}{83}
\newcommand{\StabPeriodSincePct}{95}
\newcommand{\StabYearMinPct}{55}
\newcommand{\StabYearMaxPct}{87}
\newcommand{\StabOccProductionPct}{81}
\newcommand{\StabOccSalesPct}{77}
\newcommand{\StabOccTeacherPct}{63}

\newcommand{\HeckmanMinGbmCorrR}{0.97}

\newcommand{\HeckmanGbmSlope}{0.89}
\newcommand{\HeckmanGovtSkillDiffBaseline}{0.217}
\newcommand{\HeckmanGovtSkillDiffHeckman}{0.192}

\newcommand{\RiskMissMinPct}{13.0}
\newcommand{\RiskMissMaxPct}{96.3}
\newcommand{\ImputedWeightRows}{173}
\newcommand{\ImputedWeightIndiv}{36}

\title{Estimating government worker skills\\[0ex] \Large{Application to worker selection and wage setting in Indonesia}\thanks{For helpful feedback and comments, we want to thank in no particular order: Matteo Bobba, Mitch Downey, Oscar Fentanes, Olivier de Groote, Christian Hellwig, Johanna Isman, Guohui Jiang, Pascal Lavergne, Thierry Magnac, Stéphane Straub, and Jens Wäckerle.
We are indebted to the feedback by seminar and conference participants at the IAST Political Economy workshops, the TSE Macroeconomics, Econometrics, and Development workshops, the EUDN development conference in Passau 2020, Stockholm University's applied micro workshop and the workshop of the World Bank's Bureaucracy Lab.
All remaining errors are our own.}}
\author{Kevin Michael Frick\thanks{Toulouse School of Economics, University of Toulouse Capitole, Toulouse, France.
\href{mailto:kevin-michael.frick@tse-fr.eu}{\nolinkurl{kevin-michael.frick@tse-fr.eu}}} \and Jonas Gathen\thanks{Tilburg University, Tilburg, Netherlands.
\href{mailto:j.gathen@tilburguniversity.edu}{\nolinkurl{j.gathen@tilburguniversity.edu}}}}
\date{\today}

\begin{document}

\pagenumbering{gobble} % suppress numbering completely
\maketitle

\begin{abstract}
We propose a new approach to estimate government worker skills, a setting where output is hard to observe and wages may be uninformative about skills.
The approach uses wages in comparable jobs in the private sector and machine learning tools to link skills to skill-related observables.
We apply the approach to rich Indonesian household-level panel data from 1988-2014, showing two main applications. 
First, government skills have continuously declined relative to the private sector, driven by the most skilled workers ending up in the private sector.
Second, the Indonesian government pays a wage premium of \WpLcJobPct\% conditional on skills.
\end{abstract}

\clearpage
\pagenumbering{arabic}
\setcounter{page}{1}

\section{Introduction}\label{sec:introduction}

How do relative skills of government workers evolve over the course of development, and how does this affect changes in the quality of government services? Does the government manage to select highly skilled workers? Are government wages informative about underlying skills and does the government pay a wage premium? A key ingredient for answering these questions is a good measure of government worker skills.
This paper provides a new approach to estimate such skills, showing how to estimate unobserved skills in potential government jobs for any worker in the economy.
The paper then applies the estimation approach to rich household-level panel data from Indonesia over a period of 25 years to study (1) systematic changes in the skills of government workers and their selection and (2) government wage setting.

Traditional approaches in economics measure a worker's skill --- or marginal productivity of labor --- either directly by observing output and making an assumption on the underlying production function
\parencite{attanasio2020estimating,chetty2014measuring1} or by drawing on observed wages assuming a wage setting process that makes wages a noisy signal of a worker's skill \parencite{polachekEarningsLifecycleMincer2008,meghirEarningsConsumptionLife2011,sandersLifecycleWageGrowth2012}.
For the skill estimation of government workers, both approaches are problematic, because government output is usually unobserved (think of the output of a bureaucracy), government worker wages often follow rigid wage-setting rules that disguise underlying skill differences \parencite{biasi2021labor} and wage competition may not force bureaucracies to change wages.
This paper provides a new approach to estimate government worker skills that is applicable in settings where government output is unobserved and government wages are uninformative about skill differences.

In summary, the approach estimates skills from wages in comparable jobs in the private sector, relates these skills to skill-related observables using machine learning tools, and then predicts government worker skills out-of-sample.
The idea is that there are jobs in the private sector in similar social service occupations and with similar task contents as government jobs but for which wages --- however imperfect --- are at least a very noisy signal of skills in these jobs.
This skill signal, extracted as an individual fixed effect, can then be regressed on a rich set of skill-related observables.
We show how machine learning algorithms are particularly useful here to disentangle actual skills from estimation noise and other wage determinants.
In the last step, we can then enforce the estimated relationship between individual-specific skills in comparable jobs and skill-related observables to obtain an out-of-sample potential skill estimate for government jobs.
This provides an estimate of skills in government jobs for any worker in the economy for whom we observe these skill-related observables.
The approach can thus also be used in other settings in which researchers need to estimate potential job-specific skills for workers who work in other jobs with very different skill requirements.

The approach relies on rich skill-related observables as it defines relevant skills solely as a function of observables. Unobserved skills pose two distinct problems for the approach. The first is a measurement limitation: To the extent that some skill dimensions are unobserved and only weakly correlated with observables, the approach will miss these skill dimensions. Importantly, this does not immediately bias the skill estimate since we are not interested in estimating the \emph{causal} effect of a change in a specific covariate $x$.
The second problem is selection. Since we use comparable jobs in the private sector to estimate skills in the government, an issue is if comparable private sector workers are systematically selected on skill-related unobservables that directly affect their wages (compared to government workers). While rich skill-related observables ameliorate selection concerns, we also show how to extend the baseline approach to correct for selection \parencite[]{heckman1979sample}.

We apply the estimation approach to Indonesia between 1988 and 2014 drawing on a large and high-quality representative panel that tracks individuals over time.
The data --- the Indonesian Family and Life Survey (IFLS) --- is particularly suited for the estimation approach as it features (1) a large sample of government and private sector workers across many different occupations, (2) one of the longest wage and employment panels in a developing country context, and (3) an exceptional breadth of skill-related variables such as educational background, national exam scores, test results for self-administered Raven tests \parencite{penrose1936new}, cognition and memory tasks, the big-5 personality traits \parencite{goldberg1993structure} and elicited risk and time preferences.
Drawing on the IFLS, we estimate individual-level skills in government-like jobs using comparable private sector workers.
We define comparable jobs in the private sector as jobs in similar occupations and sectors as
government jobs, and with a similar set of worker demographics, which are enforced using propensity score weighting.
The approach draws on a noisy skill signal from comparable private sector workers, which is the worker wage fixed effect after purging the wage from (1) experience effects using a standard Mincerian regression and (2) from changes in the equilibrium skill price using the flat spot identification approach in \textcite{bowlusHumanCapitalPrices2012}.
To identify skills, we use off-the-shelf machine learning algorithms to predict skills from these noisy skill signals. We find a Gradient Boosted Model (GBM) to perform best, far outperforming more traditional methods like a basis expansion, precisely because machine learning algorithms are trained to not overfit, which is crucial in our prediction exercise with noisy input data.

Having estimated government worker skills, we then illustrate two main applications for which skill estimates are key.
In the first application, we look at changes in government worker skills, the selection of government workers based on skills and how relative skills of government workers have evolved.
We find that Indonesian government workers are strongly positively selected on skills and that their skills have also increased over time as new cohorts with higher skills entered the labor force.
However, we find that relative skills of government workers actually declined over time in comparison to the private sector.
We show that a big part of this is driven by an increasing right tail of workers who would be highly skilled in government jobs but who the government does not manage to hire.
Further, we find evidence that changes in the hiring practices of Indonesian government workers in the process of democratization, decentralization and civil service reforms after the year 2000 have at best led to small improvements in how the government \emph{de facto} selects government workers.
Finally, we show evidence for the detrimental effect of hiring waves for the selection of talent in the government. Uneven hiring across years lead to differences in cohort-specific government employment shares, which have an adverse effect on government worker selection: in years in which the government hires more, average government worker skills decline, which holds conditional on the jobs that the government opens.
This is consistent with the idea that more hiring forces the government to eventually move down the distribution as the top of the distribution thins out.

In the second main application, we show how to use the skill estimates to look at government wage setting and the wage premium of government workers.
First, we find evidence that Indonesian government wages are indeed less informative about underlying skills than private sector wages in comparable jobs.
We find this by predicting government wages using estimated skills from comparable private sector jobs and by re-estimating skills using government wages instead.
However, due to more predictable life-cycle wage progression in the public sector, government wages become almost as predictive of skills when accounting for experience.
Second, we find that the Indonesian government pays a large wage premium of about \WpLcJobPct\% compared to similar jobs in the private sector.
Does this mean that the Indonesian government is overpaying for workers? While this could be true, we also find strong evidence for more discriminatory wage setting in the private sector.
A large gender wage gap in the private sector of about $ \GgWithinJobPct\%$ even after controlling for job, skills and experience accounts for about \GgExplainsWpPct\% of the government wage premium.

In the last part of the paper, we test a number of technical assumptions in the paper and extend the approach. We show evidence that the estimated mapping from observables to skills is relatively stable over time and across different occupations, consistent with interpreting our skill estimates as ``general skills''. We show that our baseline results hold when controlling for selection. As an instrument that drives selection into the public sector, we use whether a worker's parents work in the public sector. The assumption is that the instrument drives selection but does not affect wages in comparable private sector jobs conditional on skills. While we find evidence for selection, we do not find that this materially affects our main empirical results. At last, we show how to extend the approach to multiple summary dimensions of skills and richer experience profiles.

\paragraph{Related literature.}
The contribution of this paper is both technical and conceptual.
Technically, the paper contributes to the literature by proposing a novel estimation approach that allows to measure government worker skills for all government workers as well as any other worker in the economy.
This differs from two alternative approaches that have previously been proposed in the literature.
One common approach -- as followed by
\textcite{dalboStrengtheningStateCapabilities2013,dalboWhoBecomesPolitician2017,besleyGenderQuotasCrisis2017,colonnelliPatronageSelectionPublic2020} -- is to use (residualized) previous private sector wages of government workers as a measure of government worker skills.
Previous private sector employment can be a misleading measure of skills in cases where government workers take their first real job in the public sector or have previously worked in a very different occupation, in which case it might be a better measure of a worker's outside option.
More technically, the approach in this paper -- similar to
\textcite{dalboWhoBecomesPolitician2017,besleyGenderQuotasCrisis2017} -- also draws on residuals from observed private sector wages but restricts to a comparable subset of government jobs and deals directly with the econometric concern that residualized wages only give a noisy estimate of individual-level skills by nonlinearly projecting skills onto skill-related observables.
Whenever rich skill-related observables are available, we thus believe that our approach provides a better way to estimate skills from wages.

The second approach followed in  \textcite{best2023individuals} is complementary to our approach.
 \textcite{best2023individuals} consider a context in which individual bureaucrats' output is observable -- prices paid by bureaucrats in the government procurement of goods.
Similar to using wages, they then also draw on a residualized measure of output and extract an individual bureaucrat fixed effect.
While  \textcite{best2023individuals} are not directly interested in the individual fixed effects, they are interested in a variance decomposition for which noisy estimates of fixed effects would bias their results.
They thus propose a ``covariance shrinkage'' approach that uses bootstrapping to separate variances in the true signal and noise.
There are two key differences. First, the approach in this paper does not require to observe government output but informative wages in the private sector instead. And second, we use rich covariates that also allow to predict skills for any worker in the economy, while the approach in  \textcite{best2023individuals} uses a shrinkage approach without covariates.

Conceptually, the paper uses estimated government worker skills to contribute to the growing literature on the workings of bureaucracies, the developmental state and how the delivery of government services can be improved \parencite{chong2014letter,finanPersonnelEconomicsDeveloping2017,rasulManagementBureaucratsPublic2018,decarolis2020bureaucratic}.
The estimation approach allows for a systematic measurement of government worker skills, a key -- but difficult to measure -- input in the production of government services.
The estimation approach thus complements a number of recent papers that have studied how the government selects government workers \parencite{colonnelliPatronageSelectionPublic2020,dalboStrengtheningStateCapabilities2013,jiaSelectionChina2015,bhavnaniDoesAffirmativeAction2019,estradaRulesDiscretionPublic2019}, who self-selects into government jobs \parencite{ashraf2020losing,hannaDishonestySelectionPublic2017,weaver2021jobs}, how the government remunerates workers \parencite{finanPersonnelEconomicsDeveloping2017} and how the government competes with alternative employers for talent \parencite{deserranno2024impact}.
In line with  \textcite{ashraf2020losing,weaver2021jobs}, we find strong positive selection of government workers based on skills.
This also means that large documented government wage premia in developing countries \parencite{finanPersonnelEconomicsDeveloping2017} are smaller after controlling for selection on skills; for Indonesia, we find a roughly \WpLcJobPct\% wage premium conditional on skills, roughly \WpCondToUncondFrac\% of the unconditional premium.

The key novelty with respect to this literature is the broader scope of the estimation approach that allows to study all government workers over a long period of time.
The approach is particularly useful in settings where government output is hard to observe, and thus especially relevant for studying higher-level bureaucracies where government output may be hard to define and measure, allowing researchers to move beyond the study of last-mile service delivery \parencite{chaudhuryMissingActionTeacher2006,banerjeePublicActionPublic2007,finanPersonnelEconomicsDeveloping2017}.
Importantly, the broader scope of the approach allows us to establish at least two novel findings in the literature.
First, we show that despite growing absolute skills, relative skills of government workers compared to the private sector systematically declined in Indonesia over the past 30 years
We link this finding to the difficulty of the Indonesian government to attract the workers with the highest skills to the government.
Second, we show evidence for the detrimental effect of government hiring cycles on the selection of government workers.
The evidence is consistent with the idea that in years of outsized hiring, the government needs to move down the skill distribution of the applicant pool to fill all government positions.

A good sign of a new estimation approach is that it raises many interesting questions that can now be studied more rigorously: For example, what are the output or welfare costs of government hiring cycles? Or what drives the relative decline in government skills and does this go in hand with a relative decline in state capacity versus private sector capacity over the course of development? These questions are particularly well-suited for future structural work, for which the estimated government skills in this paper can function as a direct input.

\paragraph{Structure of the paper.} In \Cref{sec:methodology}, we explain the procedure to estimate government worker skills.
We apply our approach to Indonesian data in \Cref{sec:application} and report main results based on the skill estimates in \Cref{sec:main-empirical-results}. \Cref{sec:robustness} shows robustness results and discusses extensions. \Cref{sec:conclusion} concludes.

\section{Methodology}
\label{sec:methodology}

\paragraph{Conceptual framework.} The approach in this paper is to identify and estimate skills of government workers using wages in private sector jobs that are comparable to government jobs.
The main benefit of this approach is that we do not need to take a stance on whether government wages reveal anything about skills.
We start with formalizing our setup, assuming for now that a set of comparable private-sector jobs is available for which we observe a representative panel of workers and their wages.
We follow a large human capital literature in assuming that the wages in comparable private-sector jobs at least partially reflect a worker's idiosyncratic human capital and the equilibrium price of human capital \parencite[e.g.][]{becker1975investment, heckman1998explaining, huggettSourcesLifetimeInequality2011}.
Human capital of a worker $i$, $H(e_{it},z_{i})$, is a flexible function of experience $e_{it}$ and permanent skills $z_i$.
The equilibrium ``skill price'' $P_t$ changes because of aggregate changes in the supply and demand for labor, including changes in government hiring policy that affects the private sector labor market.
Individual wages can systematically deviate from worker's human capital and the skill price for a variety of reasons, including imperfect competition, dynamic contracts and information asymmetries.
This nests most models of the labor market, but rules out settings in which wages permanently deviate from skills \parencite[see][]{taberEstimationRoySearch2020}.
More formally, we assume the following setup:

\vspace{3pt}

\begin{assumption}[Conceptual framework]\label{as:conceptual}
For any worker \(i\) in a private sector job comparable to government work, real wages $W_{it}$ are given by:
\begin{equation}
W_{it} = \exp \left( w_{it} \right) = P_t \cdot H(e_{it}, z_i) \cdot \exp \left(\varepsilon_{it}\right),
\label{eq:pvt-wage-setting}
\end{equation} where:
\begin{enumerate}
\item Human capital is log-additively separable in experience and a time-invariant skill component at labor market entry $z_i$, that is
\begin{equation}
    \log H(e_{it}, z_i) = h(e_{it}) + z_i.
    \label{eq:zi-def}
\end{equation} with $h(\cdot)$ assumed to be a known functional that is linear-in-parameters (e.g. a polynomial).
\item The error $\varepsilon_{it}$ has zero mean, finite variance, and is mean-independent of $P_t$ and $H(e_{it}, z_i)$ but can feature serial correlation, compensating differentials, and job-specific components as long as these are stationary.
\item An unbiased estimator of the time-varying (log) skill price is available: $\widehat{p_t}$.
\end{enumerate}
\end{assumption}

\Cref{as:conceptual} defines worker-specific skills $z_i$ as our main object of interest in this paper.\footnote{Permanent characteristics at labor market entry are generally found to be by far the main drivers of earnings inequality \parencite[e.g.][]{keaneCareerDecisionsYoung1997, huggettSourcesLifetimeInequality2011, lamadon2022imperfect, taberEstimationRoySearch2020}, making it a natural starting point.}
By definition, these are permanent skills that are priced by and thus relevant to the set of jobs used in the estimation.
Both in this section and the empirical application, we focus on $z_i$ as a single-dimensional summary measure of ``general skills'' that are important across different jobs.
As we formalize below, $z_i$ is thus a summary of potentially many different dimensions of skills.
In the final part of the paper, we still show how to extend the setup to multiple summary dimensions of skills.

To be able to isolate $z_i$, one first requires to net out the effect of the skill price $P_t$.
Netting out the effect of the skill price is important since we are interested in identifying changes in skills over time and thus need to control for changes in wages that are simply driven by changes in the demand and supply for human capital.
A tempting alternative may be to simply include time fixed effects, but these will net out any average changes across time, whether they come from changes in the price or the quantity of skills, making it impossible to separately identify changes in skills.
Identifying and estimating the skill price is a known problem for which standard estimators already exist, which is why Assumption \ref{as:conceptual} treats this estimator as given.
Specifically, we believe there are two natural estimators for the skill price $P_t$ in settings like ours.
The first approach, which we call the ``hedonic price regression approach" is to estimate Equation \ref{eq:pvt-wage-setting} on a single cross-section, normalize $P_t$ for that year and identify changes in $P_t$ from average changes in wages after netting out the effect of changes in (estimated) skills $\widehat{H(e_{it},z_i)}$.
The second approach, which we follow for our baseline results in the empirical application, is to estimate Equation \ref{eq:pvt-wage-setting} using the full panel and identify changes in the skill price $P_t$ from the flat-spot identification approach of \textcite{bowlusHumanCapitalPrices2012}.
We provide further details in \Cref{sec:application}.

The main idea in this paper is to identify skills $z_i$ for comparable private-sector jobs, flexibly project them onto a large set of observable individual-specific and skill-related characteristics and then use this estimated mapping to predict expected skills for government workers.
We now formalize the mapping between skills and characteristics.

\vspace{3pt}

\begin{assumption}[Mapping skills to observable characteristics]
Let $x_i \in \mathbb{R}^p$ be a vector of $p$ time-invariant observables for worker $i$.
Define $f(x) \equiv \mathbb{E}[z_i \mid x_i = x]$.
We assume that the conditional expectation $f(\cdot)$ is time-invariant, is independent of whether $i$ works in the public sector or in a comparable private-sector job, and satisfies the minimal regularity condition of being Borel-measurable.
\label{as:time-inv-mapping}
\end{assumption}

While we allow considerable flexibility on the function $f$ that maps from skills to characteristics, Assumption \ref{as:time-inv-mapping} does make a number of important restrictions.
One restriction is time-invariance, which means that different dimensions of skills as captured by $x$ are valued in the same way over time.
This could be violated in case cognitive components of skills become more important over time.
Importantly, this assumption can be tested by estimating $f$ for different time windows on the underlying data.
Another key restriction is that the same $f$ is assumed to hold for both government jobs and comparable private-sector jobs, which allows us to estimate $f$ on comparable private-sector jobs. We discuss further below how unobservable skills and selection into the private sector and government jobs based on unobservables could bias the estimation and how we address this.

\paragraph{Identification and prediction.}

Given the conceptual framework, we can now formalize identification in a single proposition that also introduces the main residual equation used for estimation:

\vspace{3pt}

\begin{proposition}[Estimating and predicting skills]\label{prop:fx}
Suppose \Cref{as:conceptual,as:time-inv-mapping} hold and let $\widehat{p_t}$ be an unbiased estimator of $p_t$ constructed from comparable private-sector jobs.
Define the price-adjusted log wage as: $\widetilde{w_{it}} \equiv w_{it} - \widehat{p_t}$.
Then for any worker observed in comparable private-sector jobs:
\begin{equation}
\widetilde{w_{it}} = f(x_i) + h(e_{it}) + u_{it},
\label{eq:main-estimating-equation}
\end{equation} where $u_{it} \equiv v_i + \widetilde{\varepsilon_{it}}$, with $v_i \equiv z_i - f(x_i)$ being the projection error of the conditional expectation function and $\widetilde{\varepsilon_{it}} \equiv \varepsilon_{it} + (p_t - \widehat{p_t})$ is a composite residual that absorbs both wage shocks and estimation error in $\widehat{p_t}$.

\Cref{eq:main-estimating-equation} is semi-parametric in that $f$ can be identified non-parametrically conditional on a functional form assumption on $h(e_{it})$. The equation can be estimated using any non-parametric estimator $\widehat{f(\cdot)}$ in combination with a simple two-step estimation approach or the semi-parametric joint estimation approach by \textcite{robinson1988root}.
Machine learning estimators are particularly suitable for obtaining $\widehat{f(\cdot)}$ because it is effectively a prediction task using covariates $x$.

The estimator $\widehat{f(\cdot)}$ obtained from comparable private-sector data can then be used to form out-of-sample predictions $\widehat{f(x_i)}$ for workers not observed in those jobs (including public-sector workers) but for whom $x_i$ is observed.
\end{proposition}

\Cref{prop:fx} formalizes that given a choice for the experience profile $h(\cdot)$, the conditional expectation $f$ can be estimated flexibly using machine learning or other non-parametric estimators.
For computational ease, we follow a simple two-step estimator.\footnote{Our results are quantitatively indistinguishable when using the joint estimation approach of \textcite{robinson1988root}. Extending the Frisch-Waugh-Lovell (FWL) theorem to a partially linear model, this method non-parametrically partials out $x_i$ from both the outcome and the linear covariates to first estimate the linear parameters $\boldsymbol{\delta}$.
Because both estimators are consistent for the same population parameters, they yield almost identical estimates in our sample.
Ultimately, both approaches recover the $\boldsymbol{\delta}$ in the exact same way, by computing average residuals, but we prefer our two-step within estimator for its computational simplicity.} In the first step, we estimate $h(e_{it})$ using $\widetilde{w_{it}}$ and controlling for individual fixed effects. In the second step, we net out $\widehat{h(e_{it})}$ and estimate $f(x_i)$ using the individual-specific average residual $\hat{z}_i$, which is a noisy estimate of true skills $z_i$: $\hat{z}_i \equiv \frac{1}{N_i}\sum_{t} \big( \widetilde{w_{it}} -\widehat{h(e_{it})} \big)$.

% This follows because standard nonparametric regression and machine learning tools target conditional means by restricting $\widehat{f(\cdot)}$ to a set of candidate predictors (a ``function class'').
%Universal approximation results state that for popular, sufficiently flexible function classes (e.g., multilayer neural networks), and broad sets of target functions $f$ (e.g., continuous functions on a compact support), the approximation error $\inf_{g\in\mathcal{F}}\mathbb{E}[(f(x_i)-g(x_i))^2]$ can be made arbitrarily small \parencite{hornik1989multilayer,athey2019machine,shen2019asymptotic}.

\paragraph{Choosing the experience profile}

An important question is what to choose for the experience profile $h(e)$.
In practice, we assume a simple quadratic experience profile as in a standard Mincerian regression, but the approach can also accommodate richer experience profiles.
In this setup, how human capital changes with experience is identified from within-worker changes in the price-adjusted wage $\widetilde{w_{it}}$.
In our context, this simple experience profile gives a reasonably good fit.

\paragraph{Prediction and omitted variable bias}

Identification of skills relies on observables.
A natural question is thus whether unobservables pose a particular problem for the approach?
For this, it is useful to distinguish two distinct issues with unobservables.
The first is a pure \emph{measurement issue}: what if there are skill-related observables that we do not observe?
The key benefit of our approach is that we use $f$ as a predictive mapping, not as a model of skill formation and we are thus not interested in the causal effect of varying a component of $x$.
Econometrically, we define $f$ as a conditional expectation, which does not impose a causal interpretation and does not require that skills are fully determined by observables.
Intuitively, omitted variables will bias our estimates of the marginal effects for each correlated component $x_i$, but we are not interested in these marginal effects to begin with.
Rather, we are interested in predicting the outcome $y$ and we thus do not care which components of $x_i$ \emph{cause} skills versus which are simply \emph{correlated} with skills, making machine learning estimators particularly useful here.
However, our predicted skills $\widehat{f(x_i)}$ are only as informative as the observables $x_i$: any wage-relevant component of skills $z_i$ not sufficiently correlated with observables $x_i$ remains in the residual and cannot be predicted. % $v_i=z_i-f(x_i)$
This reduces predictive power and matters for interpretation of our skill estimates.

%Writing $z_i = f(x_i) + v_i$ defines a residual $v_i$ with $\mathbb{E}[v_i \mid x_i]=0$ by construction.
% Importantly, omitting determinants of $z_i$ from $x_i$ does not create an ``omitted variable bias'' for $f$: it simply makes $v_i$ more variable and thus makes $f(x_i)$ a noisier (i.e. worse) predictor of the underlying but unobservable individual skill $z_i$.

\paragraph{Selection on unobservables}

A distinctly separate issue of unobservables is selection.
Since we use comparable private-sector jobs to estimate skills in government jobs, an important concern is if workers select into the government versus the private sector based on unobservables.
Selection on unobservables is not generally an issue for our approach but only if selection is on unobservable wage components that are both correlated with $x_i$ and directly affect wages in comparable private-sector jobs (compared to government workers).
While we believe that our rich set of observables in the empirical application including variables on cognitive abilities already strongly ameliorates the selection issue, the selection assumption can be relaxed.
Specifically, in \Cref{sec:robustness} we show how to allow for selection using a standard \parencite{heckman1979sample} selection model where the probability of choosing a comparable private sector job is given by:
\begin{equation}
\Pr[s_{it} = 1 | x_i, r_i, e_{it}] = \Phi\bigl(g(x_i) + h(e_{it}) + \boldsymbol{\beta}' r_i + \pi_t\bigr),
\label{eq:selection-extension}
\end{equation}
where $g(\cdot)$ and $h(\cdot)$ are general functions and $r$ is a vector of instruments that affect selection into private sector jobs (versus government jobs) but does not affect wages directly.
In \Cref{sec:robustness} we show that parental private-sector employment is an appropriate instrument for selection into a private-sector job and that our main results are robust to controlling for selection.

\paragraph{Common support}

Technically, our approach does not require common support over $x_i$ across workers in comparable private-sector jobs versus government workers.
However, outside the support of the training data, the estimate $\widehat{f(\cdot)}$ necessarily relies on extrapolation over the chosen functional class for $f$.
To minimize reliance on functional forms, it is thus still good practice to verify support.
\Cref{fig:support-barcharts-dummies,fig:support-relative-density} in \Cref{appendix-common-support} show that this is the case.

%The approach identifies skills only up to normalization (an additive constant), so a wage markup or markdown that is constant across workers in the comparable private sector is absorbed by the intercept/normalization and does not affect relative skill comparisons.
% This follows because adding the same constant to all $z_i$ leaves \Cref{eq:main-estimating-equation} unchanged once the intercept is renormalized.

\paragraph{Limitation of using wages to predict skills}

Since our underlying skill estimator relies on wages, standard critiques of wage-based skill estimates apply.
For example, we interpret estimated skills as ``productive government-worker skills'' throughout, however, wages can also contain persistent non-skill premia that are correlated with observables (e.g., sorting as in \textcite{abowd1999high}, bargaining, or gender discrimination), in which case estimated skills will also reflect these premia.
The added problem for our estimator is that these non-skill premia could be specific to the private sector, in which case our estimated skills would not only reflect government-worker skills.
We return to this issue when discussing our results.

\section{Empirical application}\label{sec:application}

In this section we apply our methodology to the case of Indonesia, the fourth most populous country in the world, over a period of almost 30 years from 1988 to 2014.
We start by giving an overview of the context and underlying data.

\subsection{Context and Data}\label{sec:ontext-data}

Between 1988 and 2014, Indonesia moved away from a military-ruled, highly centralized authoritarian government under General Suharto (1967-1998) to a relatively consolidated, highly decentralized democracy.
Incomes per capita increased roughly 7-fold, poverty dramatically declined and Indonesia transitioned to become a middle-income country.
The Suharto regime was characterized by extensive cronyism and patronage under which the bureaucracy was greatly expanded but also seen as a direct political instrument to collect votes and offer political support \parencite{fismanEstimatingValuePolitical2001a,hadizPoliticalEconomyOligarchy2013,martinez-bravoNonDemocraticRootsElite2017,robisonReorganisingPowerIndonesia2004}.
Civil servants were obliged to become members of the political apparatus, support the party in power and hiring and promotion decisions were made to support regime stability \parencite{kristiansenBuyingIncomeMarket2006,mcleodInadequateBudgetsSalaries2008}.
With the Asian Financial Crisis in 1997/1998 and the subsequent fall of the Suharto regime, Indonesia embarked on the \emph{Era Reformasi}, targeting constitutional, judicial, public financial management and privatization reforms as well as an unprecedented decentralization process.
In this new, decentralized system, three-fourths of the Civil Service (including teachers and health workers) are assigned to local governments in contrast to a fraction of this under the Suharto regime.
Since the decentralization reforms, the central government can steer civil service hiring by setting overall quotas for the number of civil service jobs, while districts are left with a high degree of discretion as they decide on applicants and applicant requirements.
Throughout the 2000s, a number of ministries started bureaucracy reform initiatives that tried to set civil service remuneration on par with the private sector and pushed for more competitive hiring and promotion practices \parencite{horhoruwTransformingPublicSector2013}.
Due to uneven adoption across ministries, in 2010, the government mandated public sector reform for all central and local governments, but only in 2014, a new Indonesian Civil Service law was passed.

We draw on nationally representative data from the Indonesian Family Life Survey -- IFLS in short -- which is particularly suited for the identification strategy in this paper.
Specifically, the IFLS is a large and high-quality household- and individual-level panel dataset that collects exceptionally detailed information on individual's occupations, wages, skills and preferences.
It is based on an initial sample of 7,224 households and 22,347 individuals tracked throughout five waves (1993, 1997-98, 2000, 2007-08, and 2014-2015) with more households and individuals being added throughout the five waves.
It represents about 83\% of the Indonesian population living in 13 of the nation's 26 provinces in 1993.
Due to an intensive focus on respondent tracking, re-contact rates between any two rounds are above 90\%, and 87\% of the original households were contacted in all five rounds (see: \textcite{straussFifthWaveIndonesia2016}).\footnote{Throughout, all results are based on weighting individual observations using provided cross-sectionally representative survey weights that also correct for attrition.} As a comparison, these re-contact rates are as high or higher than most longitudinal surveys in the United States and Europe.

\paragraph{Employment and wages.} The IFLS data includes detailed employment data for each survey round.
In addition to current employment, the survey included questions on previous employment.
As in \textcite{hamory2021reevaluating}, this allows to create up to a 27-year annual individual employment panel from 1988 to 2014, making it one of the longest employment panel datasets available for developing countries and uniquely positioned to study life cycle wage growth (see: \textcite{lagakos2023role}).\footnote{Employment status and sector of employment are available for each year, but in the fourth and fifth IFLS round, earnings were collected only for the current job.
Following  \textcite{hamory2021reevaluating}, the earnings measure in this paper is the sum of all wages, profits, and benefits.}
%{Across all jobs?}}
Employment information captures principal and secondary employment including government jobs.
We focus on principal jobs as the job classification throughout this paper.
For wages we use real hourly income based on total wages and total hours worked across all jobs, and by deflating nominal values using Indonesia-wide inflation as measured by the CPI.\footnote{To best deal with periods of high inflation, we construct annual inflation for the 12-month period preceding the start of survey collection for each wave.}

\paragraph{Skill variables.} The IFLS data also contains an exceptional breadth of skill-related variables, including detailed information on education, test results for self-administered Raven tests, cognition and memory tasks, literacy and language skills, results in standard Indonesian exams, the big-5 personality traits and elicited risk and time preferences.
Unfortunately, some of these variables are not consistently measured over time and have too many missings.
In our main analysis, we thus restrict to 25 main skill-related variables, dropping variables with more than 50\% missings (e.g.\ risk and time preferences as well as national exam scores) and variables that are inconsistently measured over time (e.g.\ the Raven math-score).
The 25 variables span four conceptual groups: (i)~the Big Five personality traits (Extraversion, Conscientiousness, Neuroticism, Openness, Agreeableness); (ii)~eight education indicators (ever attended school, five educational attainment dummies for less than primary, primary, junior secondary, senior secondary, and higher education, an education missing indicator, and Islamic education); (iii)~five literacy measures (writes letters in Indonesian, writes letters in another language, reads newspapers in another language, speaks Indonesian, speaks Bugis);\footnote{Later on, we will difference out geographic factors to make sure we capture language skills rather than which part of Indonesia a person is from.} and (iv)~seven cognitive test scores (word recall, delayed word recall, counting backwards, date recall at two-of-three and all-three thresholds, working ability, and Raven IQ).
We z-standardize the continuous skill measures (Big Five personality traits, Raven IQ score, word recall, delayed word recall, counting backwards, and working ability score) pooling across all individuals.
In case variables are measured multiple times for the same individual, we use the average across waves.
For remaining missing values in continuous variables, we impute with their post-standardization mean of zero; for dummy variables, we set missing values to zero and, for dummies with at least 1,000 missing observations, add a separate binary missing indicator.
Appendix \Cref{sec:missingness-diagnostics} provides detailed information on when we measure which variable, how missingness compares across time and variables, and how we verify consistency of variables over time.

\paragraph{Definition of government worker.} In Indonesia, there are permanent civil service positions called \emph{Pegawai Negeri Sipil (PNS)} as well as temporary civil servant positions (non-PNS).
The latter are for example common in the educational sector where 60\% of new teacher hires are on temporary contracts \parencite{pierskallaPersonnelPoliticsElections2018}.
The IFLS includes both permanent and temporary government workers, defining government jobs broadly as jobs with any government office for which one receives remuneration in money or in kind.
\Cref{sec:govt-hiring} provides further details on the hiring process of government workers in Indonesia.

\paragraph{Job categories, sectors and occupations.} The IFLS distinguishes broad job categories (including government workers) from the job's sector at the 1-digit level and an occupational classification at the 2-digit level.
To allow for sufficiently large occupational groups for government workers, we recode the occupational classification into nine different 1- to 2-digit occupations that captures most variation in government sector jobs.
For example, we leave the 2-digit codes for ``teachers'' and for ``government officials and executives'', while we keep the 1-digit code ``other service areas''.
%{Give full list here}

\subsection{Descriptives}\label{descriptives}

\begin{figure}[t]
\caption{Employment shares by broad job category \label{fig:barplot-job-category-pooled}}

{\centering

\resizebox{0.75\textwidth}{!}{
% !TEX encoding = UTF-8 Unicode
\begin{tikzpicture}[x=1pt,y=1pt]
\definecolor{fillColor}{RGB}{255,255,255}
\path[use as bounding box,fill=fillColor,fill opacity=0.00] (0,0) rectangle (505.89,289.08);
\begin{scope}
\path[clip] (  0.00,  0.00) rectangle (505.89,289.08);
\definecolor{drawColor}{RGB}{255,255,255}
\definecolor{fillColor}{RGB}{255,255,255}

\path[draw=drawColor,line width= 0.6pt,line join=round,line cap=round,fill=fillColor] (  0.00,  0.00) rectangle (505.89,289.08);
\end{scope}
\begin{scope}
\path[clip] (130.12, 33.48) rectangle (499.89,283.08);
\definecolor{fillColor}{RGB}{255,255,255}

\path[fill=fillColor] (130.12, 33.48) rectangle (499.89,283.08);
\definecolor{drawColor}{gray}{0.80}

\path[draw=drawColor,line width= 0.5pt,line join=round] (130.12, 57.63) --
	(499.89, 57.63);

\path[draw=drawColor,line width= 0.5pt,line join=round] (130.12, 97.89) --
	(499.89, 97.89);

\path[draw=drawColor,line width= 0.5pt,line join=round] (130.12,138.15) --
	(499.89,138.15);

\path[draw=drawColor,line width= 0.5pt,line join=round] (130.12,178.41) --
	(499.89,178.41);

\path[draw=drawColor,line width= 0.5pt,line join=round] (130.12,218.67) --
	(499.89,218.67);

\path[draw=drawColor,line width= 0.5pt,line join=round] (130.12,258.92) --
	(499.89,258.92);
\definecolor{fillColor}{gray}{0.35}

\path[fill=fillColor] (146.93, 39.51) rectangle (176.81, 75.75);

\path[fill=fillColor] (146.93, 79.77) rectangle (188.71,116.01);

\path[fill=fillColor] (146.93,120.03) rectangle (202.66,156.26);

\path[fill=fillColor] (146.93,160.29) rectangle (257.40,196.52);

\path[fill=fillColor] (146.93,200.55) rectangle (431.02,236.78);

\path[fill=fillColor] (146.93,240.81) rectangle (483.08,277.04);
\definecolor{drawColor}{RGB}{255,255,255}

\node[text=drawColor,anchor=base,inner sep=0pt, outer sep=0pt, scale=  0.91] at (161.87, 54.50) {3\%};

\node[text=drawColor,anchor=base,inner sep=0pt, outer sep=0pt, scale=  0.91] at (167.82, 94.75) {5\%};

\node[text=drawColor,anchor=base,inner sep=0pt, outer sep=0pt, scale=  0.91] at (174.80,135.01) {6\%};

\node[text=drawColor,anchor=base,inner sep=0pt, outer sep=0pt, scale=  0.91] at (202.17,175.27) {13\%};

\node[text=drawColor,anchor=base,inner sep=0pt, outer sep=0pt, scale=  0.91] at (288.98,215.53) {33\%};

\node[text=drawColor,anchor=base,inner sep=0pt, outer sep=0pt, scale=  0.91] at (315.00,255.79) {39\%};
\end{scope}
\begin{scope}
\path[clip] (  0.00,  0.00) rectangle (505.89,289.08);
\definecolor{drawColor}{RGB}{0,0,0}

\path[draw=drawColor,line width= 0.6pt,line join=round] (130.12, 33.48) --
	(130.12,283.08);
\end{scope}
\begin{scope}
\path[clip] (  0.00,  0.00) rectangle (505.89,289.08);
\definecolor{drawColor}{gray}{0.30}

\node[text=drawColor,anchor=base east,inner sep=0pt, outer sep=0pt, scale=  0.96] at (124.72, 54.32) {Casual emp. (agric.)};

\node[text=drawColor,anchor=base east,inner sep=0pt, outer sep=0pt, scale=  0.96] at (124.72, 94.58) {Casual emp. (non-agric.)};

\node[text=drawColor,anchor=base east,inner sep=0pt, outer sep=0pt, scale=  0.96] at (124.72,134.84) {Government worker};

\node[text=drawColor,anchor=base east,inner sep=0pt, outer sep=0pt, scale=  0.96] at (124.72,175.10) {Unpaid family worker};

\node[text=drawColor,anchor=base east,inner sep=0pt, outer sep=0pt, scale=  0.96] at (124.72,215.36) {Private worker};

\node[text=drawColor,anchor=base east,inner sep=0pt, outer sep=0pt, scale=  0.96] at (124.72,255.62) {Self-employed};
\end{scope}
\begin{scope}
\path[clip] (  0.00,  0.00) rectangle (505.89,289.08);
\definecolor{drawColor}{gray}{0.20}

\path[draw=drawColor,line width= 0.6pt,line join=round] (127.12, 57.63) --
	(130.12, 57.63);

\path[draw=drawColor,line width= 0.6pt,line join=round] (127.12, 97.89) --
	(130.12, 97.89);

\path[draw=drawColor,line width= 0.6pt,line join=round] (127.12,138.15) --
	(130.12,138.15);

\path[draw=drawColor,line width= 0.6pt,line join=round] (127.12,178.41) --
	(130.12,178.41);

\path[draw=drawColor,line width= 0.6pt,line join=round] (127.12,218.67) --
	(130.12,218.67);

\path[draw=drawColor,line width= 0.6pt,line join=round] (127.12,258.92) --
	(130.12,258.92);
\end{scope}
\begin{scope}
\path[clip] (  0.00,  0.00) rectangle (505.89,289.08);
\definecolor{drawColor}{RGB}{0,0,0}

\path[draw=drawColor,line width= 0.6pt,line join=round] (130.12, 33.48) --
	(499.89, 33.48);
\end{scope}
\begin{scope}
\path[clip] (  0.00,  0.00) rectangle (505.89,289.08);
\definecolor{drawColor}{gray}{0.20}

\path[draw=drawColor,line width= 0.6pt,line join=round] (146.93, 30.48) --
	(146.93, 33.48);

\path[draw=drawColor,line width= 0.6pt,line join=round] (232.74, 30.48) --
	(232.74, 33.48);

\path[draw=drawColor,line width= 0.6pt,line join=round] (318.55, 30.48) --
	(318.55, 33.48);

\path[draw=drawColor,line width= 0.6pt,line join=round] (404.37, 30.48) --
	(404.37, 33.48);

\path[draw=drawColor,line width= 0.6pt,line join=round] (490.18, 30.48) --
	(490.18, 33.48);
\end{scope}
\begin{scope}
\path[clip] (  0.00,  0.00) rectangle (505.89,289.08);
\definecolor{drawColor}{gray}{0.30}

\node[text=drawColor,anchor=base,inner sep=0pt, outer sep=0pt, scale=  0.96] at (146.93, 21.46) {0\%};

\node[text=drawColor,anchor=base,inner sep=0pt, outer sep=0pt, scale=  0.96] at (232.74, 21.46) {10\%};

\node[text=drawColor,anchor=base,inner sep=0pt, outer sep=0pt, scale=  0.96] at (318.55, 21.46) {20\%};

\node[text=drawColor,anchor=base,inner sep=0pt, outer sep=0pt, scale=  0.96] at (404.37, 21.46) {30\%};

\node[text=drawColor,anchor=base,inner sep=0pt, outer sep=0pt, scale=  0.96] at (490.18, 21.46) {40\%};
\end{scope}
\begin{scope}
\path[clip] (  0.00,  0.00) rectangle (505.89,289.08);
\definecolor{drawColor}{RGB}{0,0,0}

\node[text=drawColor,anchor=base,inner sep=0pt, outer sep=0pt, scale=  1.20] at (315.00,  8.33) {Employment share by work category};
\end{scope}
\end{tikzpicture}
}

\par}
\input{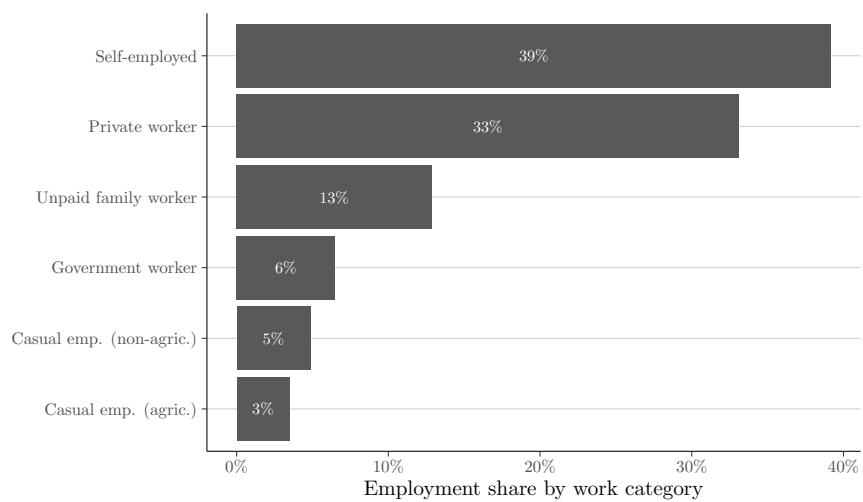}
\end{figure}

\Cref{fig:barplot-job-category-pooled} plots the share of government workers compared to the employment share of the other most common categories of work in developing countries, pooling across all workers and years.
On average, about \DescGovtSharePct\% of the workforce are government workers, around \DescPrivateFormalSharePct\% of workers have permanent, formal jobs in the private sector, roughly \DescSelfEmployedSharePct\% are self-employed, \DescUnpaidSharePct\% are unpaid family workers and the remaining \DescCasualSharePct\% are casually employed.
%{Numbers in line with govt labor market statistics?}
Compared internationally, government employment in Indonesia is relatively low as it often exceeds 10\% of the workforce for middle-income countries and may easily exceed 20\% for high-income countries \parencite{finanPersonnelEconomicsDeveloping2017}.
Based on complementary government statistics, about 1/3 of the government employees captured in the IFLS data should be non-permanent government workers.\footnote{Specifically, permanent government workers make up roughly 3.5-4\% of all employees in Indonesia based on a Civil Service census from the year 2015 reported in  \parencite{pierskallaDemocratizationRepresentativeBureaucracy2020} and the total number of employees for the same year taken from Statistics Indonesia.}

\begin{figure}[t]
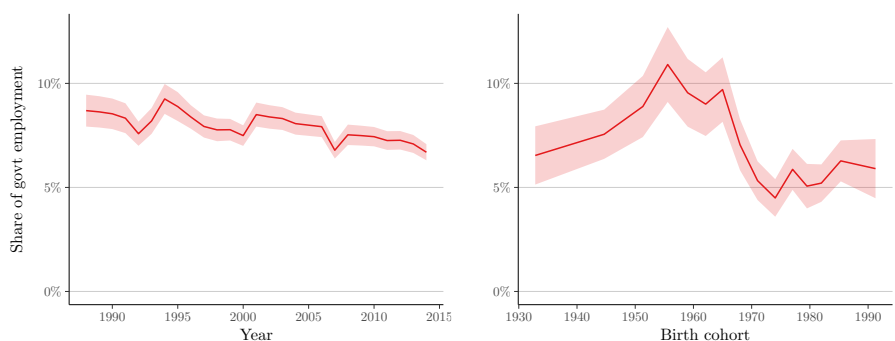

\caption{Evolution of government employment shares \label{fig:plot-govt-employment}}
{\centering
\resizebox{0.75\textwidth}{!}{
    \input{out/img/evolution_govt_employment_share.tex}
}
\par} % End the paragraph and close the group to stop centering
\input{out/fignotes/evolution_govt_employment_share.tex}
\end{figure}

\begin{figure}[t]
\caption{Employment shares across sectors of work for government vs.
private sector workers \label{fig:barplot-sectors-govt}}

{\centering

\resizebox{0.75\textwidth}{!}{
% !TEX encoding = UTF-8 Unicode
\begin{tikzpicture}[x=1pt,y=1pt]
\definecolor{fillColor}{RGB}{255,255,255}
\path[use as bounding box,fill=fillColor,fill opacity=0.00] (0,0) rectangle (505.89,289.08);
\begin{scope}
\path[clip] (  0.00,  0.00) rectangle (505.89,289.08);
\definecolor{drawColor}{RGB}{255,255,255}
\definecolor{fillColor}{RGB}{255,255,255}

\path[draw=drawColor,line width= 0.6pt,line join=round,line cap=round,fill=fillColor] (  0.00,  0.00) rectangle (505.89,289.08);
\end{scope}
\begin{scope}
\path[clip] ( 42.59, 71.93) rectangle (499.89,283.08);
\definecolor{fillColor}{RGB}{255,255,255}

\path[fill=fillColor] ( 42.59, 71.93) rectangle (499.89,283.08);
\definecolor{drawColor}{gray}{0.80}

\path[draw=drawColor,line width= 0.5pt,line join=round] ( 42.59, 81.53) --
	(499.89, 81.53);

\path[draw=drawColor,line width= 0.5pt,line join=round] ( 42.59,105.46) --
	(499.89,105.46);

\path[draw=drawColor,line width= 0.5pt,line join=round] ( 42.59,129.40) --
	(499.89,129.40);

\path[draw=drawColor,line width= 0.5pt,line join=round] ( 42.59,153.34) --
	(499.89,153.34);

\path[draw=drawColor,line width= 0.5pt,line join=round] ( 42.59,177.27) --
	(499.89,177.27);

\path[draw=drawColor,line width= 0.5pt,line join=round] ( 42.59,201.21) --
	(499.89,201.21);

\path[draw=drawColor,line width= 0.5pt,line join=round] ( 42.59,225.15) --
	(499.89,225.15);

\path[draw=drawColor,line width= 0.5pt,line join=round] ( 42.59,249.08) --
	(499.89,249.08);

\path[draw=drawColor,line width= 0.5pt,line join=round] ( 42.59,273.02) --
	(499.89,273.02);
\definecolor{fillColor}{RGB}{228,26,28}

\path[fill=fillColor] (363.15, 81.53) rectangle (383.32, 97.23);

\path[fill=fillColor] (273.48, 81.53) rectangle (293.66, 84.55);

\path[fill=fillColor] (183.82, 81.53) rectangle (203.99, 86.75);

\path[fill=fillColor] (407.98, 81.53) rectangle (428.16, 89.74);

\path[fill=fillColor] ( 94.15, 81.53) rectangle (114.33, 83.07);

\path[fill=fillColor] ( 49.32, 81.53) rectangle ( 69.49, 82.69);

\path[fill=fillColor] (318.32, 81.53) rectangle (338.49, 86.16);

\path[fill=fillColor] (452.82, 81.53) rectangle (472.99,273.48);

\path[fill=fillColor] (228.65, 81.53) rectangle (248.83, 86.87);

\path[fill=fillColor] (138.98, 81.53) rectangle (159.16, 84.09);
\definecolor{fillColor}{RGB}{55,126,184}

\path[fill=fillColor] (383.32, 81.53) rectangle (403.50,125.40);

\path[fill=fillColor] (293.66, 81.53) rectangle (313.83, 98.96);

\path[fill=fillColor] (203.99, 81.53) rectangle (224.17, 88.64);

\path[fill=fillColor] (428.16, 81.53) rectangle (448.33,145.71);

\path[fill=fillColor] (114.33, 81.53) rectangle (134.50, 83.85);

\path[fill=fillColor] ( 69.49, 81.53) rectangle ( 89.67, 83.67);

\path[fill=fillColor] (338.49, 81.53) rectangle (358.67,119.81);

\path[fill=fillColor] (472.99, 81.53) rectangle (493.17,134.72);

\path[fill=fillColor] (248.83, 81.53) rectangle (269.00, 90.96);

\path[fill=fillColor] (159.16, 81.53) rectangle (179.33, 82.91);
\end{scope}
\begin{scope}
\path[clip] (  0.00,  0.00) rectangle (505.89,289.08);
\definecolor{drawColor}{RGB}{0,0,0}

\path[draw=drawColor,line width= 0.6pt,line join=round] ( 42.59, 71.93) --
	( 42.59,283.08);
\end{scope}
\begin{scope}
\path[clip] (  0.00,  0.00) rectangle (505.89,289.08);
\definecolor{drawColor}{gray}{0.30}

\node[text=drawColor,anchor=base east,inner sep=0pt, outer sep=0pt, scale=  0.96] at ( 37.19, 78.22) {0\%};

\node[text=drawColor,anchor=base east,inner sep=0pt, outer sep=0pt, scale=  0.96] at ( 37.19,102.16) {10\%};

\node[text=drawColor,anchor=base east,inner sep=0pt, outer sep=0pt, scale=  0.96] at ( 37.19,126.09) {20\%};

\node[text=drawColor,anchor=base east,inner sep=0pt, outer sep=0pt, scale=  0.96] at ( 37.19,150.03) {30\%};

\node[text=drawColor,anchor=base east,inner sep=0pt, outer sep=0pt, scale=  0.96] at ( 37.19,173.97) {40\%};

\node[text=drawColor,anchor=base east,inner sep=0pt, outer sep=0pt, scale=  0.96] at ( 37.19,197.90) {50\%};

\node[text=drawColor,anchor=base east,inner sep=0pt, outer sep=0pt, scale=  0.96] at ( 37.19,221.84) {60\%};

\node[text=drawColor,anchor=base east,inner sep=0pt, outer sep=0pt, scale=  0.96] at ( 37.19,245.78) {70\%};

\node[text=drawColor,anchor=base east,inner sep=0pt, outer sep=0pt, scale=  0.96] at ( 37.19,269.71) {80\%};
\end{scope}
\begin{scope}
\path[clip] (  0.00,  0.00) rectangle (505.89,289.08);
\definecolor{drawColor}{gray}{0.20}

\path[draw=drawColor,line width= 0.6pt,line join=round] ( 39.59, 81.53) --
	( 42.59, 81.53);

\path[draw=drawColor,line width= 0.6pt,line join=round] ( 39.59,105.46) --
	( 42.59,105.46);

\path[draw=drawColor,line width= 0.6pt,line join=round] ( 39.59,129.40) --
	( 42.59,129.40);

\path[draw=drawColor,line width= 0.6pt,line join=round] ( 39.59,153.34) --
	( 42.59,153.34);

\path[draw=drawColor,line width= 0.6pt,line join=round] ( 39.59,177.27) --
	( 42.59,177.27);

\path[draw=drawColor,line width= 0.6pt,line join=round] ( 39.59,201.21) --
	( 42.59,201.21);

\path[draw=drawColor,line width= 0.6pt,line join=round] ( 39.59,225.15) --
	( 42.59,225.15);

\path[draw=drawColor,line width= 0.6pt,line join=round] ( 39.59,249.08) --
	( 42.59,249.08);

\path[draw=drawColor,line width= 0.6pt,line join=round] ( 39.59,273.02) --
	( 42.59,273.02);
\end{scope}
\begin{scope}
\path[clip] (  0.00,  0.00) rectangle (505.89,289.08);
\definecolor{drawColor}{RGB}{0,0,0}

\path[draw=drawColor,line width= 0.6pt,line join=round] ( 42.59, 71.93) --
	(499.89, 71.93);
\end{scope}
\begin{scope}
\path[clip] (  0.00,  0.00) rectangle (505.89,289.08);
\definecolor{drawColor}{gray}{0.20}

\path[draw=drawColor,line width= 0.6pt,line join=round] ( 69.49, 68.93) --
	( 69.49, 71.93);

\path[draw=drawColor,line width= 0.6pt,line join=round] (114.33, 68.93) --
	(114.33, 71.93);

\path[draw=drawColor,line width= 0.6pt,line join=round] (159.16, 68.93) --
	(159.16, 71.93);

\path[draw=drawColor,line width= 0.6pt,line join=round] (203.99, 68.93) --
	(203.99, 71.93);

\path[draw=drawColor,line width= 0.6pt,line join=round] (248.83, 68.93) --
	(248.83, 71.93);

\path[draw=drawColor,line width= 0.6pt,line join=round] (293.66, 68.93) --
	(293.66, 71.93);

\path[draw=drawColor,line width= 0.6pt,line join=round] (338.49, 68.93) --
	(338.49, 71.93);

\path[draw=drawColor,line width= 0.6pt,line join=round] (383.32, 68.93) --
	(383.32, 71.93);

\path[draw=drawColor,line width= 0.6pt,line join=round] (428.16, 68.93) --
	(428.16, 71.93);

\path[draw=drawColor,line width= 0.6pt,line join=round] (472.99, 68.93) --
	(472.99, 71.93);
\end{scope}
\begin{scope}
\path[clip] (  0.00,  0.00) rectangle (505.89,289.08);
\definecolor{drawColor}{gray}{0.30}

\node[text=drawColor,anchor=base,inner sep=0pt, outer sep=0pt, scale=  0.96] at ( 69.49, 59.92) {Other};

\node[text=drawColor,anchor=base,inner sep=0pt, outer sep=0pt, scale=  0.96] at (114.33, 59.92) {Mining};

\node[text=drawColor,anchor=base,inner sep=0pt, outer sep=0pt, scale=  0.96] at (159.16, 59.92) {Util.};

\node[text=drawColor,anchor=base,inner sep=0pt, outer sep=0pt, scale=  0.96] at (203.99, 59.92) {Finance};

\node[text=drawColor,anchor=base,inner sep=0pt, outer sep=0pt, scale=  0.96] at (248.83, 59.92) {Transport};

\node[text=drawColor,anchor=base,inner sep=0pt, outer sep=0pt, scale=  0.96] at (293.66, 59.92) {Constr.};

\node[text=drawColor,anchor=base,inner sep=0pt, outer sep=0pt, scale=  0.96] at (338.49, 59.92) {Retail};

\node[text=drawColor,anchor=base,inner sep=0pt, outer sep=0pt, scale=  0.96] at (383.32, 59.92) {Agric.};

\node[text=drawColor,anchor=base,inner sep=0pt, outer sep=0pt, scale=  0.96] at (428.16, 59.92) {Manuf.};

\node[text=drawColor,anchor=base,inner sep=0pt, outer sep=0pt, scale=  0.96] at (472.99, 59.92) {Social};
\end{scope}
\begin{scope}
\path[clip] (  0.00,  0.00) rectangle (505.89,289.08);
\definecolor{drawColor}{RGB}{0,0,0}

\node[text=drawColor,anchor=base,inner sep=0pt, outer sep=0pt, scale=  1.20] at (271.24, 46.79) {Broad area of work};
\end{scope}
\begin{scope}
\path[clip] (  0.00,  0.00) rectangle (505.89,289.08);
\definecolor{drawColor}{RGB}{0,0,0}

\node[text=drawColor,rotate= 90.00,anchor=base,inner sep=0pt, outer sep=0pt, scale=  1.20] at ( 14.26,177.50) {Employment share by group};
\end{scope}
\begin{scope}
\path[clip] (  0.00,  0.00) rectangle (505.89,289.08);
\definecolor{fillColor}{RGB}{255,255,255}

\path[fill=fillColor] (170.66,  6.00) rectangle (371.82, 32.45);
\end{scope}
\begin{scope}
\path[clip] (  0.00,  0.00) rectangle (505.89,289.08);
\definecolor{drawColor}{RGB}{0,0,0}

\node[text=drawColor,anchor=base west,inner sep=0pt, outer sep=0pt, scale=  1.20] at (176.66, 15.09) {comparing};
\end{scope}
\begin{scope}
\path[clip] (  0.00,  0.00) rectangle (505.89,289.08);
\definecolor{fillColor}{RGB}{255,255,255}

\path[fill=fillColor] (237.35, 12.00) rectangle (251.80, 26.45);
\end{scope}
\begin{scope}
\path[clip] (  0.00,  0.00) rectangle (505.89,289.08);
\definecolor{fillColor}{RGB}{228,26,28}

\path[fill=fillColor] (238.06, 12.71) rectangle (251.09, 25.74);
\end{scope}
\begin{scope}
\path[clip] (  0.00,  0.00) rectangle (505.89,289.08);
\definecolor{fillColor}{RGB}{255,255,255}

\path[fill=fillColor] (315.08, 12.00) rectangle (329.54, 26.45);
\end{scope}
\begin{scope}
\path[clip] (  0.00,  0.00) rectangle (505.89,289.08);
\definecolor{fillColor}{RGB}{55,126,184}

\path[fill=fillColor] (315.79, 12.71) rectangle (328.83, 25.74);
\end{scope}
\begin{scope}
\path[clip] (  0.00,  0.00) rectangle (505.89,289.08);
\definecolor{drawColor}{RGB}{0,0,0}

\node[text=drawColor,anchor=base west,inner sep=0pt, outer sep=0pt, scale=  0.96] at (257.80, 15.92) {Government};
\end{scope}
\begin{scope}
\path[clip] (  0.00,  0.00) rectangle (505.89,289.08);
\definecolor{drawColor}{RGB}{0,0,0}

\node[text=drawColor,anchor=base west,inner sep=0pt, outer sep=0pt, scale=  0.96] at (335.54, 15.92) {Private};
\end{scope}
\end{tikzpicture}
}

\par}
\input{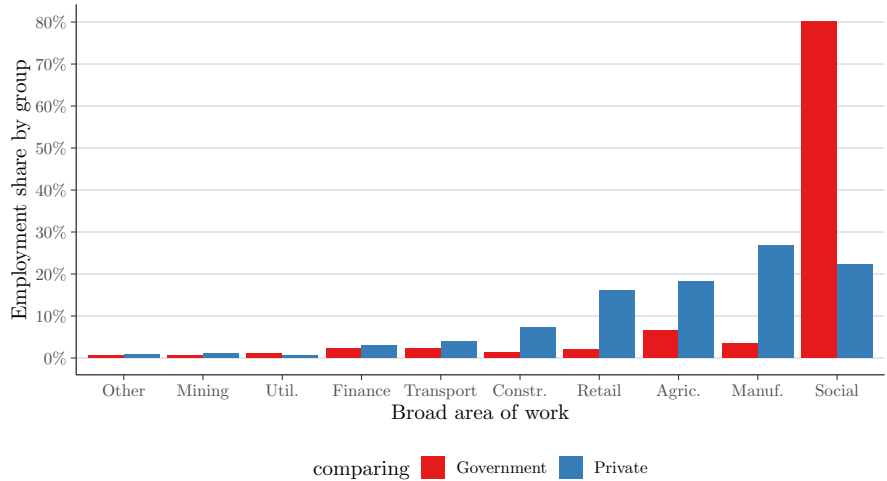}
\end{figure}

\Cref{fig:plot-govt-employment} shows the evolution of the government workforce and hiring over time.
In line with the stated government policy of reducing government employment, the left panel shows that the share of government workers has been slowly declining since the early 1990s.
The decline also shows up in the data through a large drop of hiring for birth cohorts after 1965 as reported in the right panel of \Cref{fig:plot-govt-employment}.
Next, we can ask what types of jobs government workers are doing in Indonesia.
\Cref{fig:barplot-sectors-govt} shows the distribution of government jobs across 10 different sectors and compares their shares to the private sector.
About \DescGovtSocialServicesPct\% of government sector jobs are in social services, including education and health, with the next biggest sector being agriculture and forestry at around \DescGovtAgriculturePct\%.
In comparison, around \DescPrivateSocialServicesPct\% of private sector jobs are in social services, while around \DescPrivateManufacturingPct\% are in manufacturing.

\begin{figure}[t]
\caption{Government employment shares by occupation \label{fig:barplot-occupation-govt}}

{\centering

\resizebox{0.75\textwidth}{!}{
% !TEX encoding = UTF-8 Unicode
\begin{tikzpicture}[x=1pt,y=1pt]
\definecolor{fillColor}{RGB}{255,255,255}
\path[use as bounding box,fill=fillColor,fill opacity=0.00] (0,0) rectangle (505.89,289.08);
\begin{scope}
\path[clip] (  0.00,  0.00) rectangle (505.89,289.08);
\definecolor{drawColor}{RGB}{255,255,255}
\definecolor{fillColor}{RGB}{255,255,255}

\path[draw=drawColor,line width= 0.6pt,line join=round,line cap=round,fill=fillColor] (  0.00,  0.00) rectangle (505.89,289.08);
\end{scope}
\begin{scope}
\path[clip] (105.94, 33.48) rectangle (499.89,283.08);
\definecolor{fillColor}{RGB}{255,255,255}

\path[fill=fillColor] (105.94, 33.48) rectangle (499.89,283.08);
\definecolor{drawColor}{gray}{0.80}

\path[draw=drawColor,line width= 0.5pt,line join=round] (105.94, 49.75) --
	(499.89, 49.75);

\path[draw=drawColor,line width= 0.5pt,line join=round] (105.94, 76.88) --
	(499.89, 76.88);

\path[draw=drawColor,line width= 0.5pt,line join=round] (105.94,104.02) --
	(499.89,104.02);

\path[draw=drawColor,line width= 0.5pt,line join=round] (105.94,131.15) --
	(499.89,131.15);

\path[draw=drawColor,line width= 0.5pt,line join=round] (105.94,158.28) --
	(499.89,158.28);

\path[draw=drawColor,line width= 0.5pt,line join=round] (105.94,185.41) --
	(499.89,185.41);

\path[draw=drawColor,line width= 0.5pt,line join=round] (105.94,212.54) --
	(499.89,212.54);

\path[draw=drawColor,line width= 0.5pt,line join=round] (105.94,239.67) --
	(499.89,239.67);

\path[draw=drawColor,line width= 0.5pt,line join=round] (105.94,266.80) --
	(499.89,266.80);
\definecolor{fillColor}{gray}{0.35}

\path[fill=fillColor] (123.85,118.94) rectangle (166.17,143.36);

\path[fill=fillColor] (123.85,173.20) rectangle (204.23,197.62);

\path[fill=fillColor] (123.85, 91.81) rectangle (164.12,116.22);

\path[fill=fillColor] (123.85,146.07) rectangle (187.34,170.49);

\path[fill=fillColor] (123.85, 64.68) rectangle (150.58, 89.09);

\path[fill=fillColor] (123.85,254.59) rectangle (481.98,279.01);

\path[fill=fillColor] (123.85, 37.54) rectangle (141.90, 61.96);

\path[fill=fillColor] (123.85,200.33) rectangle (214.54,224.75);

\path[fill=fillColor] (123.85,227.46) rectangle (363.30,251.88);
\end{scope}
\begin{scope}
\path[clip] (  0.00,  0.00) rectangle (505.89,289.08);
\definecolor{drawColor}{RGB}{0,0,0}

\path[draw=drawColor,line width= 0.6pt,line join=round] (105.94, 33.48) --
	(105.94,283.08);
\end{scope}
\begin{scope}
\path[clip] (  0.00,  0.00) rectangle (505.89,289.08);
\definecolor{drawColor}{gray}{0.30}

\node[text=drawColor,anchor=base east,inner sep=0pt, outer sep=0pt, scale=  0.96] at (100.54, 46.45) {Sales};

\node[text=drawColor,anchor=base east,inner sep=0pt, outer sep=0pt, scale=  0.96] at (100.54, 73.58) {Military and Police};

\node[text=drawColor,anchor=base east,inner sep=0pt, outer sep=0pt, scale=  0.96] at (100.54,100.71) {Health};

\node[text=drawColor,anchor=base east,inner sep=0pt, outer sep=0pt, scale=  0.96] at (100.54,127.84) {Agriculture};

\node[text=drawColor,anchor=base east,inner sep=0pt, outer sep=0pt, scale=  0.96] at (100.54,154.97) {Laborer};

\node[text=drawColor,anchor=base east,inner sep=0pt, outer sep=0pt, scale=  0.96] at (100.54,182.10) {Bureaucrat};

\node[text=drawColor,anchor=base east,inner sep=0pt, outer sep=0pt, scale=  0.96] at (100.54,209.23) {Service};

\node[text=drawColor,anchor=base east,inner sep=0pt, outer sep=0pt, scale=  0.96] at (100.54,236.36) {Teacher};

\node[text=drawColor,anchor=base east,inner sep=0pt, outer sep=0pt, scale=  0.96] at (100.54,263.50) {Other};
\end{scope}
\begin{scope}
\path[clip] (  0.00,  0.00) rectangle (505.89,289.08);
\definecolor{drawColor}{gray}{0.20}

\path[draw=drawColor,line width= 0.6pt,line join=round] (102.94, 49.75) --
	(105.94, 49.75);

\path[draw=drawColor,line width= 0.6pt,line join=round] (102.94, 76.88) --
	(105.94, 76.88);

\path[draw=drawColor,line width= 0.6pt,line join=round] (102.94,104.02) --
	(105.94,104.02);

\path[draw=drawColor,line width= 0.6pt,line join=round] (102.94,131.15) --
	(105.94,131.15);

\path[draw=drawColor,line width= 0.6pt,line join=round] (102.94,158.28) --
	(105.94,158.28);

\path[draw=drawColor,line width= 0.6pt,line join=round] (102.94,185.41) --
	(105.94,185.41);

\path[draw=drawColor,line width= 0.6pt,line join=round] (102.94,212.54) --
	(105.94,212.54);

\path[draw=drawColor,line width= 0.6pt,line join=round] (102.94,239.67) --
	(105.94,239.67);

\path[draw=drawColor,line width= 0.6pt,line join=round] (102.94,266.80) --
	(105.94,266.80);
\end{scope}
\begin{scope}
\path[clip] (  0.00,  0.00) rectangle (505.89,289.08);
\definecolor{drawColor}{RGB}{0,0,0}

\path[draw=drawColor,line width= 0.6pt,line join=round] (105.94, 33.48) --
	(499.89, 33.48);
\end{scope}
\begin{scope}
\path[clip] (  0.00,  0.00) rectangle (505.89,289.08);
\definecolor{drawColor}{gray}{0.20}

\path[draw=drawColor,line width= 0.6pt,line join=round] (123.85, 30.48) --
	(123.85, 33.48);

\path[draw=drawColor,line width= 0.6pt,line join=round] (219.80, 30.48) --
	(219.80, 33.48);

\path[draw=drawColor,line width= 0.6pt,line join=round] (315.76, 30.48) --
	(315.76, 33.48);

\path[draw=drawColor,line width= 0.6pt,line join=round] (411.71, 30.48) --
	(411.71, 33.48);
\end{scope}
\begin{scope}
\path[clip] (  0.00,  0.00) rectangle (505.89,289.08);
\definecolor{drawColor}{gray}{0.30}

\node[text=drawColor,anchor=base,inner sep=0pt, outer sep=0pt, scale=  0.96] at (123.85, 21.46) {0\%};

\node[text=drawColor,anchor=base,inner sep=0pt, outer sep=0pt, scale=  0.96] at (219.80, 21.46) {10\%};

\node[text=drawColor,anchor=base,inner sep=0pt, outer sep=0pt, scale=  0.96] at (315.76, 21.46) {20\%};

\node[text=drawColor,anchor=base,inner sep=0pt, outer sep=0pt, scale=  0.96] at (411.71, 21.46) {30\%};
\end{scope}
\begin{scope}
\path[clip] (  0.00,  0.00) rectangle (505.89,289.08);
\definecolor{drawColor}{RGB}{0,0,0}

\node[text=drawColor,anchor=base,inner sep=0pt, outer sep=0pt, scale=  1.20] at (302.91,  8.33) {Employment share by occupation};
\end{scope}
\begin{scope}
\path[clip] (  0.00,  0.00) rectangle (505.89,289.08);
\definecolor{drawColor}{RGB}{0,0,0}

\node[text=drawColor,rotate= 90.00,anchor=base,inner sep=0pt, outer sep=0pt, scale=  1.20] at ( 14.26,158.28) {Occupation};
\end{scope}
\end{tikzpicture}
}

\par}
\input{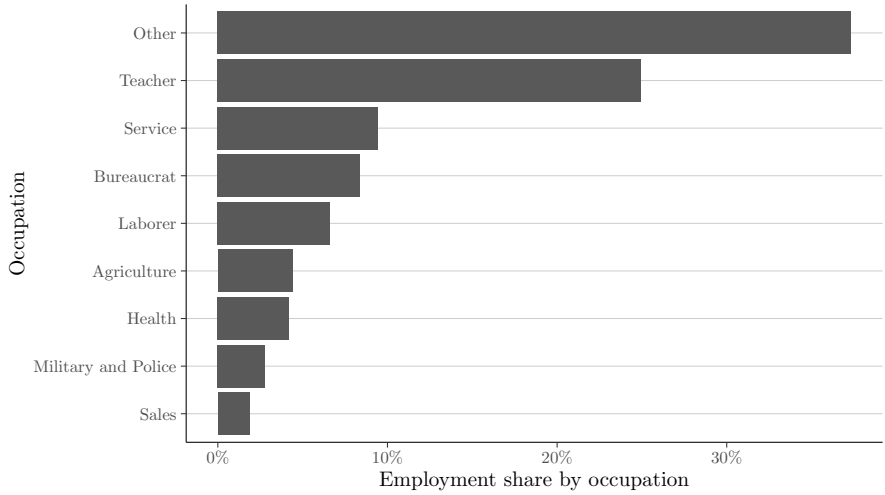}
\end{figure}

In \Cref{fig:barplot-occupation-govt}, we provide further details on which occupations government workers hold.
About \DescGovtTeachersPct\% of government workers are teachers, about \DescGovtOfficialsPct\% are government officials and executives, and about \DescGovtOtherServicePct\% of government workers work in other service areas.
Not surprisingly, the prevalence of the same occupations among private sector workers looks very different: while there are almost no private military/police nor bureaucrats and about twice as many government teachers than private teachers, there are $\DescLaborerRatio{}\times$ more private laborers and private sales workers and $\DescAgricRatio{}\times$ more private agricultural workers.

Another important feature of the data on government employment are transition rates between the private and the public sector, defined as individuals who are observed switching either from a non-government job to a government job (in-mover) or the reverse (out-mover).
About \DescMoverSharePct\% of person-year observations with government work history are movers, split almost evenly between in- and out-movers (\DescInMoverSharePct\% vs.~\DescOutMoverSharePct\%).
Among unique movers, about \DescBothWayMoverPct\% are observed moving both in and out of the government.

At last, \Cref{tab:table-main-skill-differences} reports how government workers differ from other workers in the economy in terms of observables, including skill variables. Specifically, we compare government workers to private sector workers, other workers (non-private and non-government), and the subsequent control group: private sector workers with similar occupations, sector of work, gender and age as government workers. The next section provides more details on how the control group is constructed.
In contrast to private sector workers, government workers are on average slightly more likely to be male, \DescAgeDiffYears{} years older, earn about \DescWageRatioPct\% higher hourly wages and work slightly less hours.
Importantly, even in contrast to the re-weighted control group, government workers are on average much more skilled: they are much more likely to have received higher education, and perform much better on self-administered memory and word ability tasks.
While not reported here, these skill measures are mostly increasing for all groups of workers over time.
%{Add graphs to Appendix, add risk aversion and Big 5}

\subsection{Defining comparable private sector jobs} \label{step-1-propensity-score}

Before turning to our estimation approach, we need to define our estimation sample and thus the notion of \emph{comparable private sector jobs}. These are jobs that are similar to the jobs public sector workers are doing in terms of the skills that are required and used on the job.
Technically, they are jobs for which we assume that the mapping $f(x)$ from observables to skills is similar to government jobs.
Our focus in the paper is on a set of general skills that hold across a set of different related government jobs.
For this reason, we define comparable jobs in the private sector as jobs in similar occupations and sectors as government jobs, and with a similar set of worker demographics.
Specifically, we use propensity score weighting, where propensity scores are derived from a logistic regression predicting the dummy ``government worker'' using as covariates the occupation, sector of work, age and sex of a worker and restricting to government and private sector workers only.
To ensure that the estimation sample is indeed comparable to public sector workers, we reweight private-sector workers using $p(X_i)/(1-p(X_i))$, where $p(X_i)$ is the propensity score.\footnote{These design-based weights are multiplied with standard survey weights, which we use throughout. } This is analogous to the estimation of an average treatment effect on the treated (ATT).
% Denoting the estimated propensity score by $\hat{p}(X_i)$, private-sector workers receive the weight $\hat{p}(X_i)/(1-\hat{p}(X_i))$ multiplied by their survey weight, while government workers keep their survey weight unchanged.
% The $p/(1-p)$ weighting ensures that the reweighted private-sector sample matches the covariate distribution of government workers, not the overall population.
% To see why, note that weighting private-sector observations by just $p$ would recover the population density of covariates, whereas dividing by the probability of being in the private sector $(1-p)$ further tilts the distribution towards workers whose characteristics make them look like government workers.
% Formally, the reweighted density of any covariate $X$ satisfies $g_{\text{control}}(X) \propto g(X \mid \text{private}) \cdot \frac{p(X)}{1-p(X)} = g(X \mid \text{govt})$, so that comparisons between government workers and the reweighted control group estimate the ATT.
Importantly, we drop from the outset all self-employed as well as casually employed and unpaid family members, as their jobs are not sufficiently comparable.

\begin{figure}[t]
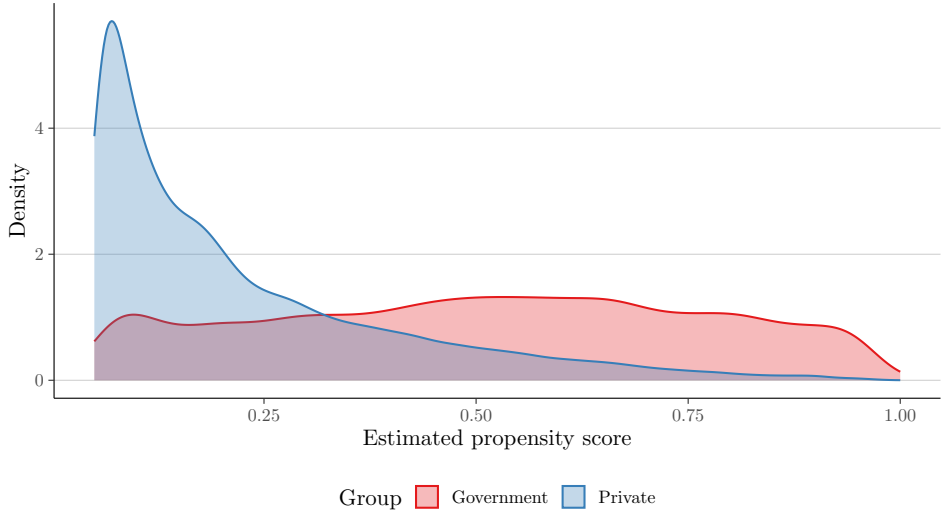

\caption{Distribution of propensity scores for being a government worker}
\label{fig:propensity-score}

{\centering

\resizebox{0.8\textwidth}{!}{
\input{out/img/propensity_score_distribution}
}

\par}
\input{out/fignotes/propensity_score_distribution.tex}
\end{figure}

\Cref{fig:propensity-score} shows the resulting distribution of the propensity score weights for public and private sector workers,
while \Cref{tab:propensity-score-logit} reports all regression coefficient estimates.
Not surprisingly, many private sector workers have very low estimated propensity scores as they work in sectors such as manufacturing, retail and construction, or in occupations that are different from government workers.
However, there is a sizeable share of private sector workers in social services and who have office and service jobs that are similar to government jobs, who have much higher estimated propensity scores and effectively form our estimation sample.
Importantly, there is common support over the entire range of propensity scores even though the right tail thins out. \Cref{fig:support-barcharts-dummies,fig:support-relative-density} in \Cref{appendix-common-support} similarly show common support over all skill variables we use.

With larger sample sizes, one could also separately apply the estimation approach to more disaggregated job categories without pooling across jobs.
The level of aggregation that we choose alleviates the difficulty of finding exact private sector counterparts to specialized jobs such as the police.
However, the underlying assumption is that even in such jobs, the general skill set needed broadly aligns with the skills needed for social service jobs in the private sector.
As explained in more detail in \Cref{sec:govt-hiring}, civil servants in Indonesia are also not trained as specialists, but rather go through a common selection and training process that emphasizes more general skills, which is exactly the skills that the estimation approach in this paper seeks to pick up.
Further below, we explicitly show that our estimation picks up on "general skills" and validate the comparison group of jobs by estimating skills for different subsets of comparable private-sector jobs.

% Another critique may be that the more multi-dimensional nature of government jobs makes performance measures difficult in the public sector \parencite{finanPersonnelEconomicsDeveloping2017}.
% However, the same argument can be made for social service jobs in the private sector.
% And importantly, the skill estimation approach here specifically allows for the idea that measuring performance in a job is difficult and that wages are only noisy signals of underlying human capital and skills.
% In the results section, we will specifically test the informativeness of wages for underlying skills and compare the private and public sector.

\subsection{Estimating skills} \label{step-2-estimating skills}

After specifying the estimation sample, we follow the estimation strategy introduced in \Cref{sec:methodology}. In all estimation steps, we additionally control for province fixed effects to ensure that estimated skills are not simply capturing geographic differences in wages.

% We start by backing out the skill price $\widehat{p_t}$ for government-like jobs in the private sector.
% We then estimate the experience profile and predict skills.
% For the baseline estimation, we abstract from selection on unobservables, but in \Cref{sec:robustness} we show that results are robust to correcting for selection.

\paragraph{Skill price estimation}

\begin{figure}[t]
\caption{Evolution of skill price for comparable jobs in the private sector \label{fig:skill-price}}

{\centering

\resizebox{0.75\textwidth}{!}{
% !TEX encoding = UTF-8 Unicode
\begin{tikzpicture}[x=1pt,y=1pt]
\definecolor{fillColor}{RGB}{255,255,255}
\path[use as bounding box,fill=fillColor,fill opacity=0.00] (0,0) rectangle (505.89,245.72);
\begin{scope}
\path[clip] (  0.00,  0.00) rectangle (505.89,245.72);
\definecolor{drawColor}{RGB}{255,255,255}
\definecolor{fillColor}{RGB}{255,255,255}

\path[draw=drawColor,line width= 0.6pt,line join=round,line cap=round,fill=fillColor] (  0.00,  0.00) rectangle (505.89,245.72);
\end{scope}
\begin{scope}
\path[clip] ( 37.26, 33.48) rectangle (499.89,239.72);
\definecolor{fillColor}{RGB}{255,255,255}

\path[fill=fillColor] ( 37.26, 33.48) rectangle (499.89,239.72);
\definecolor{drawColor}{gray}{0.80}

\path[draw=drawColor,line width= 0.5pt,line join=round] ( 37.26, 74.88) --
	(499.89, 74.88);

\path[draw=drawColor,line width= 0.5pt,line join=round] ( 37.26,132.38) --
	(499.89,132.38);

\path[draw=drawColor,line width= 0.5pt,line join=round] ( 37.26,189.88) --
	(499.89,189.88);
\definecolor{fillColor}{RGB}{153,153,153}

\path[fill=fillColor,fill opacity=0.30] ( 58.29,132.38) --
	( 75.11,125.51) --
	( 91.94,116.02) --
	(108.76,113.11) --
	(125.58,111.43) --
	(142.40,103.88) --
	(159.23, 93.79) --
	(176.05, 88.37) --
	(192.87, 81.67) --
	(209.70,105.41) --
	(226.52, 82.48) --
	(243.34, 78.44) --
	(361.10,132.86) --
	(478.86,230.34) --
	(478.86, 52.20) --
	(361.10, 45.19) --
	(243.34, 42.85) --
	(226.52, 44.98) --
	(209.70, 55.32) --
	(192.87, 45.45) --
	(176.05, 48.47) --
	(159.23, 51.08) --
	(142.40, 55.61) --
	(125.58, 60.62) --
	(108.76,100.58) --
	( 91.94,104.52) --
	( 75.11,114.67) --
	( 58.29,132.38) --
	cycle;

\path[] ( 58.29,132.38) --
	( 75.11,125.51) --
	( 91.94,116.02) --
	(108.76,113.11) --
	(125.58,111.43) --
	(142.40,103.88) --
	(159.23, 93.79) --
	(176.05, 88.37) --
	(192.87, 81.67) --
	(209.70,105.41) --
	(226.52, 82.48) --
	(243.34, 78.44) --
	(361.10,132.86) --
	(478.86,230.34);

\path[] (478.86, 52.20) --
	(361.10, 45.19) --
	(243.34, 42.85) --
	(226.52, 44.98) --
	(209.70, 55.32) --
	(192.87, 45.45) --
	(176.05, 48.47) --
	(159.23, 51.08) --
	(142.40, 55.61) --
	(125.58, 60.62) --
	(108.76,100.58) --
	( 91.94,104.52) --
	( 75.11,114.67) --
	( 58.29,132.38);
\definecolor{drawColor}{RGB}{228,26,28}
\definecolor{fillColor}{RGB}{228,26,28}

\path[draw=drawColor,line width= 0.4pt,line join=round,line cap=round,fill=fillColor] ( 58.29,132.38) circle (  2.50);

\path[draw=drawColor,line width= 0.4pt,line join=round,line cap=round,fill=fillColor] ( 75.11,119.95) circle (  2.50);

\path[draw=drawColor,line width= 0.4pt,line join=round,line cap=round,fill=fillColor] ( 91.94,110.10) circle (  2.50);

\path[draw=drawColor,line width= 0.4pt,line join=round,line cap=round,fill=fillColor] (108.76,106.63) circle (  2.50);

\path[draw=drawColor,line width= 0.4pt,line join=round,line cap=round,fill=fillColor] (125.58, 81.15) circle (  2.50);

\path[draw=drawColor,line width= 0.4pt,line join=round,line cap=round,fill=fillColor] (142.40, 74.89) circle (  2.50);

\path[draw=drawColor,line width= 0.4pt,line join=round,line cap=round,fill=fillColor] (159.23, 68.12) circle (  2.50);

\path[draw=drawColor,line width= 0.4pt,line join=round,line cap=round,fill=fillColor] (176.05, 64.36) circle (  2.50);

\path[draw=drawColor,line width= 0.4pt,line join=round,line cap=round,fill=fillColor] (192.87, 59.86) circle (  2.50);

\path[draw=drawColor,line width= 0.4pt,line join=round,line cap=round,fill=fillColor] (209.70, 75.17) circle (  2.50);

\path[draw=drawColor,line width= 0.4pt,line join=round,line cap=round,fill=fillColor] (226.52, 59.77) circle (  2.50);

\path[draw=drawColor,line width= 0.4pt,line join=round,line cap=round,fill=fillColor] (243.34, 56.82) circle (  2.50);

\path[draw=drawColor,line width= 0.4pt,line join=round,line cap=round,fill=fillColor] (361.10, 74.05) circle (  2.50);

\path[draw=drawColor,line width= 0.4pt,line join=round,line cap=round,fill=fillColor] (478.86,103.49) circle (  2.50);

\path[draw=drawColor,line width= 1.1pt,line join=round] ( 58.29,132.38) --
	( 75.11,119.95) --
	( 91.94,110.10) --
	(108.76,106.63) --
	(125.58, 81.15) --
	(142.40, 74.89) --
	(159.23, 68.12) --
	(176.05, 64.36) --
	(192.87, 59.86) --
	(209.70, 75.17) --
	(226.52, 59.77) --
	(243.34, 56.82) --
	(361.10, 74.05) --
	(478.86,103.49);
\definecolor{drawColor}{RGB}{0,0,0}

\path[draw=drawColor,line width= 1.1pt,dash pattern=on 4pt off 4pt ,line join=round] ( 37.26,132.38) -- (499.89,132.38);
\end{scope}
\begin{scope}
\path[clip] (  0.00,  0.00) rectangle (505.89,245.72);
\definecolor{drawColor}{RGB}{0,0,0}

\path[draw=drawColor,line width= 0.6pt,line join=round] ( 37.26, 33.48) --
	( 37.26,239.72);
\end{scope}
\begin{scope}
\path[clip] (  0.00,  0.00) rectangle (505.89,245.72);
\definecolor{drawColor}{gray}{0.30}

\node[text=drawColor,anchor=base east,inner sep=0pt, outer sep=0pt, scale=  0.96] at ( 31.86, 71.57) {0.5};

\node[text=drawColor,anchor=base east,inner sep=0pt, outer sep=0pt, scale=  0.96] at ( 31.86,129.07) {1.0};

\node[text=drawColor,anchor=base east,inner sep=0pt, outer sep=0pt, scale=  0.96] at ( 31.86,186.57) {1.5};
\end{scope}
\begin{scope}
\path[clip] (  0.00,  0.00) rectangle (505.89,245.72);
\definecolor{drawColor}{gray}{0.20}

\path[draw=drawColor,line width= 0.6pt,line join=round] ( 34.26, 74.88) --
	( 37.26, 74.88);

\path[draw=drawColor,line width= 0.6pt,line join=round] ( 34.26,132.38) --
	( 37.26,132.38);

\path[draw=drawColor,line width= 0.6pt,line join=round] ( 34.26,189.88) --
	( 37.26,189.88);
\end{scope}
\begin{scope}
\path[clip] (  0.00,  0.00) rectangle (505.89,245.72);
\definecolor{drawColor}{RGB}{0,0,0}

\path[draw=drawColor,line width= 0.6pt,line join=round] ( 37.26, 33.48) --
	(499.89, 33.48);
\end{scope}
\begin{scope}
\path[clip] (  0.00,  0.00) rectangle (505.89,245.72);
\definecolor{drawColor}{gray}{0.20}

\path[draw=drawColor,line width= 0.6pt,line join=round] ( 75.11, 30.48) --
	( 75.11, 33.48);

\path[draw=drawColor,line width= 0.6pt,line join=round] (159.23, 30.48) --
	(159.23, 33.48);

\path[draw=drawColor,line width= 0.6pt,line join=round] (243.34, 30.48) --
	(243.34, 33.48);

\path[draw=drawColor,line width= 0.6pt,line join=round] (327.46, 30.48) --
	(327.46, 33.48);

\path[draw=drawColor,line width= 0.6pt,line join=round] (411.57, 30.48) --
	(411.57, 33.48);

\path[draw=drawColor,line width= 0.6pt,line join=round] (495.68, 30.48) --
	(495.68, 33.48);
\end{scope}
\begin{scope}
\path[clip] (  0.00,  0.00) rectangle (505.89,245.72);
\definecolor{drawColor}{gray}{0.30}

\node[text=drawColor,anchor=base,inner sep=0pt, outer sep=0pt, scale=  0.96] at ( 75.11, 21.46) {1990};

\node[text=drawColor,anchor=base,inner sep=0pt, outer sep=0pt, scale=  0.96] at (159.23, 21.46) {1995};

\node[text=drawColor,anchor=base,inner sep=0pt, outer sep=0pt, scale=  0.96] at (243.34, 21.46) {2000};

\node[text=drawColor,anchor=base,inner sep=0pt, outer sep=0pt, scale=  0.96] at (327.46, 21.46) {2005};

\node[text=drawColor,anchor=base,inner sep=0pt, outer sep=0pt, scale=  0.96] at (411.57, 21.46) {2010};

\node[text=drawColor,anchor=base,inner sep=0pt, outer sep=0pt, scale=  0.96] at (495.68, 21.46) {2015};
\end{scope}
\begin{scope}
\path[clip] (  0.00,  0.00) rectangle (505.89,245.72);
\definecolor{drawColor}{RGB}{0,0,0}

\node[text=drawColor,anchor=base,inner sep=0pt, outer sep=0pt, scale=  1.20] at (268.58,  8.33) {Year};
\end{scope}
\begin{scope}
\path[clip] (  0.00,  0.00) rectangle (505.89,245.72);
\definecolor{drawColor}{RGB}{0,0,0}

\node[text=drawColor,rotate= 90.00,anchor=base,inner sep=0pt, outer sep=0pt, scale=  1.20] at ( 14.26,136.60) {Skill price};
\end{scope}
\end{tikzpicture}
}

\par}
\input{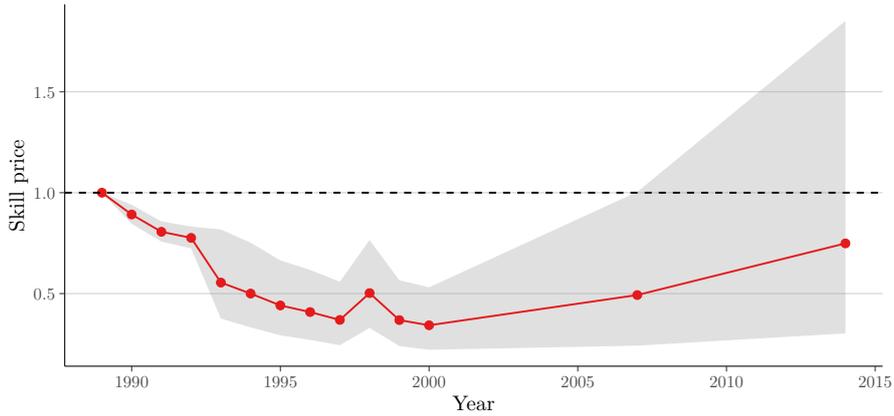}
\end{figure}

We start by purging the observed wage from changes in the equilibrium skill price for government-like jobs in the private sector.
The skill price captures the systematic component in wages that varies over time due to changes in supply and demand for skills.
Purging these price changes is important to be able to compare skills across time; i.e.
to estimate the mapping from observables to skills, $f(x)$, on wage data that spans across time.
As described in \Cref{sec:methodology}, we do so using the popular flat-spot identification approach by \textcite{bowlusHumanCapitalPrices2012}, but in \Cref{sec:skill-generality} we show that our final skill estimates are very similar when using the alternative "hedonic price regression approach".
\textcite{bowlusHumanCapitalPrices2012} propose to use within-individual wage changes for workers who are around their "flat-spot region", the experience range in which workers' human capital accumulation plateaus.
The idea is that in this experience range, average or median changes to worker hourly wages are only driven by changes in the price of skills rather than changes in their quantity.
We implement the flat spot using experience as measured by years since labor market entry. We choose a flat spot region from 22 to 34 years of experience; for a worker entering the labor force at age 22, the flat spot is then between the ages of 44 and 56 years. Importantly, this choice of flat spot is consistent with the wage experience profile estimated below (see \Cref{fig:wage-profile}), which implies that growth in human capital is indeed close to zero within this range.
We use median instead of average within-individual wage changes to reduce the role of outliers, but show in Appendix \Cref{fig:skill-price-comparison} that the skill price is statistically indistinguishable when using average changes. To control for potential differences in the equilibrium skill price across geography we include province fixed-effects when estimating within-worker changes.

\Cref{fig:skill-price} plots our estimated skill price. Overall, we find the skill price follows a U-shaped pattern.
There is a relative decline until the Asian Financial Crisis with a rebound since then.
The initial relative decline is in line with an increase in labor supply due to the large school expansion programs that Indonesia started in the 1970s and whose treated cohorts entered the labor market in the late 80s and early 1990s \parencite{dufloSchoolingLaborMarket2001,dufloMediumRunEffects2004}.
There is also a sizable drop in 1998 due to the Asian Financial Crisis that hit in late 1997 and depressed economy-wide demand.
With the recovery from the Asian Financial Crisis after the year 2000, the relative skill price increases again, in line with strong economy-wide demand outpacing supply as the economy grows and increasingly shifts towards services.
Standard errors are increasing because we show the cumulative estimate.
Due to larger standard errors, the evolution for later years is statistically less clear, but we show in  \Cref{sec:skill-generality} that our subsequent results seem not to be sensitive to different skill price estimates.

\paragraph{Experience profile estimates.}

Next, we form the price-adjusted log wage $\widetilde{w_{it}}$ by differencing out changes in the skill price $\widehat{p_t}$, and estimate the experience profile.
As in a standard Mincerian regression, we assume a quadratic experience profile:
\begin{equation}
  \log h_{it} = \delta_0 e_{it} + \delta_1 e_{it}^2,
  \label{eq:experience-profile}
\end{equation}
We then estimate this experience profile on $\widetilde{w_{it}}$, controlling for province and individual fixed-effects to ensure our estimates are consistent for any $f(x_i)$.
\Cref{tab:table-skill-fe} reports our main estimates for \((\delta_0,\delta_1)\) and \Cref{fig:wage-profile} shows that the quadratic experience profile fits (residualized) wage changes reasonably well.
The estimated parameters imply standard concave experience profiles that are close to flat around the ``flat spot range'' assumed in the previous step.
Within-individual variation is key to ensure that estimated experience profiles are not biased by systematic composition changes in skills at labor market entry, such as less experienced cohorts entering the labor market with better education.

\paragraph{Estimation of skills.}

In the last step, we flexibly estimate true skills $f(x_i)$ from the noisy average individual wage residual $\hat{z}_i = \frac{1}{N_i}\sum_{t} \big( \widetilde{w_{it}} -\widehat{h(e_{it})} \big)$. $\hat{z}_i$ is net of  experience effects and equilibrium changes in the skill price, but only a noisy observation of true skills given the importance of other non-skill-related determinants of wages.\footnote{As for the skill price estimation, we additionally residualize the individual fixed effect $\hat{z}_i$ on province fixed effects to ensure that we predict skills and do not simply pick up on wage differences across provinces.}
Technically, the individual average only nets out the error $u_{it}$ in expectation, which is noisy given that most individuals are not observed for many periods.
To correctly predict true skills $f(x_i)$ and separate signal from noise, we draw on standard off-the-shelf machine learning algorithms.
Specifically, we compare three different machine learning algorithms with two benchmark algorithms.
For the benchmarks, we use (1) a standard OLS regression with dummies for the educational background (less than primary, primary, junior secondary, senior secondary, higher education, missing, and Islamic education), and (2) a full basis expansion including all variables, their squared terms, and all pairwise first-order interaction terms.
We use the basis expansion to show the high risk of overfitting in our context with limited observations and many possible variables.
The three machine learning algorithms we draw on are LASSO (similarly using all variables and their first-order interaction terms), Random Forest and a Gradient-boosted model (GBM).

Due to the high flexibility of machine learning algorithms, it is important to avoid overfitting, which leads to picking up noise in small sample estimation.
We use 10-fold cross validation for training the hyperparameters of the machine learning algorithms using a simple grid-search over the possible hyperparameters.
As the objective, we use the average out-of-fold $R^2$, and we use the same folds across all methods to ensure comparability.
Given the limited sample size, we do not additionally split the data into training and evaluation sets.
The final individual-level dataset for \emph{training} the skill estimation algorithm includes \TrainingSampleSize{} unique individuals who carry different weights based on their jobs' similarity to government worker jobs.

% We use a single 10-fold split of the private-sector training sample at the worker level and compute out-of-fold predictions for all methods.
% For LASSO, we set the penalty to $\lambda_{\min}$ from 10-fold CV; RF and GBM are tuned by an expanded grid search. In all cases, tuning maximizes out-of-fold \emph{outcome} $R^2$ for $\widetilde{w_{it}}$ at the observation level.
% Reported Total and fixed-effect $R^2$ and RMSE values are computed out-of-fold using the same folds for all methods. Coefficients in the experience-profile table are estimated on the full sample.

% We report two relevant $R^2$ measures and their RMSE counterparts: (i) \emph{Total $R^2$} (and Total RMSE) are the overall (weighted) out-of-fold fit metrics for predicting $\widetilde{w_{it}}$; and (ii) \emph{Fixed-effect (skill) $R^2$} (and Fixed-effect RMSE) are out-of-fold diagnostics from predicting worker-level skills (computed as price-deflated wages net of the parametric terms and aggregated to the worker level) using $\widehat{f(x)}$, and therefore directly measure how informative observables are about permanent skills.

\begin{table}
\centering
\begin{talltblr}[         %% tabularray outer open
caption={Skill estimate correlations and prediction accuracy by method},
note{}={\textbf{Notes:} \textit{Correlations are Pearson correlations. The in-sample R2 uses fitted values for the model trained on the full sample. The reported out-of-fold R2 is the average out-of-fold R2 across all folds. All R2 are computed using: R2 = 1 - SSE/TSS. Correlations and R2 use survey weights.}},
label={tab:table-correlation-skill-estimates},
]                     %% tabularray outer close
{                     %% tabularray inner open
width={1\linewidth},
colspec={X[]X[]X[]X[]X[]X[]X[]X[]},
column{1,2,3,4,5,6,7,8}={}{halign=l,},
colspec={X[2]X[]X[]X[]X[]X[]X[]X[]},
hline{1}={1-8}{solid, black, 0.08em},hline{7}={1-8}{solid, black, 0.08em},
}                     %% tabularray inner close
\toprule
Method & Edu OLS & Basis & LASSO & RF & GBM & In-sample R2 & Out-of-fold R2 \\ \midrule %% TinyTableHeader
Edu OLS & 1.0 & 0.55 & 0.76 & 0.63 & 0.77 & 0.11 & 0.088 \\
Basis & 0.55 & 1.0 & 0.79 & 0.78 & 0.67 & 0.34 & -0.060 \\
LASSO & 0.76 & 0.79 & 1.0 & 0.77 & 0.88 & 0.22 & 0.10 \\
RF & 0.63 & 0.78 & 0.77 & 1.0 & 0.72 & 0.60 & 0.097 \\
GBM & 0.77 & 0.67 & 0.88 & 0.72 & 1.0 & 0.19 & 0.11 \\
\bottomrule
\end{talltblr}
\end{table}

\Cref{tab:table-correlation-skill-estimates} reports the performance of each of the different algorithms in terms of the average out-of-fold $R^2$.
We find the GBM algorithm to perform the best with an \(R^2\) of \GbmOofRsqPct\%, while LASSO achieves \LassoOofRsqPct\% and the Random Forest \RfOofRsqPct\%.
In comparison to the simple OLS algorithm, the best-performing machine learning algorithms explain about \MlVsOlsGainPct\% more variance in the noisy individual fixed effects. Additional variables and non-linear models clearly improve on the basic OLS model, but the basic educational background still explains most of the variation in skills in the data.
We can further decompose the gains that machine learning algorithms achieve.
Noting that LASSO is simply OLS with many more variables while avoiding overfitting, we can interpret the difference between LASSO and GBM as the importance of non-linearities. Most of the predictive gains compared to OLS come from having additional variables (about \RsqGainMoreVarsPct\% of the total $R^2$ gain), with non-linearities playing a limited additional role.

How should we judge the overall level of the $R^2$? At first look, an out-of-fold $R^2$ of \GbmOofRsqPct\% may seem small.
However, note that this is in a setting in which we expect wages to be only imperfectly correlated with skills; our wage residuals will for example capture compensating differentials and idiosyncratic luck in the job matching process that persists over time, as well as temporary shocks that are only netted out in expectation.
Our results imply that these factors explain about \GbmOofRsqComplementPct\% of (residual) wage variation.
Preventing overfitting is key to derive this conclusion. If we had instead looked at the in-sample $R^2$ of our algorithms, we could have wrongly concluded that most (residual) wage variation is due to skills.
For example, the Random Forest algorithm manages to achieve an in-sample $R^2$ of \RfInRsqPct\%.
The basis expansion further exemplifies the important difference between in-sample and out-of-sample fit. Its in-sample $R^2$ is \BasisInRsqPct\% (see \Cref{tab:table-correlation-skill-estimates}), however, this is purely because of overfitting; the true out-of-fold $R^2$ for the basis expansion is in fact negative (\BasisOofRsqPct\%), i.e. worth than a simple mean.
Since the exercise is precisely about predicting skills for workers for whom we do not observe them, the relevant metric is the out-of-fold predictive power, and our machine learning algorithms dominate here.

The predictive $R^2$ is obviously also a function of how noisy the initial individual (residual) fixed effect is. If we had a longer worker panel, one could directly extract a more informative $\hat{z}_i$ and the corresponding $R^2$ would be higher. \Cref{tab:r2-by-years-observed,tab:pred-by-years-observed} show that the $R^2$ is indeed increasing when restricting to workers that we observe more periods. For example, restricting to workers with at least \SensMinYears{} years of observations more than doubles the out-of-fold $R^2$ to \SensGbmOofRsqPct\% (GBM) and \SensLassoOofRsqPct\% (LASSO) for the machine learning algorithms.

So what information are the machine learning algorithms additionally picking up? Focussing on the GBM estimate as the best-performing baseline estimate, \Cref{tab:gbm-variable-importance} in \Cref{sec:tables} ranks variables by variable importance, which weights the relative contribution of each covariate in the prediction task taking into account non-linearities and interactions.
We can see that higher education is by far the most important predictor of skills, but many other variables are also important. Among the 15 most important variables are other educational variables (senior secondary and less than primary), literacy skills (like reading a newspaper in another language, speaking Bahasa Indonesia, and writing letters), all Big Five personality traits, and cognitive skills (Raven's IQ score, counting backwards, and word ability).
Another way to see the importance of using additional covariates is to look at the correlation of predictions across different algorithms as reported in \Cref{tab:table-correlation-skill-estimates}.
Comparing the predictions of the three different machine learning algorithms with the OLS predictions, we can note that the correlation is at most \CorrMaxOlsPct\%, indicating that the OLS predictions are missing important variation that the different machine learning algorithms take into account.
Secondly, we can compare the different predictions of the machine learning algorithms among each other.
For example, comparing LASSO with GBM, we find that their correlation is \CorrLassoGbmPct\%, indicating that they both capture similar additional variation (due to the additional covariates).

\paragraph{Predicting skills out of sample}

In the last step, we use the estimated $\widehat{f(x)}$ from \cref{eq:main-estimating-equation} to get an estimate $\widehat{z_i}$ of ``skills'' for all workers, including those who never worked in the public sector.

\section{Main empirical results}\label{sec:main-empirical-results}

Taking the individual skill estimates based on the best-performing machine learning algorithm, we now show two main empirical applications using estimates of government worker skills.

\subsection{Application 1: Changes in skills and selection}\label{application-1-changes-in-skills-selection}

In the first application of the skill estimation approach, we zone in on the selection of government workers and changes in skills of government workers versus the overall population.
We show how relative skills declined over time and point to large costs of fluctuations in government hiring for the attraction of talent for the government.

\subsubsection*{Evolution of absolute and relative skills}\label{evolution-of-absolute-and-relative-skills}
\addcontentsline{toc}{subsubsection}{Evolution of absolute and relative skills}

\begin{figure}[t]
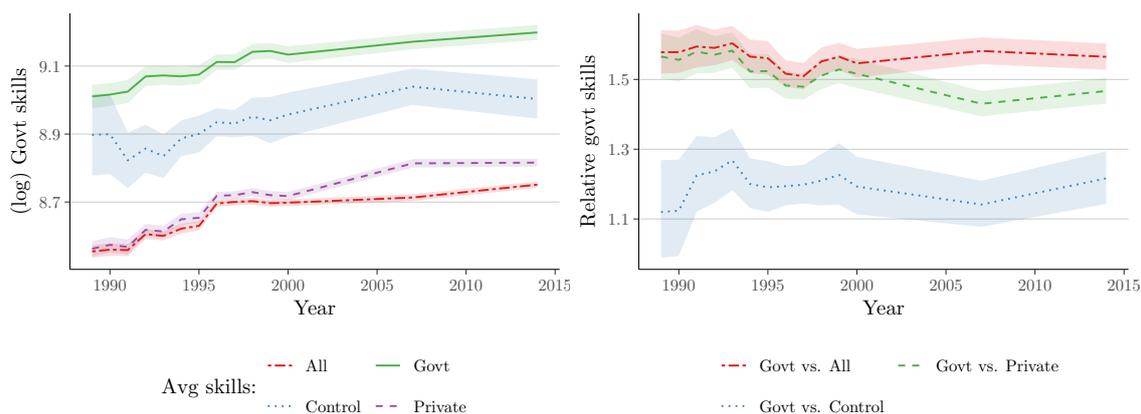

\caption{Evolution of government worker skills \label{fig:plot-evolution-govt-skills}}
{\centering
\resizebox{0.95\textwidth}{!}{
\input{out/img/evolution_govt_skills.tex}
}
\par}
\input{out/fignotes/evolution_govt_skills.tex}
\end{figure}

Can the government attract the workers who would be most skilled at government jobs? And how does this change over time as the growing private sector competes for talent? \Cref{fig:plot-evolution-govt-skills} plots the evolution of skills in government jobs for government workers, private sector workers, private sector workers in jobs comparable to the government (the control group) and all other workers, revealing three important facts.
First, government workers are on average much more skilled: in \SkillFirstYear, they are about \RelGovtControlFirstPct\% more skilled than comparable private sector workers and about \RelGovtAllFirstPct\% more skilled than the overall working population and private sector workers, in line with large observable skill differences reported in \Cref{tab:table-main-skill-differences}.
Second, as visible on the left plot, average skills increased across all groups of workers since \SkillFirstYear, in line with general increases in education: government worker skills grew by about \GrowthGovtPct\%, while private sector worker skills grew by about \GrowthPrivatePct\% over the \SkillFirstYear--\SkillLastYear{} period.
Third, relative skills of government workers -- as reported in the right panel -- declined compared to private sector workers.

\begin{figure}[t]
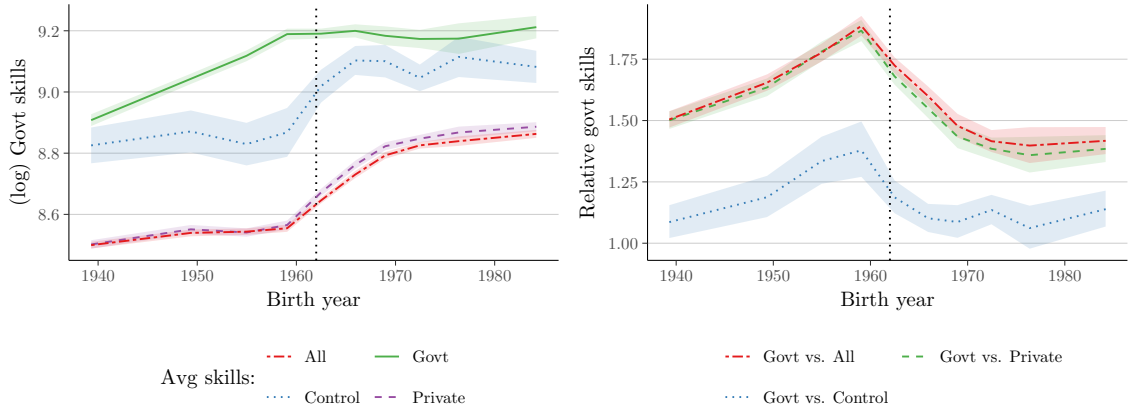

\caption{Government worker skills across birth cohorts \label{fig:plot-evolution-govt-skills-cohorts}}
{\centering
\resizebox{0.95\textwidth}{!}{
\input{out/img/evolution_govt_skills_by_cohort_bin.tex}
}
\par}
\input{out/fignotes/evolution_govt_skills_by_cohort_bin.tex}
\end{figure}

What drives this decline in government worker skills compared to the private sector? We find support for the idea that over the period 1988 to 2014, the government has been able to compete for talent with comparable jobs in the private sector, but increasingly failed to hire skilled workers from new entering cohorts who instead went to the general private sector.
To show this, \Cref{fig:plot-evolution-govt-skills-cohorts} reports skills by birth cohort.
Overall, more recent cohorts have higher skills, in line with improvements in general education.
However, the big increases in education (and skills) starting for cohorts in the 1960s as Indonesia launched its largest school construction program (see: \textcite{dufloSchoolingLaborMarket2001}) more strongly benefited skills in the private sector.
As shown in the right plot of \Cref{fig:plot-evolution-govt-skills-cohorts}, relative skills strongly declined starting for cohorts in the 1960s.
These strong cohort-level declines in relative skills are much more muted in the aggregate, because the cohort composition only changes slowly over time; however, these results suggest that relative government worker skills will decline (much) more in the future.
Importantly, as we show in \Cref{fig:plot-evolution-govt-skills-cohorts-jobfix}, the relative decline for cohorts after 1960 is not driven by changes in the type of government jobs offered, with \Cref{fig:plot-evolution-govt-skills-cohorts} looking almost indistinguishable after holding the occupation and sector composition of government jobs fixed over time.

At the same time, \Cref{fig:plot-govt-employment} already showed that government hiring strongly declined in birth cohorts shortly after 1960.
In the case that government jobs are in high demand, one would have expected that a decline in supply would have led to an increase in relative skills as long as the government is good at selecting workers on skills. While this effect can indeed be seen for private sector workers in jobs similar to government jobs (the control group), it does not seem to play out for the rest of the private sector.
So is the Indonesian government simply becoming worse at selecting other private sector workers? And are there actionable ways to improve the selection of government workers? The next two subsections zone in on these questions in turn.

\subsubsection*{The government selection rule}\label{the-government-selection-rule}
\addcontentsline{toc}{subsubsection}{The government selection rule}

Motivated by the decline in relative skills of government workers, we now turn to studying the selection of government workers across the entire skill distribution.
For example, are changes in average skills of government workers driven by the entry of low skilled workers or the failure to attract the most skilled workers into the government?
We show that the Indonesian government is generally successful at selecting more skilled workers, but that this selection fails at the top, with the most skilled workers not ending up in the government.

We show this by defining a simple reduced-form \emph{selection rule} that denotes the relative probability of being selected into the government conditional on one's skills, formally defined as:
\begin{equation}
s_t(z) \equiv \frac{f_t(z | \text{selected})}{f_t(z)}
\label{eq:selection-rule}
\end{equation} where \(f_t(z)\) gives the skill distribution in the entire population and \(f_t(z | \text{selected})\) the skill distribution of selected government workers.
The selection rule is above unity at any skill level \(z\) if the state selects more people with these skills than would be expected under uniform drawing from the conditional skill distribution in the population.
%This selection rule is also particularly informative in settings where most workers would in principle be willing to take a government job, which is the empirically relevant setting for developing countries and many developed countries.\footnote{In 2014, there were around 2.6 million applicants for 100k open civil service positions in Indonesia (see \href{https://www.vice.com/en_asia/article/yp9jnb/why-are-so-many-indonesians-obsessed-with-becoming-civil-servants}{here} as of April 22, 2024).}
The rule captures a combination of selection into who applies for a government job and selection of applicants via a multi-stage application process.\footnote{With direct data on who applies for a government job, one could also directly define the \emph{selection rule} \(s_t(z)\) as:
\begin{equation}
s_t(z) \equiv f_t(z | (\text{selected} | \text{applied})) = \frac{f_t(z | \text{applied} \; \cap \; \text{selected})}{f_t(z|\text{applied})}
\label{eq:selection-rule-applied}
\end{equation} where \(f_t(z | \text{applied})\) gives the skill distribution of the applicant pool and \(f_t(z | \text{applied} \; \cap \; \text{selected})\) the skill distribution of newly selected government workers. Unfortunately, we do not directly observe who applies for government jobs.}
Importantly, the selection rule $s_t(z)$ will also capture corruption, clientelism and politically-motivated hiring that influence selection of applicants and hires \parencite{colonnelliPatronageSelectionPublic2020,hannaDishonestySelectionPublic2017,jiaSelectionChina2015,weaver2021jobs}, which is important in the Indonesian context.
Specifically, there is good evidence for widespread corruption among bureaucrats \parencite{valsecchiCorruptBureaucratsResponse2016}, politics influencing selection into and promotion patterns within the Civil Service \parencite{pierskallaPersonnelPoliticsElections2018}, and civil service jobs being sold in an auction-like fashion in the past \parencite{kristiansenBuyingIncomeMarket2006}.

\begin{figure}[t]
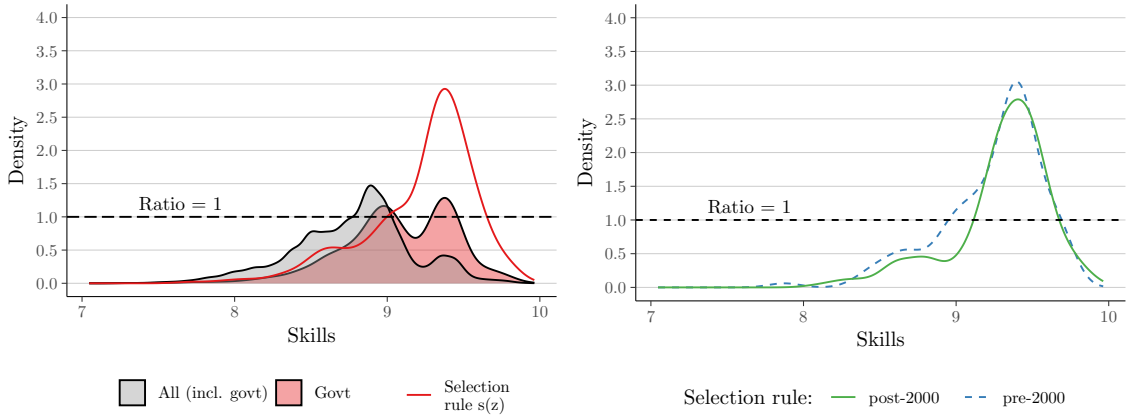

\caption{Estimated government selection rule and changes over time \label{fig:plot-selection-rule-joint}}
{\centering
\resizebox{0.95\textwidth}{!}{
\input{out/img/government_selection_rule_joint_plot.tex}
}
\par}
\input{out/fignotes/government_selection_rule_joint_plot.tex}
\end{figure}

\Cref{fig:plot-selection-rule-joint} plots the estimated selection rule comparing  government workers to all workers using a direct estimator of the density ratio.\footnote{Specifically, we estimate the ratio using unconstrained Least-Squares Importance Fitting (uLSIF) proposed by  \parencite{hidoStatisticalOutlierDetection2011}, which is more robust than separately estimating skill distributions for nominator and denominator and then forming the ratio of the two.
Similar to density estimation, there are many different direct estimators for density ratios proposed in the literature, which we found to perform very similarly for our application.} As we show in the Appendix \Cref{fig:plot-selection-rule-joint-appendix}, selection rule results are very similar when comparing only to private sector workers.
The left plot shows the estimated selection rule and separate skill densities for the two groups of workers.
Two main points stand out: First, the government selection rule is mostly increasing.
The Indonesian government is much less likely to pick low skilled workers as civil servants compared to their density in the data, which translates into a ratio below 1, and is more likely to pick high skilled workers leading to density ratios above 1.
Second, the selection rule drops for the most skilled workers.
If everyone in the economy would be willing to take a government job, the estimates imply that the government becomes worse at selecting the most qualified workers in the economy.
The more likely explanation in this case is that the workers who would be most skilled in a government sector job are actually not interested in taking government jobs.

The right plot in \Cref{fig:plot-selection-rule-joint} tests whether the selection rule has changed over time and whether this may explain the decline in relative skills of government workers.
Quantitatively, we find that the estimated \emph{selection rule} is remarkably constant over time, despite Indonesia moving from a highly centralized autocratic government with a regime-aligned bureaucracy \parencite{hadizPoliticalEconomyOligarchy2013,robisonReorganisingPowerIndonesia2004} to a democratic system with a highly decentralized bureaucracy \parencite{bluntPatronageProgressPostSoeharto2012,brinkerhoffPerformancebasedPublicManagement2013} over the entire period.
The selection rule shifted downwards over time, in line with a general decline in government hiring (see \Cref{fig:plot-govt-employment}).
If anything, the \emph{selection rule} after the year 2000 is more selective at the lower tail and only starts to decrease at a higher level of skills, potentially indicating a success of numerous bureaucracy reforms implemented across ministries in the 2000s and a change to a computer-assisted selection that has been shown to have reduced corruption at the selection stage \parencite{kuipers2023failing}.
In \Cref{fig:plot-selection-rule-jobfix}, we show that the changes in the selection rule are almost unchanged when holding the occupation and sector composition of government jobs fixed over time.

What then explains the decrease in relative skills of government workers if it is not changes in the selection rule? The main driving force is a shift in the underlying skill distribution.
While the selection rule stays roughly constant, the right tail of the skill distribution grows and more highly skilled workers end up working in the private sector and the rest of the economy.

\subsubsection*{Skill selection and the costs of government hiring waves}\label{skill-selection-the-costs-of-government-hiring-waves}
\addcontentsline{toc}{subsubsection}{Skill selection and the costs of government hiring waves}

In this final subsection, we highlight the potential costs of fluctuations in government hiring for the skill selection of government workers.
Besides its important policy implications, this final application is also a test of the skill selection mechanism.
The idea is that there is strong year-to-year variation in how much the government is hiring, driven by political cycles, the discrete nature of legislation and fluctuations in government revenue. For example, Indonesia has a history of announcing government hiring freezes in years where government finances are strained and strongly expanding hiring in years where the government sees financing windfalls (e.g. from oil revenue).
If hiring by the government happens disproportionately for workers at labor market entry, then one would expect differential exposure of birth cohorts to government hiring, as evidenced by observed differences in cohort-specific government employment shares.
If skill selection as a mechanism has bite, then conditional on hiring for the same type of job and having similar cohort-to-cohort skill distributions, cohorts that enter the labor market in years where the government is hiring disproportionately more should see lower average skills in the government.
That is, in years of disproportionately high hiring, the government is more likely to draw from lower parts of the skill distribution because the hiring pool is finite.
If the hiring pool replenishes over time as new cohorts enter, the government could thus higher better workers by smoothing government hiring over time.

To test this idea formally, we construct the empirical cohort-level government employment share (denoted by \(GES\)) in our pooled panel. To focus on the effects of hiring \emph{waves}, we study the effects of \emph{deviations} in the government employment share \(GES\) from its trend on the relative skills of government workers (compared to the private sector). While this exercise has a strong theoretical foundation, the empirical implementation suffers from an obvious and big limitation: since this exploits cohort-to-cohort variation in hiring and there are only about 50 cohorts in our data, the sample size for this exercise is greatly restricted and results are naturally noisy. Partly for this reason, we do not additionally try to isolate purely exogenous variation in the government employment share and leave more rigorous causal evidence to future work.

With these limitations in mind, we run cohort-level regressions of the form:
\begin{equation}
\text{Relative skills}_c = \alpha + \beta^C \, \text{GES}_c + \text{Trend controls}_c + \zeta_{c}
\label{eq:cohort-level-reg-govt-hiring}
\end{equation}
where the dependent variable denotes average relative skills of government workers (versus private sector workers) in a cohort. To capture the effect of \emph{deviations} in the government employment share \(GES\) from its trend, we consider two main specifications: one specification in first differences (looking at the effect of \emph{changes} in the \(GES\) on \emph{changes} in relative skills) and one specification in levels where we additionally control for a (cubic) trend across cohorts to capture the effects of deviations from this trend.
We further residualize government worker skills on changes in the composition of jobs to ensure that our results are not driven by hiring waves being correlated with the opening of specific types of jobs. We also restrict to cohorts with at least 25 government-worker observations, with similar results when varying the cutoff (see \Cref{tab:table-govt-hiring-nodrop,tab:table-govt-hiring-min10,tab:table-govt-hiring-min50}).

\begin{table}
\centering
\begin{talltblr}[         %% tabularray outer open
caption={The government skill premium and government hiring intensity\label{tab:table-govt-hiring}},
note{}={\textbf{Notes:} \textit{All data restricted to workers between the age of 25 and 58, the official age span of government workers. The estimation sample drops all birth-year cohorts with fewer than 25 government worker-year observations in the analysis sample. Cohort-level regressions of the relative skills of government workers vs. private sector workers (``Rel. govt skills'') on government hiring intensity. Columns 1--2 use the level of the government employment share; columns 3--6 use first differences. Columns 2, 5, and 6 hold government jobs fixed: government worker skills are residualized out of occupation and sector effects from a government-only regression, with a counterfactual reference at (Teacher, Social services), and rescaled so the weighted government mean matches the raw government mean. Non-government worker skills are unchanged. The cohort sample is identical across all six columns. The `Cohort trend' row indicates whether a cubic birth-cohort polynomial is included as an additional control; its coefficients and the constant are suppressed. Heteroskedasticity-robust standard errors in parentheses.}},
]                     %% tabularray outer close
{                     %% tabularray inner open
width={1\linewidth},
colspec={X[]X[]X[]X[]X[]X[]X[]},
column{2,3,4,5,6,7}={}{halign=c,},
column{1}={}{halign=l,},
hline{6}={1,2,3,4,5,6,7}{solid, black, 0.05em},
colspec={X[2]X[]X[]X[]X[]X[]X[]},
hline{1}={1-7}{solid, black, 0.08em},hline{10}={1-7}{solid, black, 0.08em},
}                     %% tabularray inner close
\toprule
& Rel. govt skills & Rel. govt skills  & $\Delta$ Rel. govt skills & $\Delta$ Rel. govt skills  & $\Delta$ Rel. govt skills   & $\Delta$ Rel. govt skills    \\ \midrule %% TinyTableHeader
Govt empl share & -0.37 & -0.42 &  &  &  &  \\
& (0.63) & (0.59) &  &  &  &  \\
$\Delta$ Govt empl share &  &  & -0.93 & -0.93 & -1.03 & -1.04 \\
&  &  & (0.48) & (0.49) & (0.44) & (0.45) \\
Observations & 50 & 50 & 49 & 49 & 49 & 49 \\
R2 & 0.49 & 0.45 & 0.11 & 0.12 & 0.15 & 0.16 \\
Cohort trend & Yes & Yes & No & Yes & No & Yes \\
Holding govt jobs fixed & No & Yes & No & No & Yes & Yes \\
\bottomrule
\end{talltblr}
\end{table}

\Cref{tab:table-govt-hiring} reports our six resulting specifications and shows that despite the limited sample size, more government hiring in a given cohort is robustly associated with lower relative skills of government workers.
We also want to emphasize that our main coefficient estimates are likely attenuated because we use the empirically observed cohort-specific government employment share \(GES\), which is only a noisy estimate of the true \(GES\).
Columns 1 and 2 show the level results relative to the cohort trend and find a negative, though insignificant result.
Columns 3--6 show results using first differences, showing consistently large negative results that are also significant at the 5\% level based on heteroscedasticity-robust standard errors. To put the magnitude of these effects in perspective, we compare it to the stark decline in the cohort-level government-to-private skill gap between 1960 and 1980 as shown in \Cref{fig:plot-evolution-govt-skills-cohorts} (and \Cref{fig:plot-evolution-govt-skills-cohorts-jobfix}). A one-standard-deviation shift in the level of \(GES\), conditional on the cubic cohort trend (column 2), accounts for \HiringShareCohortChangePct\% of that narrowing, while a one-standard-deviation change in \(\Delta GES\) in the first-difference specification (column 6) accounts for \HiringShareCohortChangeDiffPct\%. 

In summary, hiring waves are a potentially important source for lower government worker skills. Given the continuing use of hiring freezes in Indonesia, as evidenced by annual hiring freezes in 2015 and 2016, fluctuations in government hiring are likely to continue to play an important role for the selection of talent for the government. 

\subsection{Application 2: Government wage setting}\label{application-2-government-wage-setting}

The second major application looks at the wages of government workers; specifically at the informativeness of government wages and whether government workers are overpriced compared to the private sector.

\subsubsection*{Are government wages uninformative?}\label{are-government-wages-uninformative}
\addcontentsline{toc}{subsubsection}{Are government wages uninformative?}

The starting point of the estimation approach was the idea that government worker wages are potentially uninformative about underlying skills, because --- as in the case of Indonesia --- they follow rigid tenure schedules and allowances with little use of performance incentives or wage dispersion.
We now show how to use our approach to actually test this.
We find more nuanced results: government wages are far from uninformative about underlying skills, but comparable private sector wages are more informative about skills.
We show this with two different exercises.

\begin{table}
\centering
\begin{talltblr}[         %% tabularray outer open
caption={Regression results: Informativeness of government wages\label{tab:table-govt-wage-informativeness}},
note{}={\textbf{Notes:} \textit{Dependent variable is log real hourly wage. Columns 1-3 focus on public worker wages only, while Columns 4-6 focus on control workers (all private sector workers reweighted by propensity scores using sector, occupation, age, and a male indicator). Columns 1 and 4 predict wages using skills only. Columns 2, 3, 5 and 6 instead use alternative lifecycle-style skill specifications.}},
]                     %% tabularray outer close
{                     %% tabularray inner open
width={1\linewidth},
colspec={X[]X[]X[]X[]X[]X[]X[]},
column{2,3,4,5,6,7}={}{halign=c,},
column{1}={}{halign=l,},
hline{12}={1,2,3,4,5,6,7}{solid, black, 0.05em},
colspec={X[2]X[]X[]X[]X[]X[]X[]},
hline{1}={1-7}{solid, black, 0.08em},hline{16}={1-7}{solid, black, 0.08em},
}                     %% tabularray inner close
\toprule
& (1) & (2) & (3) & (4) & (5) & (6) \\ \midrule %% TinyTableHeader
Constant & 2.27 & -0.47 & 0.58 & -0.89 & -2.06 & -1.86 \\
& (0.23) & (0.22) & (0.22) & (0.10) & (0.10) & (0.10) \\
Skill & 0.77 &  & 0.87 & 1.08 &  & 1.13 \\
& (0.03) &  & (0.02) & (0.01) &  & (0.01) \\
Life-cycle skill &  & 1.04 &  &  & 1.17 &  \\
&  & (0.02) &  &  & (0.01) &  \\
Experience &  &  & 0.07 &  &  & 0.05 \\
&  &  & (0.00) &  &  & (0.00) \\
Experience square &  &  & -0.00 &  &  & -0.00 \\
&  &  & (0.00) &  &  & (0.00) \\
\emph{Fit statistics} &  &  &  &  &  &  \\
Observations & 12068 & 12068 & 12068 & 50057 & 50057 & 50057 \\
R2 & 0.07 & 0.14 & 0.18 & 0.15 & 0.18 & 0.19 \\
SEs & IID & IID & IID & IID & IID & IID \\
\bottomrule
\end{talltblr}
\end{table}

First, we test the relative informativeness of government wages by predicting real hourly wages for government workers and comparable private sector workers using estimated skills.
\Cref{tab:table-govt-wage-informativeness} reports these results.
Comparing Columns 1 and 4 shows that government worker wages are indeed far less informative about underlying skills than wages in comparable private sector jobs: the \(R^2\) is \RsqGovtSkillPct\% vs.~\RsqCtrlSkillPct\%; i.e. comparable private sector wages are about \RsqCtrlVsGovtRatio{} times more informative.
However, this gap narrows substantially once one controls for experience profiles, given a far more deterministic government wage experience schedule.
For example, when additionally controlling for experience (columns 3 and 6), the \(R^2\) rises to \RsqGovtExperiencePct\% for government workers, close to the \RsqCtrlExperiencePct\% for comparable private sector workers.

\begin{figure}[t]
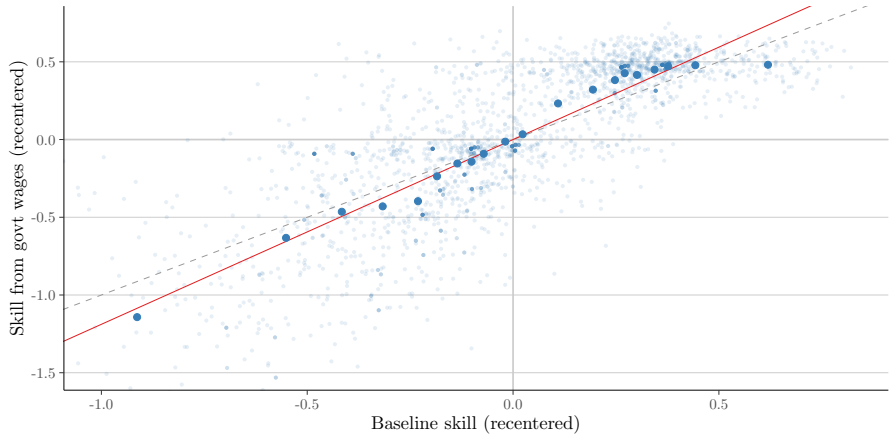

\caption{Correlation of baseline and government-wage-based skill estimates \label{fig:plot-correlation-skills-govt-wages}}
{\centering
\resizebox{0.75\textwidth}{!}{
\input{out/img/correlation_skills_govt_wages.tex}
}
\par}
\input{out/fignotes/correlation_skills_govt_wages.tex}
\end{figure}

The second test of the informativeness of government worker wages is to re-estimate government worker skills following all estimation stages but using government wages instead and then looking at the correlation across the two skill measures.
As shown in \Cref{fig:plot-correlation-skills-govt-wages}, the correlation between the two skill measures is high, with both the Pearson and Spearman rank correlations being around 83\%. In line with this, the slope of the correlation is close to one. % correlation of \GovtWagePearsonR{} and Spearman rank correlation of \GovtWageSpearmanR.
Assuming that baseline skills are estimated correctly, government wages are thus clearly informative about underlying skills.
However, as also visible from \Cref{fig:plot-correlation-skills-govt-wages}, the correlation clearly flattens out for the most skilled workers; the binned scatter points are almost flat at the very right tail. For example, restricting to the top quartile based on the baseline skill estimates only gives a correlation of 0.121. This means that government wages are particularly uninformative for the most skilled workers.

\subsubsection*{Is there a government wage premium?}\label{is-there-a-government-wage-premium}
\addcontentsline{toc}{subsubsection}{Is there a government wage premium?}

\begin{table}
\centering
\begin{talltblr}[         %% tabularray outer open
caption={Regression results: Government wage premium\label{tab:table-govt-wage-premium}},
note{}={\textbf{Notes:} \textit{Sample restricted to government workers and control group (all private sector workers reweighted by propensity scores using sector, occupation, age, and a male indicator). Adding occupation and sector fixed effects reduces the sample size slightly due to missings in the occupation variable.}},
]                     %% tabularray outer close
{                     %% tabularray inner open
width={1\linewidth},
colspec={X[]X[]X[]X[]X[]X[]X[]X[]X[]},
column{2,3,4,5,6,7,8,9}={}{halign=c,},
column{1}={}{halign=l,},
hline{12}={1,2,3,4,5,6,7,8,9}{solid, black, 0.05em},
colspec={X[2]X[]X[]X[]X[]X[]X[]X[]X[]},
hline{1}={1-9}{solid, black, 0.08em},hline{20}={1-9}{solid, black, 0.08em},
}                     %% tabularray inner close
\toprule
& Base & Year FE & Job FE & Skill & Skill+Job & LC & LC+Job & LC+T \\ \midrule %% TinyTableHeader
Constant & 8.80 &  &  &  &  &  &  &  \\
& (0.01) &  &  &  &  &  &  &  \\
Govt worker & 0.50 & 0.54 & 0.51 & 0.37 & 0.40 & 0.34 & 0.36 & 0.14 \\
& (0.01) & (0.05) & (0.05) & (0.04) & (0.04) & (0.04) & (0.04) & (0.04) \\
Skill &  &  &  & 0.92 & 0.72 &  &  &  \\
&  &  &  & (0.05) & (0.08) &  &  &  \\
Life-cycle skill &  &  &  &  &  & 1.09 & 0.94 & 0.93 \\
&  &  &  &  &  & (0.03) & (0.05) & (0.05) \\
Govt worker x Year &  &  &  &  &  &  &  & 0.02 \\
&  &  &  &  &  &  &  & (0.00) \\
\emph{Fit statistics} &  &  &  &  &  &  &  &  \\
Observations & 62125 & 62125 & 60081 & 62125 & 60081 & 62125 & 60081 & 60081 \\
R2 & 0.05 & 0.08 & 0.16 & 0.18 & 0.21 & 0.22 & 0.24 & 0.25 \\
R2 (within) &  & 0.06 & 0.06 & 0.16 & 0.11 & 0.20 & 0.15 & 0.15 \\
SEs & IID & by: year & by: year & by: year & by: year & by: year & by: year & by: year \\
Year dummies &  & X & X & X & X & X & X & X \\
Occupation dummies &  &  & X &  & X &  & X & X \\
Sector dummies &  &  & X &  & X &  & X & X \\
\bottomrule
\end{talltblr}
\end{table}

The estimation approach allows for a direct measurement of the wage premium of government workers compared to similar jobs in the private sector: conditional on the same level of skills, do government workers earn higher wages? The answer is: yes.
Government workers earn a large wage premium of about \WpLcJobPct\%.
Specifically, \Cref{tab:table-govt-wage-premium} reports different estimates for the wage premium restricting to government workers and the control group of comparable private sector workers.
Column 1 documents a large unconditional wage premium of \WpUncondLogPts{} log points, which translates to roughly \WpUncondPct\% higher real hourly wages.
The wage premium substantially decreases when controlling for worker skills (column 4), pointing to strong positive selection on skills.
The wage premium also generally reduces when focussing on within-job comparisons by including occupation and sector dummies (columns 3, 5 and 7).
To avoid biasing estimated wage premia due to compositional changes over time -- for example, if there are more control group workers in more recent periods -- results for columns (2) to (8) include year dummies.
At last, one may be interested in capturing the wage premium conditional on human capital that incorporates experience.
Columns (6) and (7) report wage premia controlling for experience using the life-cycle skill measure \(h_{i,e,t}\) that incorporates estimated human capital experience profiles.
These estimates lead to the most conservative wage premia of around \WpLcJobLogPts{} log points, or a wage premium of roughly \WpLcJobPct\%.
The last column considers changes in the wage premium over time and finds that the government wage premium is clearly increasing over time.
The increasing wage premium for similar jobs in the private sector may explain the results of the previous section, namely why relative skills have not declined in these jobs as for the overall private sector.

\begin{table}
\centering
\begin{talltblr}[         %% tabularray outer open
caption={Regression results: Government wage premium and gender gap\label{tab:table-govt-wage-premium-gender-gap}},
note{}={\textbf{Notes:} \textit{Dependent variable is always log real hourly wage. Sample restricted to government workers and control group (all private sector workers reweighted by propensity scores using sector, occupation, age, and a male indicator). Adding occupation and sector fixed effects reduces the sample size slightly due to missings in the occupation variable.}},
]                     %% tabularray outer close
{                     %% tabularray inner open
width={1\linewidth},
colspec={X[]X[]X[]X[]X[]X[]},
column{2,3,4,5,6}={}{halign=c,},
column{1}={}{halign=l,},
hline{18}={1,2,3,4,5,6}{solid, black, 0.05em},
colspec={X[2]X[]X[]X[]X[]X[]},
hline{1}={1-6}{solid, black, 0.08em},hline{26}={1-6}{solid, black, 0.08em},
}                     %% tabularray inner close
\toprule
& (1) & (2) & (3) & (4) & (5) \\ \midrule %% TinyTableHeader
Constant & 8.50 &  &  &  &  \\
& (0.01) &  &  &  &  \\
Govt worker & 0.86 & 0.78 & 0.65 & 0.60 & 0.38 \\
& (0.02) & (0.04) & (0.03) & (0.04) & (0.04) \\
Male & 0.42 & 0.49 & 0.46 & 0.41 & 0.41 \\
& (0.01) & (0.04) & (0.03) & (0.02) & (0.03) \\
Govt worker x Male & -0.50 & -0.39 & -0.36 & -0.34 & -0.33 \\
& (0.02) & (0.03) & (0.02) & (0.02) & (0.02) \\
Skill &  &  & 0.71 &  &  \\
&  &  & (0.08) &  &  \\
Life-cycle skill &  &  &  & 0.90 & 0.90 \\
&  &  &  & (0.06) & (0.06) \\
Govt worker x Year &  &  &  &  & 0.02 \\
&  &  &  &  & (0.00) \\
Male x Year &  &  &  &  & -0.00 \\
&  &  &  &  & (0.00) \\
\emph{Fit statistics} &  &  &  &  &  \\
Observations & 62125 & 60081 & 60081 & 60081 & 60081 \\
R2 & 0.06 & 0.18 & 0.23 & 0.26 & 0.26 \\
R2 (within) &  & 0.08 & 0.13 & 0.16 & 0.16 \\
SEs & IID & by: year & by: year & by: year & by: year \\
Year dummies &  & X & X & X & X \\
Occupation dummies &  & X & X & X & X \\
Sector dummies &  & X & X & X & X \\
\bottomrule
\end{talltblr}
\end{table}

What is driving this large observed wage premium? It turns out that an important driver of the government wage premium is the absence of a gender wage gap in the public sector.
Women in Indonesia's private sector face a large real hourly wage penalty compared to men in the private sector.
\Cref{tab:table-govt-wage-premium-gender-gap} reports a raw gender wage gap of \GgWithinJobLogPts{} log points, or a roughly \GgWithinJobPct\% wage cut for women compared to men in comparable jobs in the private sector (column 2), which decreases slightly after controlling for the same skills (column 3) or life-cycle skills (column 4).
This gender wage gap is also stable over time (coefficient ``Male x year'' in column 5).
The gender wage gap in the public sector, on the other hand, is much smaller, at less than \GgGovtVsPrivateRatio{} of the wage gap in the private sector after controlling for the same job, experience and life-cycle skills (column 4 and using the sum of the coefficients for ``Male'' and ``Govt worker x Male'').
That is, in the absence of a differential gender wage gap in the private sector, the government wage premium would be roughly \GgExplainsWpPct\% lower (the sum of coefficients ``Govt job'' and ``Govt job x Male'' in column 4 in comparison to the coefficient for ``Govt job'' in \Cref{tab:table-govt-wage-premium} column 7).

\section{Robustness \& Extensions}\label{sec:robustness}

This section goes through the main robustness tests and extensions.

\subsection{How general are estimated skills?}
\label{sec:skill-generality}

\subsubsection*{Are the type of skills changing over time?}
\label{sec:time-stability}

One important assumption we made is that the skill mapping $f(X_i)$ is time-invariant. This assumption could be violated if different skill dimensions become meaningfully more or less important over time.
Fortunately, one can test this assumption directly by re-estimating $f(X_i)$ for different time periods.
Consistent with our assumption, we find that estimated skills are fairly stable over time, which we take as a sign that we pick up on more general skills rather than specific skills (like IT skills) that have changed in importance over time.

\begin{figure}[t]
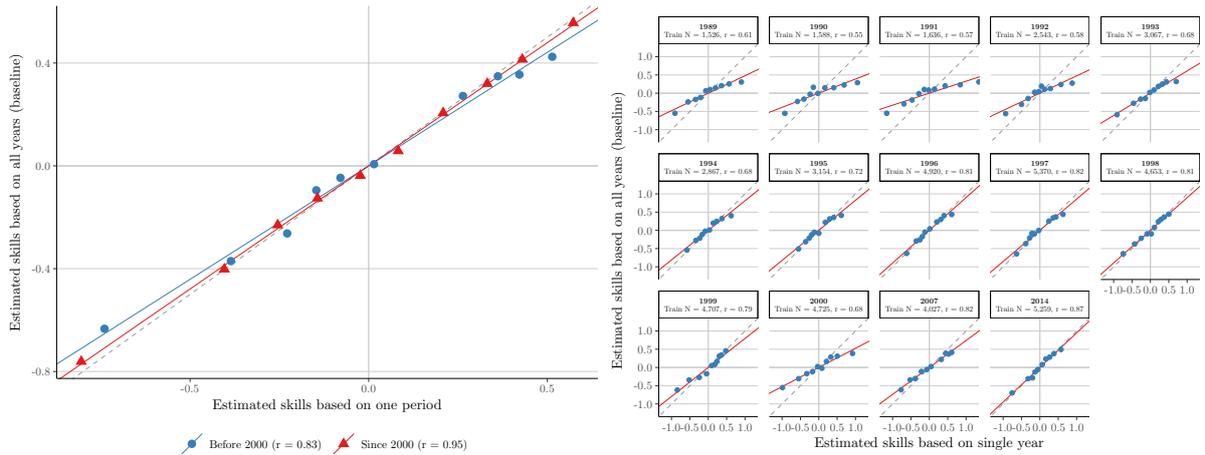

\caption{Testing the time stability of skill estimates \label{fig:fx-stability-year-period}}
{\centering
\resizebox{\textwidth}{!}{
\input{out/img/fx_stability_year_period}
}
\par}
\input{out/fignotes/fx_stability_year_period}
\end{figure}

To show this, we consider two tests.
First, we do a simple sample split; estimating skills on data from before 2000 only and on observations since the year 2000.
For each of the two samples, we train a GBM algorithm as for the baseline (though we do not re-tune the hyperparameters), we then use the trained model to predict skills for all workers and compare these skills with the baseline skill estimates. The left panel of \Cref{fig:fx-stability-year-period} compares these estimated skills with our baseline estimates using a binned scatter plot. Correlations are very high (\StabPeriodBeforePct\% for the before 2000 sample, \StabPeriodSincePct\% for the since 2000 sample), and the estimated slopes are both close to a perfect 45-degree-line.

Next, we consider a more systematic exercise where we estimate skills using data from a single year only, re-training the GBM on comparable private-sector observations from that year alone, then predict skills for all workers and compare these predictions with our baseline skill estimates.
The right panel of \Cref{fig:fx-stability-year-period} displays binned scatters of baseline skills versus skill estimates based on each year, showing again almost perfect 45-degree-line fits for most years.
Corresponding (Pearson) correlations are not perfect but consistently high, ranging from \StabYearMinPct\% to \StabYearMaxPct\%. Correlations are higher for later years that provide larger training samples in line with lower correlations simply being driven by less precise estimation rather than true differences. Whether any downstream results are affected by these changes are a function of both the slope and the correlation and given the strong overlap in both, we interpret the results from this exercise as providing strong evidence that the mapping $f(X_i)$ is relatively stable over time. Note also that the single-year estimates are simply the "hedonic price regression approach" mentioned in \Cref{sec:methodology}, which effectively shows that our results are also not sensitive to the specific skill-price estimation method used.

% While our results are consistent with the time-invariability assumption, one might actually be surprised about the stability, given that some specific skills are likely to have changed in importance over time (e.g. IT skills).
% We believe the right way to think about these results is that our approach captures general rather than specific skills like IT skills.

\subsubsection*{Are our estimated skills occupation specific?}
\label{sec:occupation-stability}

Throughout, we argued that our estimated skills are picking up ``general'' skills, rationalizing the use of comparable private sector jobs to infer skills in the public sector, and also rationalizing pooling across different types of jobs in the private sector.
The previous subsection already provided some evidence consistent with the idea that our approach picks up on ``general'' skills.
Here, we provide further evidence by re-estimating skills across different (comparable) occupations in the private sector, showing that estimated skills are broadly similar, apart from a few exceptions.
That is consistent with estimated skills mostly capturing general skills applicable across different occupations, rather than specific skills required in specific occupations.

\begin{figure}[t]
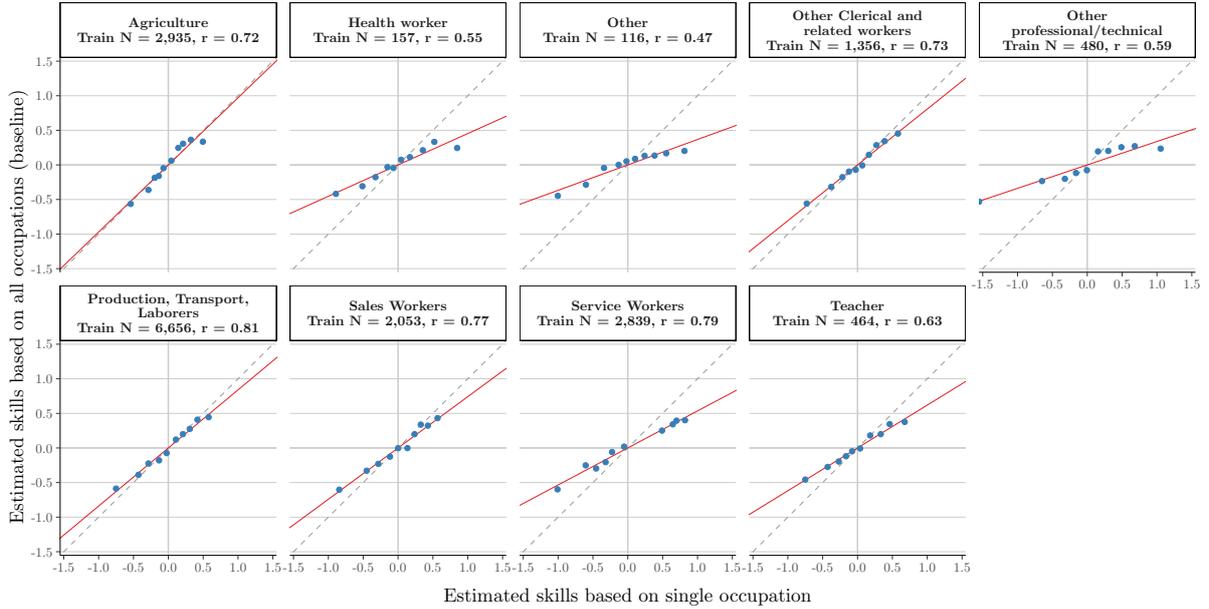

\caption{Testing the stability of skill estimates across different occupations \label{fig:fx-stability-occupation}}
{\centering
\resizebox{\textwidth}{!}{
\input{out/img/fx_stability_occupation}
}
\par}
\input{out/fignotes/fx_stability_occupation}
\end{figure}

To show this, we follow our previous approach for testing the time stability by systematically re-training our model for each comparable private sector occupation separately. That is, for each occupation $g$, the GBM algorithm is trained only on workers in occupation $g$ and then used to predict skills for all workers in all occupations.
We restrict attention to occupations with more than 100 observations, as skill estimates are hard to interpret otherwise.
\Cref{fig:fx-stability-occupation} correlates occupation-specific skill estimates with our baseline skill estimates pooled across all occupations.
Correlations are generally high and estimated skills are reasonably close to the 45-degree-line for the majority of occupations.
For example, the correlation for  "Production workers, transport workers and laborers" is \StabOccProductionPct\% and \StabOccSalesPct\% for "Sales workers", with a fit very close to the 45-degree-line for both.
There are, however, some exceptions. For example, for teachers, where we estimate skills solely based on private-sector teachers, the correlation is only \StabOccTeacherPct\%, potentially indicating that our skill estimation approach captures more specific teaching skills if estimated purely on teachers.
Still, correlations are particularly high for the occupations for whom we observe the most workers, indicating that lower correlations may simply be driven by noisier estimates due to smaller estimation sample sizes.
We thus conclude from this exercise that our approach seems to mostly pick up on general skills that are applicable across many different occupations rather than occupation-specific skills that may also be particular to the private sector.

\subsection{Selection on unobservables}\label{sec:robustness-selection}

One main concern about the estimation approach is that identification is entirely based on observables.
While we observe a large set of variables based on which workers do differentially sort into government and comparable private sector jobs, a valid concern remains that workers still sort into jobs based on remaining unobservables.
While sorting based on unobservables is not generally a concern for the approach, it is a concern in the case where workers sort into comparable private sector jobs based on persistent unobservables that are correlated with their wages.
In this case, we would estimate $f(X_i)$ with a systematic bias and thus wrongly infer skills.\footnote{It is important to note that since we estimate skills from comparable private sector jobs, the primary concern is also about selection on unobservables for these comparable private sector jobs, and not selection into the government. However, these are clearly related to the extent that selection on unobservables into the government mechanically affects selection into other jobs that were not chosen.}
This subsection takes this concern seriously and proposes an extension that allows for selection on unobservables based on the Heckman selection approach as outlined in \Cref{sec:methodology} and \Cref{eq:selection-extension}. By proposing a plausible and strong instrument, we show that there is indeed evidence for selection on unobservables but this does not significantly change our main skill estimates nor quantitative results.

The methodological approach as outlined in \Cref{sec:methodology} is to specifically model selection into comparable private-sector jobs using a first-stage probit regression where we allow selection to be a function of all the covariates that predict skills as well as an additional instrument that affects selection, but not wages (or skills) directly. Following the Heckman selection approach, we then correct for this selection bias by including the inverse Mills ratio (IMR) of the probit regression in the outcome equation where we estimate skills.
Full details and all additional tables are in \Cref{sec:heckman-appendix}; here we summarise our key findings.\footnote{The probit regression includes the instrument, all skill covariates, experience and experience squared. We then include the individual-specific average IMR in the outcome regression. Here, we slightly deviate from the baseline estimation approach in that we cannot estimate individual fixed-effects first, as these would absorb the IMR. Instead, we estimate skills based on the joint estimation approach by \textcite{robinson1988root} in combination with a Mundlak device \parencite{mundlak1978pooling}, adding as controls within-individual means of the time-varying regressors (experience and experience squared). These means absorb the component of individual-level unobservables that correlates with the regressors -- replicating the bias-reduction of individual fixed effects -- while leaving time-invariant variables identified.}

As an instrument for selection into a comparable private-sector job, we use whether either parent of the worker has ever worked in a comparable private-sector job. Having a parent in a comparable private-sector job plausibly raises a worker's own propensity to enter jobs in the comparable private sector through information, networks, and role models \parencite{alesina2001redistribution}. The assumption is that conditional on individual skills $z_i$, this selection based on the parent's background does not directly affect the worker's wage in these types of jobs.
Given that we focus mostly on social service jobs in the private sector, we believe this to be a plausible assumption; these are specifically not sons and daughters who end up working for their parent's own company.
The first-stage Wald statistic is $\HeckmanFirstWald$ ($p \HeckmanFirstP$), confirming that the instrument is strongly relevant.
To provide further evidence in line with the exclusion restriction, we follow standard practice for Heckman selection models and check whether the instrument is predictive of wages in the outcome equation after controlling for the inverse Mills ratio and observable skills $f(X_i)$.
Reassuringly, the instrument is not significant in this augmented equation (Wald statistic $\HeckmanExclWald$, $p = \HeckmanExclP$), supporting the exclusion restriction.

What do our results imply for our skill estimates?
We indeed find evidence for selection on unobservables: the inverse Mills ratio is positive in the outcome equation
($\hat{\gamma} = \HeckmanImrCoef$), implying that workers with higher unobserved propensity to enter the (comparable) private sector also earn higher wages conditional
on observables.
However, controlling for this selection does not materially change our main skill estimates: Pearson correlations between baseline and Heckman-corrected skills exceed $\HeckmanMinGbmCorrR$ for the GBM estimator across all subsamples.
The reason is that controlling for selection on unobservables mostly just uniformly scales down the effect of observables without materially affecting the ranking of (observable) skills.
This means, for example, that we find a slightly smaller (observable) skill gap between government and comparable private sector workers, but changes in relative skills over time are still almost unchanged.
Given the very high correlation of the two sets of skill estimates, all of our main results go through.
Full results including the Heckman-corrected skill estimates and their comparison with
baseline estimates are reported in \Cref{sec:heckman-appendix}.

\subsection{Multi-dimensional skills and more flexible skill-experience profiles}\label{more-flexible-skill-experience-profiles}

At last, we note that the estimation approach can readily be extended to multi-dimensional individual skill estimates and more flexible skill-experience profiles. One natural example would be to consider not only individual skills at labor market entry but also \emph{skills at learning over the life cycle}.
In these cases, the key to estimation is to construct multiple individual-specific fixed effects and then separately regress these on skill- or learning-related observables.
To see how this would look, assume that human capital \(H_{i,e,t}\) evolves according to the following factor structure:
\begin{equation}
H_{i,e,t} = exp(z_i) \times exp(g_i \times \delta_e)
\end{equation} where \(z_i\) are individual time-fixed skills at labor market entry as before, \(\delta_e\) are arbitrary experience dummies to flexibly capture experience profiles, and \(g_i\) denotes new individual-specific learning capabilities. The factor structure allows individual-specific skills and individual-specific wage-experience profiles, but restricts the shape of the latter to a factor structure.
This is a strict generalization of our baseline model and can be an important extension in cases where empirically observed concave wage-experience profiles differ in slope across individuals \parencite[e.g. see:][]{primiceriHeterogeneousLifecycleProfiles2009,lagakosLifeCycleWage2018}. Given the additional steps in our estimation, we believe this would overfit on the data in our context, but we provide further details in Appendix \Cref{appendix-multidim-skills} on how to implement this extension using a simple two-step estimator based on  \textcite{pesaranEstimationInferenceLarge2006}. Multiple dimensions of skills $z_i$ can also be incorporated using multiple orthogonalized factors following \textcite{baiPanelDataModels2009,ahnPanelDataModels2013}.

\section{Conclusion}\label{sec:conclusion}

This paper provided a new approach to estimate government worker skills using residualized wages in comparable jobs in the private sector, relating these to skill-related observables using machine learning tools and then predicting government worker skills out-of-sample.
We then showed two main applications drawing on rich Indonesian household-level panel data.
First, we showed evidence for the selection of government workers.
Despite growing absolute skills, relative skills of government workers compared to private sector workers declined consistently between 1989 and 2014.
We linked this finding to the difficulty of the Indonesian government to attract the workers with the highest skills to the government.
Furthermore, we showed evidence for the detrimental effect of government hiring cycles on the selection of government workers.
The evidence is consistent with the idea that in years of outsized hiring, the government is forced to move down the skill distribution of the applicant pool to fill all government positions, leading to lower average skills.
The clear policy implication is that if the government wants to maximize talent in the public sector, it should smooth government hiring over time to not be forced to hire lower skilled workers in years with outsized demand. In the second main application, we looked at government wage setting and showed that the Indonesian government pays a wage premium of about \WpLcJobPct\% conditional on skills, about \GgExplainsWpPct\% of which is driven by the large gender wage gap in Indonesia's private sector.

A good sign of a new estimation approach is that it raises many interesting questions --- both conceptual and theoretical --- that can now be studied more rigorously: For example, what are the output or welfare costs of government hiring cycles? Or what drives the relative decline in government skills and does the relative decline in government skills hold across countries? Does the relative decline in skills go in hand with a relative decline in state capacity (compared to private sector capacity) over the course of development? All of these questions are particularly well-suited for future structural work on the functioning of bureaucracies, for which the estimated government skills in this paper can directly be used as inputs.
This is just one promising direction where the novel method proposed in this paper could be used to answer open questions in the literature.

The estimation approach allows for readily available extensions and further applications, which we leave for future work.
For example, in many settings with increasingly detailed administrative data, one will also be able to observe applicants directly in which case the selection rule proposed in this paper can be studied for different selection margins.
Furthermore, future research could utilize administrative data on promotion patterns to study whether the most skilled workers are more likely to be promoted.

\clearpage
\printbibliography
\clearpage

\appendix
% Resets the figure counter at every section and adds the section number
\counterwithin{figure}{section}
\counterwithin{table}{section}

\clearpage
\section{Further details on context and data (cleaning)}

\subsection{Details on the selection and hiring process of government workers}\label{sec:govt-hiring}

Entering the Civil Service is remarkably well-defined in Indonesia despite large changes to the Civil Service over the time period of interest as applicants run through a centralized application process.
Applicants apply to the specific position or district they are interested in, but still run through a centralized application process and due to frequent rotations and across-country stationing, are likely to end up with a position somewhere else than where they applied if they are admitted.
Formal requirements of applying to the Civil Service are that individuals have to be between 18-35 years old, never been imprisoned, not be a member of a political party, be in good physical and mental health and be willing to work in any region in Indonesia.
For each job opening there are then additional educational requirements, which are set by the district and which often mean an undergraduate diploma.
Since 2012,
%{Check this!!}
All applicants have to go through a civil servant enrollment test (\emph{CPNS}), which includes three parts: an administrative selection, a basic competence test (\emph{Seleksi Kompetensi Dasar}) recently administered via a computer-assisted test and a specific field competence selection (\emph{Seleksi Kompetensi Bidang}).\footnote{Under the Suharto regime, the Civil Service system was organized as a military-type organisation where new recruits were not differentiated other than by level of education, which duced generalist civil servants and abolished further specializations within the bureaucracy \parencite{mcleodPrivateSectorLessons2006}.
Recent reforms have tried to reverse this.}
%{Add ADB 2004, p. 58}

Solely based on aggregate numbers, obtaining a Civil Service job is difficult.
In 2014, prior to a 4-year public sector employment moratorium, there were more than 2.6 million applicants for 100,000 available positions, which translates into an acceptance rate of slightly below 4\% (see: \textcite{anandari2019}).
This is similar to the 1-5\% acceptance rates reported in Kristiansen and Ramli \parencite{kristiansenBuyingIncomeMarket2006} for two Indonesian regions in the early 2000s.\footnote{Note, the percentage of acceptances has increased in the author's sample, which would be in line with the model on the evolution of state capacity.}

In practice, it is unclear how well the recruitment system in place selects qualified candidates and how this changed over time.

\parencite{horhoruwTransformingPublicSector2013} notes that it is unclear whether reform processes since 2001 have actually led to an improvement of hiring practices beyond just a few reform-minded institutions.
In Pierskalla and Sacks \parencite{pierskallaPersonnelPoliticsElections2018}, the authors draw on teacher censuses to show that changes in the political system after 1998 actually had negative effects on public hiring.
They find that increased political competition gave local elites an incentive to use their discretionary control over state hiring to increase patronage efforts as evidenced by election-related increases in the number of contract teachers on local payrolls and increases in civil service teacher certifications.
At the same time,  \parencite{pierskallaDemocratizationRepresentativeBureaucracy2020} use data on the universe of civil servants to show that civil servants with a postgraduate education are twice as likely to be promoted after 1999 in comparison to before, indicating a combination of composition changes and more performance-related promotion patterns.

%{the selection process is not always competitive, with the recruitment system said to be characterised by informal payments for entry into the system and for promotions (World Bank, 2009a)}

Kristiansen and Ramli \parencite{kristiansenBuyingIncomeMarket2006} draw on in-depth qualitative and quantitative evidence from interviews and focus groups with a non-representative sample of 60 civil servants in two areas of Indonesia to document that personal ties and nepotism are often named as primary reasons for hiring.
Moreover, the selling of government jobs is widely practiced.
\textcite{kristiansenBuyingIncomeMarket2006} document that all respondents paid for their first Civil Service position and that the average reported price for these jobs is around 2.5 times the official annual initial salary offered.
This is slightly higher than the 17 months of salary reported recently in  \textcite{weaver2021jobs}.\footnote{The author does not share the country of study to provide additional security for their survey respondents.} There is also some evidence that the average real price for a government position has slowly increased between 1995-2004.
Prices are positively correlated with the salary of the job (which in turn is mechanically tied to the education level of the civil servant) and seem to be positively correlated with the ease of rent-seeking possibilities in the specific job offered.\footnote{Among the usual rent-seeking possibilities are various forms of contract kickbacks, payment from staff in exchange for positions and hiring on projects, loan accounts structured to earn interest by the agency, provision of ghost services, inflated invoicing in collusion with contractors, procedures for tax avoidance, irregular payments for health and education services, bribes to police officers and judges, and speed money to obtain formal papers and permits (World Bank 2003; Vian 2005; Chapman 2005; Azfar 2005).} This evidence on prices is in line with a competitive auction price for government sector jobs as found in  \textcite{weaver2021jobs}.
In the end, it is unclear how these unlawful hiring practices perform with respect to selecting the most qualified candidates as this depends on the correlation between quality and the ability to pay for a job or the probability of knowing someone important in the bureaucracy.
Interestingly,  \textcite{weaver2021jobs} finds that for the context they look at, this correlation is highly positive so that the selling of government sector jobs actually leads to a good selection rule in terms of quality of the new hires.

\subsection{Survey weights and imputation of weights}\label{sec:weights}

We use survey weight in our analysis.
Weights can improve descriptiveness by aligning the sample with a target population, and they can address selection when sampling or attrition is correlated with outcomes; however, they typically increase variance, and if effects are fairly homogeneous, weighted and unweighted estimates should be similar \parencite{roodman2025schoolinglabormarketconsequences}.
The IFLS documentation emphasizes that panel weights are intended only for the IFLS1 roster sample, while cross-sectional weights can be appropriate in longitudinal analyses when the target population is not restricted to IFLS1 \parencite{ifls_datanotes}.

We therefore use cross-sectional person weights to keep the target population aligned with each wave rather than restricting attention to the IFLS1 cohort.
For example, if one is interested in outcomes by wave $\tau$ for those present at wave $\tau - 1$, the IFLS documentation indicates that the wave $\tau - 1$ cross-sectional weight is appropriate even though the analysis uses longitudinal information \parencite{ifls_datanotes}.

Because we use person-years that fall between survey waves, we assign each observation the closest observed wave weight with a non-missing value.
For example, a 1996 person-year receives the 1997 cross-sectional weight when that weight is available, while a 1999 person-year receives the 2000 cross-sectional weight.
We do not use last-observation-carried-forward filling (e.g., filling downward and then upward) because that can systematically select the earlier wave even when the later wave is closer in time.
This approach assumes that wave weights are a reasonable proxy for nearby non-wave years, meaning that representativeness does not shift sharply between waves and selection into the wave sample is not correlated with outcomes in adjacent non-wave years beyond observed covariates.
Some individuals have no cross-sectional weight in any wave, so a small number of person-years are imputed to the overall median weight after this procedure.
Specifically, the issue concerns \ImputedWeightRows{} rows in the panel, or \ImputedWeightIndiv{} individuals, and those rows are imputed to the median weight.

\clearpage

\section{Further estimation details}

\subsection{Grid search parameters for machine learning methods}\label{sec:grid-search}

\begin{table}[t]
\centering
\caption{Hyperparameter grids used for cross-validated tuning}
\label{tab:grid-search-params}
\begin{threeparttable}
\begin{tabular}{lll}
\toprule
\toprule
Method & Hyperparameter & Values in grid \\
\midrule
LASSO & $\lambda$ & regularization path; choose $\lambda_{\min}$ by out-of-fold outcome $R^2$ \\
Random Forest & Number of trees & 300, 500, 800 \\
 & Variables per split & $\max(1,\lfloor\sqrt{p}\rfloor)$, $\max(1,\lfloor p/3\rfloor)$, $\max(1,\lfloor p/2\rfloor)$ \\
 & Min. node size & 5, 10 \\
GBM & Interaction depth & 4, 6 \\
 & Learning rate & 0.05, 0.1 \\
 & Bag fraction & 0.6, 0.8 \\
 & Number of trees & 200, 400 \\
\bottomrule
\bottomrule
\end{tabular}
\begin{tablenotes}[flushleft]
\footnotesize
\item \textit{\textbf{Notes:} $p$ is the number of predictors in the flexible component $x$.
Grid search uses 10-fold CV and selects the configuration that maximizes out-of-fold outcome $R^2$ (for $\widetilde{w_{it}}$).
Fixed-effect $R^2$ is reported separately as an out-of-fold worker-level diagnostic.}
\end{tablenotes}
\end{threeparttable}
\end{table}

\FloatBarrier
\subsection{Leave-one-group-out stability of skill predictions}\label{appendix-fx-stability}

The table below reports, for each leave-one-group-out scenario in \Cref{sec:robustness}, the Pearson correlation between centered full-sample skills and centered leave-out GBM predictions, the OLS slope through the origin, the number of worker-year observations in the prediction sample, and the number of distinct individuals.
For the leave-year-out scenarios, training sample size increases over time as later survey waves have larger panels, which accounts for the generally higher correlations in later years.
For the leave-occupation-out scenarios, the very low correlations for Military/Police and School Principal reflect the small size of those occupation training samples rather than evidence against skill generality.

\begin{table}
\centering
\begin{talltblr}[         %% tabularray outer open
entry=none,label=none,
note{}={Each row reports agreement between full-sample skills and reverse leave-one-group-out GBM skill predictions on the Private/public workers comparison sample.},
note{ }={For each row, the GBM is trained only on the indicated year or occupation and then used to predict skills for the selected comparison sample.},
note{  }={Correlations, recentering, and bins use the same propensity-adjusted worker-level survey weights as the public/private estimation sample.},
note{   }={The figures use 10 within-sample quantile bins of the centered leave-out skill.},
]                     %% tabularray outer close
{                     %% tabularray inner open
width={1\linewidth},
colspec={X[]X[]X[]X[]X[]X[]X[]},
column{1,2,3,4,5,6,7}={}{halign=l,},
}                     %% tabularray inner close
\toprule
Comparison & Group & Pearson r & Spearman rho & Slope & Training workers & People compared \\ \midrule %% TinyTableHeader
Train left-out year & 1989 & \num{0.607} & \num{0.630} & \num{0.480} & \num{1526.000} & \num{14950.000} \\
Train left-out year & 1990 & \num{0.547} & \num{0.586} & \num{0.389} & \num{1588.000} & \num{14950.000} \\
Train left-out year & 1991 & \num{0.572} & \num{0.623} & \num{0.329} & \num{1636.000} & \num{14950.000} \\
Train left-out year & 1992 & \num{0.584} & \num{0.609} & \num{0.475} & \num{2543.000} & \num{14950.000} \\
Train left-out year & 1993 & \num{0.681} & \num{0.712} & \num{0.611} & \num{3067.000} & \num{14950.000} \\
Train left-out year & 1994 & \num{0.684} & \num{0.708} & \num{0.811} & \num{2867.000} & \num{14950.000} \\
Train left-out year & 1995 & \num{0.716} & \num{0.721} & \num{0.825} & \num{3154.000} & \num{14950.000} \\
Train left-out year & 1996 & \num{0.806} & \num{0.818} & \num{0.898} & \num{4920.000} & \num{14950.000} \\
Train left-out year & 1997 & \num{0.817} & \num{0.805} & \num{0.856} & \num{5370.000} & \num{14950.000} \\
Train left-out year & 1998 & \num{0.814} & \num{0.841} & \num{0.895} & \num{4653.000} & \num{14950.000} \\
Train left-out year & 1999 & \num{0.785} & \num{0.826} & \num{0.795} & \num{4707.000} & \num{14950.000} \\
Train left-out year & 2000 & \num{0.679} & \num{0.705} & \num{0.526} & \num{4725.000} & \num{14950.000} \\
Train left-out year & 2007 & \num{0.821} & \num{0.825} & \num{0.739} & \num{4027.000} & \num{14950.000} \\
Train left-out year & 2014 & \num{0.870} & \num{0.858} & \num{0.947} & \num{5259.000} & \num{14950.000} \\
Train left-out occupation & Agriculture & \num{0.716} & \num{0.750} & \num{0.971} & \num{2935.000} & \num{14950.000} \\
Train left-out occupation & Health worker & \num{0.549} & \num{0.581} & \num{0.453} & \num{157.000} & \num{14950.000} \\
Train left-out occupation & Other & \num{0.475} & \num{0.470} & \num{0.367} & \num{116.000} & \num{14950.000} \\
Train left-out occupation & Other Clerical and related workers & \num{0.734} & \num{0.735} & \num{0.807} & \num{1356.000} & \num{14950.000} \\
Train left-out occupation & Other professional/technical & \num{0.594} & \num{0.646} & \num{0.339} & \num{480.000} & \num{14950.000} \\
Train left-out occupation & Production, Transport, Laborers & \num{0.812} & \num{0.810} & \num{0.840} & \num{6656.000} & \num{14950.000} \\
Train left-out occupation & Sales Workers & \num{0.766} & \num{0.766} & \num{0.742} & \num{2053.000} & \num{14950.000} \\
Train left-out occupation & Service Workers & \num{0.795} & \num{0.786} & \num{0.535} & \num{2839.000} & \num{14950.000} \\
Train left-out occupation & Teacher & \num{0.631} & \num{0.619} & \num{0.619} & \num{464.000} & \num{14950.000} \\
\bottomrule
\end{talltblr}
\end{table}

\FloatBarrier
\subsection{Common support for skill variables}\label{appendix-common-support}

\begin{figure}[t]
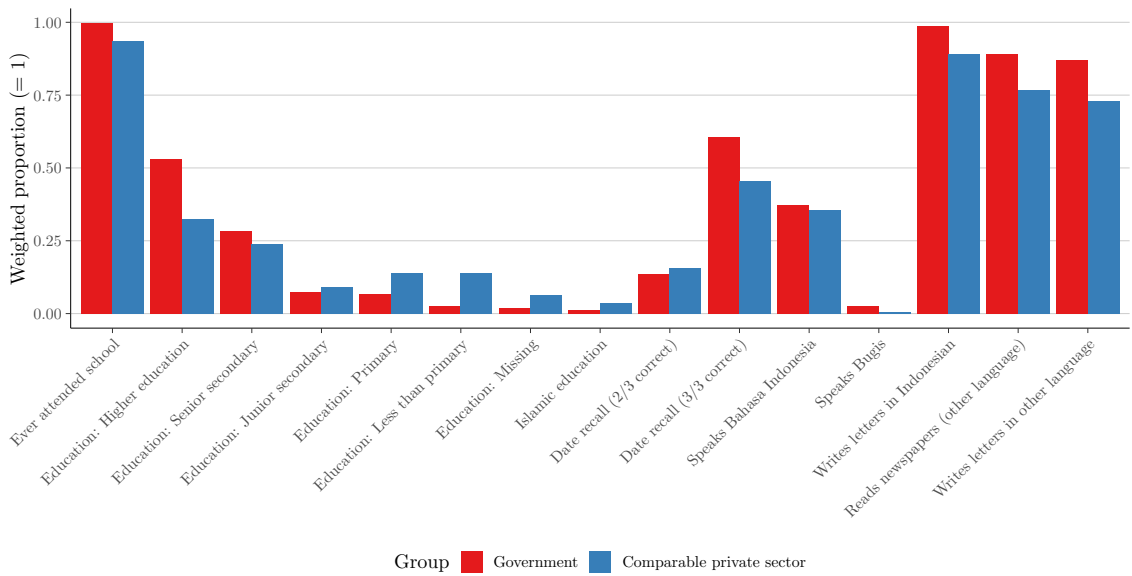

\caption{Common support: binary skill variables \label{fig:support-barcharts-dummies}}
{\centering
\resizebox{0.95\textwidth}{!}{
\input{out/img/support_barcharts_dummies.tex}
}
\par}
\input{out/fignotes/support_barcharts_dummies.tex}
\end{figure}

\begin{figure}[t]
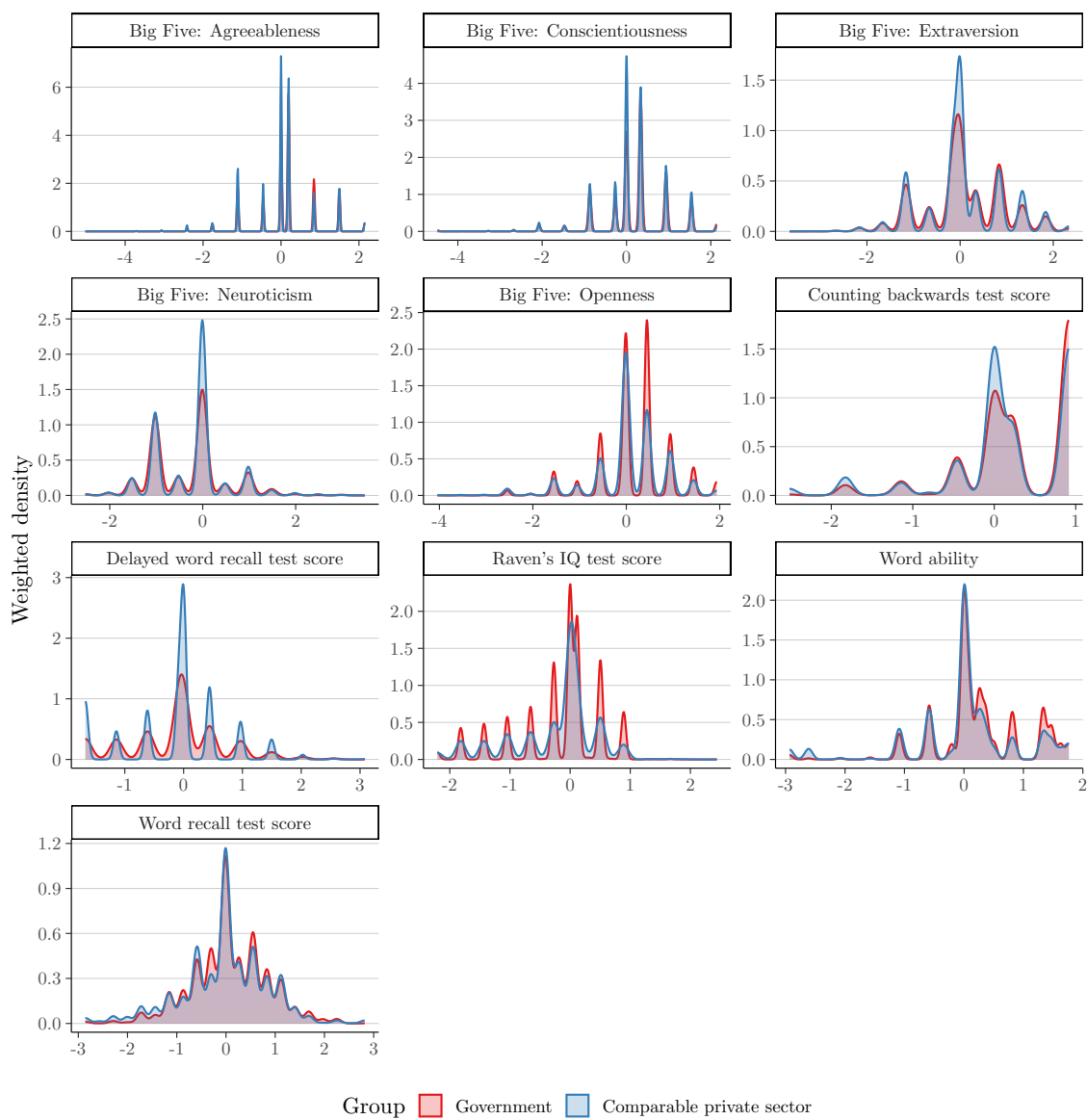

\caption{Common support: continuous skill variables \label{fig:support-relative-density}}
{\centering
\resizebox{0.95\textwidth}{!}{
\input{out/img/support_relative_density.tex}
}
\par}
\input{out/fignotes/support_relative_density.tex}
\end{figure}

\Cref{fig:support-barcharts-dummies} plots the weighted proportion equal to 1 for each of the 15 binary skill covariates, covering education levels, literacy and language skills, and date recall tasks, comparing government workers to the propensity-score-weighted private-sector comparison group.
Both groups have non-trivial shares at both zero and one for every variable.
\Cref{fig:support-relative-density} shows kernel density estimates for the 10 standardized continuous covariates: the five cognitive scores (word recall, delayed word recall, word ability, Raven IQ, and backward counting) and the Big Five personality traits.
The density curves of the two groups overlap across the full range of each variable.
Together, the figures confirm that the skill estimation does not rely on extrapolation beyond the support of the training data.

\FloatBarrier
\subsection{Missingness diagnostics for individual covariates}\label{sec:missingness-diagnostics}

\begin{figure}[t]
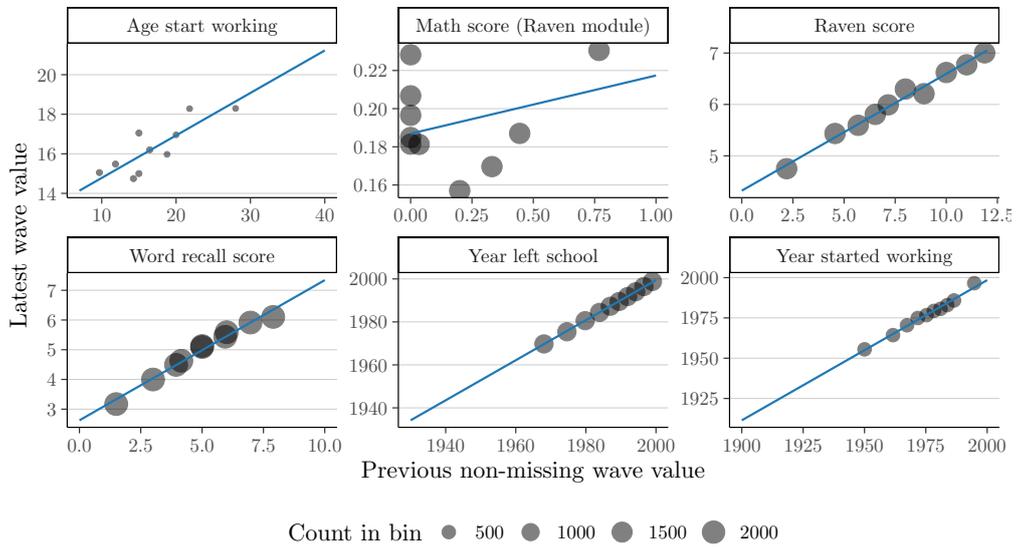

\caption{Cross-wave consistency of conflicting scores \label{fig:plot-conflict-score-correlations}}

{\centering

\resizebox{0.85\textwidth}{!}{
\input{out/img/conflict_score_correlations.tex}
}

\par}
\input{out/fignotes/conflict_score_correlations.tex}
\end{figure}

\Cref{tab:missingness-indiv-permanent-overall} reports missingness rates for the skill-related covariates used in the ML stage, computed on the worker-panel overlap sample.
The largest missingness rates are concentrated in the risk-preference measures (\RiskMissMinPct\% to \RiskMissMaxPct\%, excluding questions that were not asked because of different paths in the survey) and in psychometric and cognition variables collected only in later waves.
By contrast, core literacy variables and education categories show substantially lower missingness, generally below 10\%.
These patterns are consistent with module coverage differences across IFLS waves and motivate our imputation-plus-missing-indicator strategy discussed in the data preparation section.
\Cref{tab:missingness-indiv-permanent-overall} decomposes missingness into two sources: individuals not tracked in the wave(s) where a variable is measured, and individuals tracked in those waves but still missing the item.
It also reports cross-wave conflicts, defined as cases where a variable is observed in at least two waves and the recorded values disagree (missing-versus-observed pairs are not treated as conflicts).
Conflicts are generally rare, and we resolve them by taking the latest available wave value.
For variables with conflicts, \Cref{fig:plot-conflict-score-correlations} shows how the latest observed score relates to the previous non-missing wave value.
On average, scores do correlate across waves, but correlations are often modest, indicating substantial measurement noise in these items.
That noise is one reason the individual-variable $R^2$ values are limited, even though the broader feature set remains informative because many variables are jointly correlated with skills (as seen in the reported Pearson/Spearman patterns and the estimated cognitive $g$ factor/common-factor structure in the data).
For variables collected only in the final wave (Big Five traits, date task, delayed recall, backward counting, word ability), the dominant source of missingness is non-tracking in that wave, with item nonresponse playing a much smaller role.
For risk preference measures, the missingness is driven largely by tracked respondents who do not complete the module, indicating item-level nonresponse rather than attrition.
For multi-wave cognitive measures such as Raven and math scores, missingness is mixed: some is due to attrition from later waves and some is due to nonresponse among tracked respondents.
By contrast, core literacy and education indicators are available in multiple waves and show comparatively low missingness, consistent with their broader coverage.
Because non-tracking reflects panel attrition and late-wave coverage, it can induce selection if early attriters differ in latent skills or wage components not captured by observables, which is most relevant for W5-only psychometric modules.
We mitigate this by using cross-sectional weights that correct for attrition, carrying forward later-wave values when available, and including missing-indicator terms so that the ML stage conditions explicitly on missingness patterns.
The remaining concern is selection on unobservables among early attriters, which would primarily affect estimates that rely on late-wave psychometric measures rather than repeatedly measured literacy and education items.

\begin{longtblr}[         %% tabularray outer open
caption={Missingness in individual permanent covariates},
note{}={\textbf{Notes:} \textit{Missingness diagnostics are computed on the worker-panel overlap sample (N = 24,020). Percent missing is defined as missing observations divided by the diagnostics-sample size. Only variables with at least 1 percent missing are reported in this table. Survey-weight variables (pwt*) are excluded from this table. For categorical questions, response dummies are collapsed into one row per question. Diagnostics include individual permanent covariates, with education reported as discretized education groups. Missing: not tracked counts individuals not tracked in any source wave for that variable; Missing: tracked counts individuals tracked in at least one source wave but still missing. Variables with less than 1 percent missing (excluded from this table): Can write letter: indo (all responses), Can write letter: other (all responses), Count languages, Ever attended school (all responses), Language spoken (all responses), Male, Reads newspaper: indo (all responses), Reads newspaper: other (all responses). Variables with 0 missing observations: Age start working, Birth year, Education group: higher education, Education group: junior secondary, Education group: less than primary, Education group: missing, Education group: primary, Education group: senior secondary, Islamic education, Year started working.}},
label={tab:missingness-indiv-permanent-overall},
]                     %% tabularray outer close
{                     %% tabularray inner open
width={1\linewidth},
colspec={X[]X[]X[]X[]X[]},
column{1,2,3,4,5}={}{halign=l,},
colspec={X[2]X[]X[]X[]X[]},
hline{1}={1-5}{solid, black, 0.08em},hline{43}={1-5}{solid, black, 0.08em},
}                     %% tabularray inner close
\toprule
Variable & Missing observations & Missing share (\%) & Missing: not tracked & Missing: tracked \\ \midrule %% TinyTableHeader
Risk 22e (all responses) & 23,120 & 96.25 & 3,116 & 20,004 \\
Risk 21e (all responses) & 22,893 & 95.31 & 3,116 & 19,777 \\
Risk 22d (all responses) & 21,896 & 91.16 & 3,116 & 18,780 \\
Risk 15 (all responses) & 21,629 & 90.05 & 3,116 & 18,513 \\
Relative english score & 20,752 & 86.39 & 13,321 & 7,431 \\
Relative math score & 19,657 & 81.84 & 13,321 & 6,336 \\
Relative indonesia score & 19,580 & 81.52 & 13,321 & 6,259 \\
Risk 21d (all responses) & 19,315 & 80.41 & 3,116 & 16,199 \\
Risk 12 (all responses) & 19,237 & 80.09 & 3,116 & 16,121 \\
Draw correct (all responses) & 18,595 & 77.41 & 6,784 & 11,811 \\
Animals mentioned & 18,588 & 77.39 & 6,784 & 11,804 \\
Relative total score & 18,061 & 75.19 & 13,321 & 4,740 \\
Risk 05 (all responses) & 15,436 & 64.26 & 3,116 & 12,320 \\
Risk 04 (all responses) & 13,619 & 56.70 & 3,116 & 10,503 \\
Year left school & 12,669 & 52.74 & 3,664 & 9,005 \\
Risk 02 (all responses) & 11,993 & 49.93 & 3,116 & 8,877 \\
Math score (Raven module) & 8,093 & 33.69 & 3,797 & 4,296 \\
Risk 03 (all responses) & 7,659 & 31.89 & 3,116 & 4,543 \\
Backward counting task score & 7,339 & 30.55 & 6,784 & 555 \\
Word ability score & 6,812 & 28.36 & 6,784 & 28 \\
Big Five (all traits) & 6,798 & 28.30 & 6,798 & 0 \\
Date correct (all responses) & 6,798 & 28.30 & 6,784 & 14 \\
Delayed word recall score & 6,784 & 28.24 & 6,784 & 0 \\
Raven score & 5,716 & 23.80 & 3,797 & 1,919 \\
Risk 21c (all responses) & 5,125 & 21.34 & 3,116 & 2,009 \\
Risk 14 (all responses) & 5,074 & 21.12 & 3,116 & 1,958 \\
Gamble loving & 4,144 & 17.25 & 3,116 & 1,028 \\
Risk 13 (all responses) & 4,144 & 17.25 & 3,116 & 1,028 \\
Count ethnicities & 4,084 & 17.00 & 2,693 & 1,391 \\
Ethnicity (all responses) & 4,084 & 17.00 & 2,693 & 1,391 \\
Risk 22c (all responses) & 3,946 & 16.43 & 3,116 & 830 \\
Reason (all responses) & 3,797 & 15.81 & 3,797 & 0 \\
Word recall score & 3,363 & 14.00 & 3,157 & 206 \\
Risk 21b (all responses) & 3,352 & 13.96 & 3,116 & 236 \\
Risk 22b (all responses) & 3,276 & 13.64 & 3,116 & 160 \\
Day correct (all responses) & 3,164 & 13.17 & 3,157 & 7 \\
Gamble averse & 3,122 & 13.00 & 3,116 & 6 \\
Risk 01 (all responses) & 3,122 & 13.00 & 3,116 & 6 \\
Risk 11 (all responses) & 3,122 & 13.00 & 3,116 & 6 \\
Risk 21a (all responses) & 3,122 & 13.00 & 3,116 & 6 \\
Risk 22a (all responses) & 3,122 & 13.00 & 3,116 & 6 \\
\bottomrule
\end{longtblr}

\begin{longtblr}[         %% tabularray outer open
caption={Cross-wave conflicts in one-answer individual covariates},
note{}={\textbf{Notes:} \textit{Conflict diagnostics are computed on the worker-panel overlap sample (N = 24,020). Conflict diagnostics are restricted to variables that should have one unique answer over time. Conflicts are counted only when a variable is non-missing in at least two waves and the values differ; missing-versus-observed pairs are not treated as conflicts. Conflict share is defined as conflicts divided by the number of individuals with non-missing observations in at least two waves. Only variables with at least one conflict are reported.}},
label={tab:conflicts-indiv-permanent-overall},
]                     %% tabularray outer close
{                     %% tabularray inner open
width={1\linewidth},
colspec={X[]X[]X[]X[]X[]X[]},
column{1,2,3,4,5,6}={}{halign=l,},
colspec={X[2]X[]X[]X[]X[]X[]},
hline{1}={1-6}{solid, black, 0.08em},hline{5}={1-6}{solid, black, 0.08em},
}                     %% tabularray inner close
\toprule
Variable & Conflicts (n) & Conflict share (\%) & Multi-wave non-missing (n) & Waves collected & Conflict handling \\ \midrule %% TinyTableHeader
Year left school & 8,026 & 68.28 & 11,754 & W2, W3, W4 & Last non-missing (later wave) \\
Year started working & 3,912 & 77.14 & 5,071 & W1, W2, W3, W4, W5 & Last non-missing (later wave) \\
Age start working & 658 & 83.72 & 786 & W1, W2, W3, W4, W5 & Last non-missing (later wave) \\
\bottomrule
\end{longtblr}

\clearpage

\clearpage

\clearpage
\FloatBarrier
\section{Additional tables}\label{sec:tables}
\begin{table}
\centering
\begin{talltblr}[         %% tabularray outer open
caption={Main observable differences: government vs. comparable private sector vs. other workers\label{tab:table-main-skill-differences}},
note{}={\textbf{Notes:} \textit{Based on pooled data and restricting to workers in working age (between 16 to 70 years old), positive work hours, and non-missings in age, experience, and annual net income. Columns include 13,546 government worker-year observations, 60,294 propensity-score-reweighted comparable private-sector observations, 60,544 all private-sector observations, and 76,732 other workers (self-employed, casual, unpaid family). The total sample comprises 150,822 worker-year observations (aged 16-70, positive hours, valid age/experience/income). Unique government workers: n = 2,894 of 31,232 (9.3\%) of total workers. Comparable private-sector workers are reweighted by propensity scores (sector, occupation, age, and a male indicator). All private workers use survey weights without propensity-score reweighting. Other includes all non-government, non-private workers (e.g., self-employed, casual, unpaid family) using survey weights. Mean relative wage is the ratio of individual hourly wages to the year-specific arithmetic mean wage, computed after winsorizing log wages at the 2.5th and 97.5th percentiles before exponentiating. Apart from dummy variables, all skill-related observables are z-standardized across all individuals.}},
]                     %% tabularray outer close
{                     %% tabularray inner open
width={1\linewidth},
colspec={X[]X[]X[]X[]X[]X[]X[]X[]X[]},
cell{2}{1}={}{halign=l,},
cell{3}{1}={}{halign=l,},
cell{4}{1}={}{halign=l,},
cell{5}{1}={}{halign=l,},
cell{6}{1}={}{halign=l,},
cell{7}{1}={}{halign=l,},
cell{8}{1}={}{halign=l,},
cell{9}{1}={}{halign=l,},
cell{10}{1}={}{halign=l,},
cell{11}{1}={}{halign=l,},
cell{12}{1}={}{halign=l,},
cell{13}{1}={}{halign=l,},
cell{1}{1}={}{halign=l, halign=c,},
cell{2}{2}={}{halign=r,},
cell{2}{3}={}{halign=r,},
cell{2}{4}={}{halign=r,},
cell{2}{5}={}{halign=r,},
cell{2}{6}={}{halign=r,},
cell{2}{7}={}{halign=r,},
cell{2}{8}={}{halign=r,},
cell{2}{9}={}{halign=r,},
cell{3}{2}={}{halign=r,},
cell{3}{3}={}{halign=r,},
cell{3}{4}={}{halign=r,},
cell{3}{5}={}{halign=r,},
cell{3}{6}={}{halign=r,},
cell{3}{7}={}{halign=r,},
cell{3}{8}={}{halign=r,},
cell{3}{9}={}{halign=r,},
cell{4}{2}={}{halign=r,},
cell{4}{3}={}{halign=r,},
cell{4}{4}={}{halign=r,},
cell{4}{5}={}{halign=r,},
cell{4}{6}={}{halign=r,},
cell{4}{7}={}{halign=r,},
cell{4}{8}={}{halign=r,},
cell{4}{9}={}{halign=r,},
cell{5}{2}={}{halign=r,},
cell{5}{3}={}{halign=r,},
cell{5}{4}={}{halign=r,},
cell{5}{5}={}{halign=r,},
cell{5}{6}={}{halign=r,},
cell{5}{7}={}{halign=r,},
cell{5}{8}={}{halign=r,},
cell{5}{9}={}{halign=r,},
cell{6}{2}={}{halign=r,},
cell{6}{3}={}{halign=r,},
cell{6}{4}={}{halign=r,},
cell{6}{5}={}{halign=r,},
cell{6}{6}={}{halign=r,},
cell{6}{7}={}{halign=r,},
cell{6}{8}={}{halign=r,},
cell{6}{9}={}{halign=r,},
cell{7}{2}={}{halign=r,},
cell{7}{3}={}{halign=r,},
cell{7}{4}={}{halign=r,},
cell{7}{5}={}{halign=r,},
cell{7}{6}={}{halign=r,},
cell{7}{7}={}{halign=r,},
cell{7}{8}={}{halign=r,},
cell{7}{9}={}{halign=r,},
cell{8}{2}={}{halign=r,},
cell{8}{3}={}{halign=r,},
cell{8}{4}={}{halign=r,},
cell{8}{5}={}{halign=r,},
cell{8}{6}={}{halign=r,},
cell{8}{7}={}{halign=r,},
cell{8}{8}={}{halign=r,},
cell{8}{9}={}{halign=r,},
cell{9}{2}={}{halign=r,},
cell{9}{3}={}{halign=r,},
cell{9}{4}={}{halign=r,},
cell{9}{5}={}{halign=r,},
cell{9}{6}={}{halign=r,},
cell{9}{7}={}{halign=r,},
cell{9}{8}={}{halign=r,},
cell{9}{9}={}{halign=r,},
cell{10}{2}={}{halign=r,},
cell{10}{3}={}{halign=r,},
cell{10}{4}={}{halign=r,},
cell{10}{5}={}{halign=r,},
cell{10}{6}={}{halign=r,},
cell{10}{7}={}{halign=r,},
cell{10}{8}={}{halign=r,},
cell{10}{9}={}{halign=r,},
cell{11}{2}={}{halign=r,},
cell{11}{3}={}{halign=r,},
cell{11}{4}={}{halign=r,},
cell{11}{5}={}{halign=r,},
cell{11}{6}={}{halign=r,},
cell{11}{7}={}{halign=r,},
cell{11}{8}={}{halign=r,},
cell{11}{9}={}{halign=r,},
cell{12}{2}={}{halign=r,},
cell{12}{3}={}{halign=r,},
cell{12}{4}={}{halign=r,},
cell{12}{5}={}{halign=r,},
cell{12}{6}={}{halign=r,},
cell{12}{7}={}{halign=r,},
cell{12}{8}={}{halign=r,},
cell{12}{9}={}{halign=r,},
cell{13}{2}={}{halign=r,},
cell{13}{3}={}{halign=r,},
cell{13}{4}={}{halign=r,},
cell{13}{5}={}{halign=r,},
cell{13}{6}={}{halign=r,},
cell{13}{7}={}{halign=r,},
cell{13}{8}={}{halign=r,},
cell{13}{9}={}{halign=r,},
cell{1}{3}={}{halign=r, halign=c,},
cell{1}{5}={}{halign=r, halign=c,},
cell{1}{7}={}{halign=r, halign=c,},
cell{1}{9}={}{halign=r, halign=c,},
cell{1}{2}={c=2,}{halign=r, halign=c, halign=c,},
cell{1}{4}={c=2,}{halign=r, halign=c, halign=c,},
cell{1}{6}={c=2,}{halign=r, halign=c, halign=c,},
cell{1}{8}={c=2,}{halign=r, halign=c, halign=c,},
colspec={X[2]X[]X[]X[]X[]X[]X[]X[]X[]},
hline{1}={1-9}{solid, black, 0.08em},hline{13}={1-9}{solid, black, 0.08em},
}                     %% tabularray inner close
\toprule
& Government &  & Comp. private &  & All private &  & Other &  \\ \cmidrule[lr]{2-3}\cmidrule[lr]{4-5}\cmidrule[lr]{6-7}\cmidrule[lr]{8-9}
& Mean & SD & Mean & SD & Mean & SD & Mean & SD \\ \midrule %% TinyTableHeader
Male & 0.68 & 0.46 & 0.68 & 0.47 & 0.65 & 0.48 & 0.63 & 0.48 \\
Age & 39 & 9.7 & 41 & 13 & 34 & 12 & 42 & 12 \\
Relative wage & 1.7 & 1.5 & 1.3 & 1.5 & 0.88 & 1.2 & 0.92 & 1.4 \\
Weekly hours & 39 & 15 & 36 & 19 & 39 & 20 & 37 & 22 \\
Higher educ & 0.53 & 0.5 & 0.35 & 0.48 & 0.086 & 0.28 & 0.033 & 0.18 \\
Word recall & 0.1 & 0.78 & 0.03 & 0.88 & -0.22 & 0.92 & -0.5 & 0.93 \\
Speak Indonesian & 0.38 & 0.49 & 0.37 & 0.48 & 0.26 & 0.44 & 0.16 & 0.37 \\
Word ability & 0.28 & 0.77 & 0.11 & 0.91 & -0.15 & 0.95 & -0.34 & 0.97 \\
Raven IQ & -0.12 & 0.69 & -0.18 & 0.74 & -0.34 & 0.82 & -0.54 & 0.8 \\
Count backwards & 0.28 & 0.68 & 0.15 & 0.76 & -0.05 & 0.86 & -0.089 & 0.85 \\
Date correct (3/3) & 0.61 & 0.49 & 0.48 & 0.5 & 0.41 & 0.49 & 0.32 & 0.46 \\
\bottomrule
\end{talltblr}
\end{table}

\begin{table}
\centering
\begin{talltblr}[         %% tabularray outer open
caption={Propensity score logit\label{tab:propensity-score-logit}},
note{}={\textbf{Notes:} \textit{Sample restricted to government and private-sector workers with non-missing sector, occupation, age, male, and survey weights. Survey weights applied; sector and occupation dummies included but omitted from the table.}},
]                     %% tabularray outer close
{                     %% tabularray inner open
width={1\linewidth},
colspec={X[]X[]},
column{2}={}{halign=c,},
column{1}={}{halign=l,},
hline{8}={1,2}{solid, black, 0.05em},
colspec={X[2]X[]},
hline{1}={1-2}{solid, black, 0.08em},hline{11}={1-2}{solid, black, 0.08em},
}                     %% tabularray inner close
\toprule
& Logit \\ \midrule %% TinyTableHeader
Constant & -5.53 \\
& (0.05) \\
Age & 0.05 \\
& (0.00) \\
Male & 0.53 \\
& (0.02) \\
\emph{Fit statistics} &  \\
Observations & 173153 \\
SEs & IID \\
\bottomrule
\end{talltblr}
\end{table}

\begin{table}
\centering
\begin{talltblr}[         %% tabularray outer open
caption={Fixed-effects wage profile used for skill residualization\label{tab:table-skill-fe}},
note{}={\textbf{Notes:} \textit{Sample restricted to private-sector observations. Reported coefficients come from a worker fixed-effects regression of log hourly wages on linear and quadratic experience with province fixed effects, weighted by survey weights. This regression is used to partial out the common wage profile before constructing worker skill as the worker-level mean residual after removing the experience profile and province effects.}},
]                     %% tabularray outer close
{                     %% tabularray inner open
width={1\linewidth},
colspec={X[]X[]},
column{2}={}{halign=c,},
column{1}={}{halign=l,},
hline{6}={1,2}{solid, black, 0.05em},
colspec={X[2]X[]},
hline{1}={1-2}{solid, black, 0.08em},hline{13}={1-2}{solid, black, 0.08em},
}                     %% tabularray inner close
\toprule
& Within estimator \\ \midrule %% TinyTableHeader
Experience & 0.031 \\
& (0.009) \\
Experience square & -0.001 \\
& (0.000) \\
\emph{Fit statistics} &  \\
Observations & 50042 \\
R2 & 0.79 \\
R2 (within) & 0.01 \\
SEs & by: pidlink \\
Individual fixed effect & X \\
Province dummies & X \\
\bottomrule
\end{talltblr}
\end{table}

\FloatBarrier
\begin{longtblr}[         %% tabularray outer open
caption={In-sample and out-of-fold R2 by minimum years observed},
note{}={\textbf{Notes:} \textit{Rows restrict the worker-level skill sample to pidlink observed for at least Min years observed private-sector years. In-sample R2 uses fitted values from each final model trained on the full worker-level sample and is computed as weighted 1 - SSE/TSS. Out-of-sample R2 uses worker-level out-of-fold predictions from the same cross-validation folds and the same weighted 1 - SSE/TSS definition.}},
label={tab:r2-by-years-observed},
]                     %% tabularray outer close
{                     %% tabularray inner open
width={1\linewidth},
colspec={X[]X[]X[]X[]X[]},
column{1,2,3,4,5}={}{halign=l,},
colspec={X[2]X[]X[]X[]X[]},
hline{1}={1-5}{solid, black, 0.08em},hline{37}={1-5}{solid, black, 0.08em},
}                     %% tabularray inner close
\toprule
Method & Min years observed & N workers & In-sample R2 & Out-of-sample R2 \\ \midrule %% TinyTableHeader
Edu OLS & 1 & 13311 & 0.114 & 0.096 \\
Edu OLS & 2 & 8423 & 0.178 & 0.164 \\
Edu OLS & 3 & 6663 & 0.202 & 0.186 \\
Edu OLS & 4 & 5580 & 0.218 & 0.204 \\
Edu OLS & 5 & 4570 & 0.210 & 0.193 \\
Edu OLS & 6 & 3317 & 0.228 & 0.221 \\
Edu OLS & 7 & 2471 & 0.247 & 0.239 \\
Basis & 1 & 13311 & 0.337 & -0.013 \\
Basis & 2 & 8423 & 0.249 & -0.002 \\
Basis & 3 & 6663 & 0.280 & 0.020 \\
Basis & 4 & 5580 & 0.252 & -0.012 \\
Basis & 5 & 4570 & 0.245 & -0.028 \\
Basis & 6 & 3317 & 0.236 & -0.035 \\
Basis & 7 & 2471 & 0.257 & -0.066 \\
LASSO & 1 & 13311 & 0.219 & 0.116 \\
LASSO & 2 & 8423 & 0.221 & 0.172 \\
LASSO & 3 & 6663 & 0.252 & 0.197 \\
LASSO & 4 & 5580 & 0.248 & 0.201 \\
LASSO & 5 & 4570 & 0.261 & 0.218 \\
LASSO & 6 & 3317 & 0.285 & 0.251 \\
LASSO & 7 & 2471 & 0.308 & 0.267 \\
RF & 1 & 13311 & 0.603 & 0.108 \\
RF & 2 & 8423 & 0.526 & 0.183 \\
RF & 3 & 6663 & 0.540 & 0.203 \\
RF & 4 & 5580 & 0.534 & 0.205 \\
RF & 5 & 4570 & 0.534 & 0.208 \\
RF & 6 & 3317 & 0.571 & 0.255 \\
RF & 7 & 2471 & 0.584 & 0.266 \\
GBM & 1 & 13311 & 0.189 & 0.118 \\
GBM & 2 & 8423 & 0.174 & 0.159 \\
GBM & 3 & 6663 & 0.201 & 0.182 \\
GBM & 4 & 5580 & 0.209 & 0.187 \\
GBM & 5 & 4570 & 0.223 & 0.201 \\
GBM & 6 & 3317 & 0.242 & 0.221 \\
GBM & 7 & 2471 & 0.267 & 0.242 \\
\bottomrule
\end{longtblr}

\begin{table}
\centering
\begin{talltblr}[         %% tabularray outer open
caption={Baseline model prediction accuracy by minimum years observed\label{tab:pred-by-years-observed}},
note{}={\textbf{Notes:} \textit{Models are trained once on the full baseline sample (all workers with 1+ private-sector years observed) and not re-estimated. We apply the fixed estimated models and report the predicted R2 for workers with at least Min years observed. R2 is computed as weighted 1 - SSE/TSS.}},
]                     %% tabularray outer close
{                     %% tabularray inner open
width={1\linewidth},
colspec={X[]X[]X[]X[]},
column{1,2,3,4}={}{halign=l,},
colspec={X[2]X[]X[]X[]},
hline{1}={1-4}{solid, black, 0.08em},hline{37}={1-4}{solid, black, 0.08em},
}                     %% tabularray inner close
\toprule
Method & Min years observed & N workers & Out-of-sample R2 \\ \midrule %% TinyTableHeader
Edu OLS & 1 & 13311 & 0.096 \\
Edu OLS & 2 & 8423 & 0.164 \\
Edu OLS & 3 & 6663 & 0.186 \\
Edu OLS & 4 & 5580 & 0.204 \\
Edu OLS & 5 & 4570 & 0.193 \\
Edu OLS & 6 & 3317 & 0.221 \\
Edu OLS & 7 & 2471 & 0.239 \\
Basis & 1 & 13311 & -0.013 \\
Basis & 2 & 8423 & -0.002 \\
Basis & 3 & 6663 & 0.020 \\
Basis & 4 & 5580 & -0.012 \\
Basis & 5 & 4570 & -0.028 \\
Basis & 6 & 3317 & -0.035 \\
Basis & 7 & 2471 & -0.066 \\
LASSO & 1 & 13311 & 0.116 \\
LASSO & 2 & 8423 & 0.172 \\
LASSO & 3 & 6663 & 0.197 \\
LASSO & 4 & 5580 & 0.201 \\
LASSO & 5 & 4570 & 0.218 \\
LASSO & 6 & 3317 & 0.251 \\
LASSO & 7 & 2471 & 0.267 \\
RF & 1 & 13311 & 0.108 \\
RF & 2 & 8423 & 0.183 \\
RF & 3 & 6663 & 0.203 \\
RF & 4 & 5580 & 0.205 \\
RF & 5 & 4570 & 0.208 \\
RF & 6 & 3317 & 0.255 \\
RF & 7 & 2471 & 0.266 \\
GBM & 1 & 13311 & 0.118 \\
GBM & 2 & 8423 & 0.159 \\
GBM & 3 & 6663 & 0.182 \\
GBM & 4 & 5580 & 0.187 \\
GBM & 5 & 4570 & 0.201 \\
GBM & 6 & 3317 & 0.221 \\
GBM & 7 & 2471 & 0.242 \\
\bottomrule
\end{talltblr}
\end{table}

\begin{table}
\centering
\begin{talltblr}[         %% tabularray outer open
caption={GBM variable importance},
note{}={\textbf{Notes:} \textit{Relative influence measure from the GBM model: sum of squared improvements from each split, averaged across trees and normalized to 100.}},
label={tab:gbm-variable-importance},
]                     %% tabularray outer close
{                     %% tabularray inner open
width={1\linewidth},
colspec={X[]X[]},
column{1,2}={}{halign=l,},
colspec={X[2]X[]},
hline{1}={1-2}{solid, black, 0.08em},hline{27}={1-2}{solid, black, 0.08em},
}                     %% tabularray inner close
\toprule
Variable & Variable importance \\ \midrule %% TinyTableHeader
Education: Higher education & 100.0 \\
Reads newspapers (other language) & 53.1 \\
Big Five: Neuroticism & 43.9 \\
Big Five: Extraversion & 39.3 \\
Raven's IQ test score & 30.3 \\
Big Five: Conscientiousness & 28.8 \\
Counting backwards test score & 25.1 \\
Word ability & 20.3 \\
Big Five: Openness & 17.8 \\
Education: Less than primary & 15.4 \\
Speaks Bahasa Indonesia & 12.8 \\
Speaks Bugis & 12.3 \\
Education: Senior secondary & 8.4 \\
Writes letters in Indonesian & 7.7 \\
Big Five: Agreeableness & 6.7 \\
Word recall test score & 6.4 \\
Delayed word recall test score & 5.6 \\
Education: Primary & 3.6 \\
Date recall (3/3 correct) & 2.6 \\
Date recall (2/3 correct) & 2.3 \\
Education: Missing & 0.9 \\
Writes letters in other language & 0.7 \\
Ever attended school & 0.7 \\
Education: Junior secondary & 0.0 \\
Islamic education & 0.0 \\
\bottomrule
\end{talltblr}
\end{table}

\clearpage
\FloatBarrier
\section{Additional empirical results}\label{sec:additional-empirical-results}

This section provides additional results for Section 4 that are in part referenced in the main text.

\begin{figure}[t]
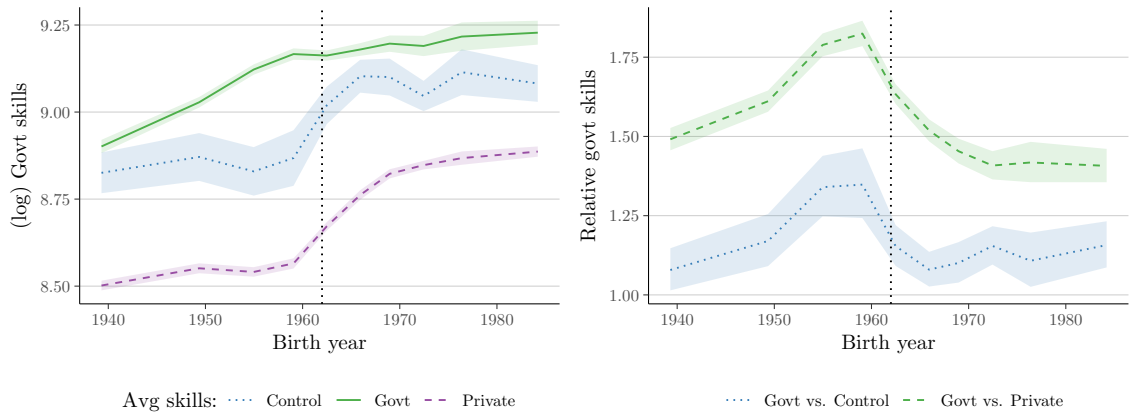

\caption{Government worker skills across birth cohorts (job composition held fixed) \label{fig:plot-evolution-govt-skills-cohorts-jobfix}}
{\centering
\resizebox{0.95\textwidth}{!}{
\input{out/img/evolution_govt_skills_by_cohort_bin_jobfix.tex}
}
\par}
\input{out/fignotes/evolution_govt_skills_by_cohort_bin_jobfix.tex}
\end{figure}

\Cref{fig:plot-evolution-govt-skills-cohorts-jobfix} shows absolute and relative government worker skills across (binned) cohorts, now additionally holding the composition of government jobs fixed.
The difference between \Cref{fig:plot-evolution-govt-skills-cohorts-jobfix} and \Cref{fig:plot-evolution-govt-skills-cohorts} thus gives the importance of compositional changes in the skill intensity of government jobs over time in explaining changes in government worker skills.
Differences are very small in magnitude, indicating that changes in the composition of government jobs in terms of sector and occupation cannot explain large changes in the relative skills of government workers.
The corresponding hiring-intensity regressions that hold government jobs fixed are reported as columns 2, 5, and 6 of \Cref{tab:table-govt-hiring} in the main text.

\Cref{tab:table-govt-hiring-nodrop,tab:table-govt-hiring-min10,tab:table-govt-hiring-min50} re-run the six specifications of \Cref{tab:table-govt-hiring} without dropping any birth-year cohorts, and after dropping all birth-year cohorts with fewer than 10 or 50 government worker-year observations, respectively, to check that the estimated relationship between the government employment share and relative government worker skills is not being driven by the specific cohort sample used in the main text.

\begin{table}
\centering
\begin{talltblr}[         %% tabularray outer open
caption={Robustness: the government skill premium and government hiring intensity, without dropping any cohorts\label{tab:table-govt-hiring-nodrop}},
note{}={\textbf{Notes:} \textit{All data restricted to workers between the age of 25 and 58, the official age span of government workers. Same specifications and column order as \Cref{tab:table-govt-hiring}, but without dropping any birth-year cohorts from the analysis sample. Cohort-level regressions of the relative skills of government workers vs. private sector workers (``Rel. govt skills'') on government hiring intensity. Columns 1--2 use the level of the government employment share; columns 3--6 use first differences. Columns 2, 5, and 6 hold government jobs fixed: government worker skills are residualized out of occupation and sector effects from a government-only regression, with a counterfactual reference at (Teacher, Social services), and rescaled so the weighted government mean matches the raw government mean. Non-government worker skills are unchanged. The cohort sample is identical across all six columns. The `Cohort trend' row indicates whether a cubic birth-cohort polynomial is included as an additional control; its coefficients and the constant are suppressed. Heteroskedasticity-robust standard errors in parentheses.}},
]                     %% tabularray outer close
{                     %% tabularray inner open
width={1\linewidth},
colspec={X[]X[]X[]X[]X[]X[]X[]},
column{2,3,4,5,6,7}={}{halign=c,},
column{1}={}{halign=l,},
hline{6}={1,2,3,4,5,6,7}{solid, black, 0.05em},
colspec={X[2]X[]X[]X[]X[]X[]X[]},
hline{1}={1-7}{solid, black, 0.08em},hline{10}={1-7}{solid, black, 0.08em},
}                     %% tabularray inner close
\toprule
& Rel. govt skills & Rel. govt skills  & $\Delta$ Rel. govt skills & $\Delta$ Rel. govt skills  & $\Delta$ Rel. govt skills   & $\Delta$ Rel. govt skills    \\ \midrule %% TinyTableHeader
Govt empl share & -0.18 & -0.16 &  &  &  &  \\
& (0.60) & (0.60) &  &  &  &  \\
$\Delta$ Govt empl share &  &  & -0.82 & -0.83 & -0.95 & -0.96 \\
&  &  & (0.47) & (0.50) & (0.43) & (0.46) \\
Observations & 57 & 57 & 55 & 55 & 55 & 55 \\
R2 & 0.58 & 0.55 & 0.07 & 0.08 & 0.11 & 0.12 \\
Cohort trend & Yes & Yes & No & Yes & No & Yes \\
Holding govt jobs fixed & No & Yes & No & No & Yes & Yes \\
\bottomrule
\end{talltblr}
\end{table}

\begin{table}
\centering
\begin{talltblr}[         %% tabularray outer open
caption={Robustness: the government skill premium and government hiring intensity, dropping cohorts with fewer than 10 government worker observations\label{tab:table-govt-hiring-min10}},
note{}={\textbf{Notes:} \textit{All data restricted to workers between the age of 25 and 58, the official age span of government workers. Same specifications and column order as \Cref{tab:table-govt-hiring}, but dropping all birth-year cohorts with fewer than 10 government worker-year observations in the analysis sample. Cohort-level regressions of the relative skills of government workers vs. private sector workers (``Rel. govt skills'') on government hiring intensity. Columns 1--2 use the level of the government employment share; columns 3--6 use first differences. Columns 2, 5, and 6 hold government jobs fixed: government worker skills are residualized out of occupation and sector effects from a government-only regression, with a counterfactual reference at (Teacher, Social services), and rescaled so the weighted government mean matches the raw government mean. Non-government worker skills are unchanged. The cohort sample is identical across all six columns. The `Cohort trend' row indicates whether a cubic birth-cohort polynomial is included as an additional control; its coefficients and the constant are suppressed. Heteroskedasticity-robust standard errors in parentheses.}},
]                     %% tabularray outer close
{                     %% tabularray inner open
width={1\linewidth},
colspec={X[]X[]X[]X[]X[]X[]X[]},
column{2,3,4,5,6,7}={}{halign=c,},
column{1}={}{halign=l,},
hline{6}={1,2,3,4,5,6,7}{solid, black, 0.05em},
colspec={X[2]X[]X[]X[]X[]X[]X[]},
hline{1}={1-7}{solid, black, 0.08em},hline{10}={1-7}{solid, black, 0.08em},
}                     %% tabularray inner close
\toprule
& Rel. govt skills & Rel. govt skills  & $\Delta$ Rel. govt skills & $\Delta$ Rel. govt skills  & $\Delta$ Rel. govt skills   & $\Delta$ Rel. govt skills    \\ \midrule %% TinyTableHeader
Govt empl share & -0.33 & -0.39 &  &  &  &  \\
& (0.63) & (0.60) &  &  &  &  \\
$\Delta$ Govt empl share &  &  & -0.82 & -0.81 & -0.98 & -0.98 \\
&  &  & (0.50) & (0.50) & (0.46) & (0.45) \\
Observations & 55 & 55 & 54 & 54 & 54 & 54 \\
R2 & 0.53 & 0.50 & 0.07 & 0.08 & 0.11 & 0.12 \\
Cohort trend & Yes & Yes & No & Yes & No & Yes \\
Holding govt jobs fixed & No & Yes & No & No & Yes & Yes \\
\bottomrule
\end{talltblr}
\end{table}

\begin{table}
\centering
\begin{talltblr}[         %% tabularray outer open
caption={Robustness: the government skill premium and government hiring intensity, dropping cohorts with fewer than 50 government worker observations\label{tab:table-govt-hiring-min50}},
note{}={\textbf{Notes:} \textit{All data restricted to workers between the age of 25 and 58, the official age span of government workers. Same specifications and column order as \Cref{tab:table-govt-hiring}, but dropping all birth-year cohorts with fewer than 50 government worker-year observations in the analysis sample. Cohort-level regressions of the relative skills of government workers vs. private sector workers (``Rel. govt skills'') on government hiring intensity. Columns 1--2 use the level of the government employment share; columns 3--6 use first differences. Columns 2, 5, and 6 hold government jobs fixed: government worker skills are residualized out of occupation and sector effects from a government-only regression, with a counterfactual reference at (Teacher, Social services), and rescaled so the weighted government mean matches the raw government mean. Non-government worker skills are unchanged. The cohort sample is identical across all six columns. The `Cohort trend' row indicates whether a cubic birth-cohort polynomial is included as an additional control; its coefficients and the constant are suppressed. Heteroskedasticity-robust standard errors in parentheses.}},
]                     %% tabularray outer close
{                     %% tabularray inner open
width={1\linewidth},
colspec={X[]X[]X[]X[]X[]X[]X[]},
column{2,3,4,5,6,7}={}{halign=c,},
column{1}={}{halign=l,},
hline{6}={1,2,3,4,5,6,7}{solid, black, 0.05em},
colspec={X[2]X[]X[]X[]X[]X[]X[]},
hline{1}={1-7}{solid, black, 0.08em},hline{10}={1-7}{solid, black, 0.08em},
}                     %% tabularray inner close
\toprule
& Rel. govt skills & Rel. govt skills  & $\Delta$ Rel. govt skills & $\Delta$ Rel. govt skills  & $\Delta$ Rel. govt skills   & $\Delta$ Rel. govt skills    \\ \midrule %% TinyTableHeader
Govt empl share & -0.47 & -0.70 &  &  &  &  \\
& (0.69) & (0.62) &  &  &  &  \\
$\Delta$ Govt empl share &  &  & -0.94 & -0.97 & -1.12 & -1.14 \\
&  &  & (0.50) & (0.51) & (0.45) & (0.47) \\
Observations & 43 & 43 & 42 & 42 & 42 & 42 \\
R2 & 0.49 & 0.47 & 0.12 & 0.14 & 0.19 & 0.21 \\
Cohort trend & Yes & Yes & No & Yes & No & Yes \\
Holding govt jobs fixed & No & Yes & No & No & Yes & Yes \\
\bottomrule
\end{talltblr}
\end{table}

\begin{figure}[t]
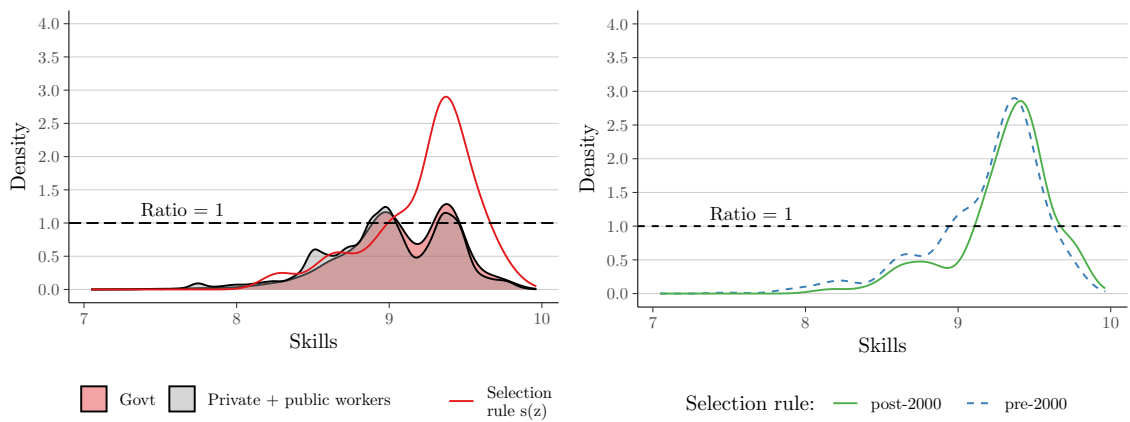

\caption{Government selection rule: government vs. private and public sector workers \label{fig:plot-selection-rule-joint-appendix}}
{\centering
\resizebox{0.95\textwidth}{!}{
\input{out/img/government_selection_rule_joint_plot_appendix.tex}
}
\par}
\input{out/fignotes/government_selection_rule_joint_plot_appendix.tex}
\end{figure}

\Cref{fig:plot-selection-rule-joint-appendix} shows the government selection rule comparing government workers to the control group (private and public sector workers only), as a robustness check to the main text version which compares to all workers.

\begin{figure}[t]
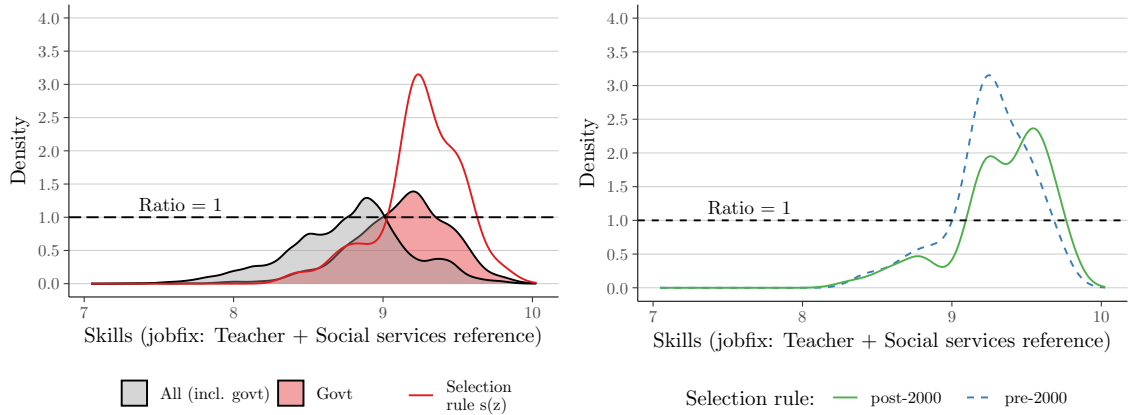

\caption{Estimated government selection rule and changes over time, holding the composition of government jobs fixed\label{fig:plot-selection-rule-jobfix}}
{\centering
\resizebox{0.95\textwidth}{!}{
\input{out/img/government_selection_rule_jobfix_joint_plot.tex}
}
\par}
\input{out/fignotes/government_selection_rule_jobfix_joint_plot.tex}
\end{figure}

\Cref{fig:plot-selection-rule-jobfix} shows the government selection rule using jobfix skills where only government workers are residualized against the govt-only regression and non-government workers keep raw skill. For government workers, we add back the predicted log skill at (Teacher, 9:Social services) for their cohort and then multiplicatively rescale so the govt weighted mean matches the raw govt weighted mean.

\begin{figure}[t]
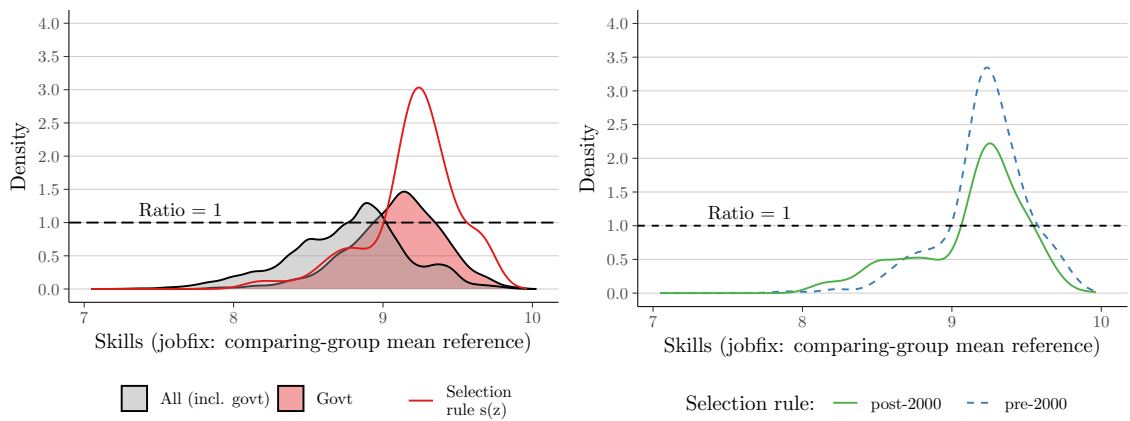

\caption{Government selection rule: jobfix skills (group-mean reference) \label{fig:plot-selection-rule-jobfix-groupmean}}
{\centering
\resizebox{0.95\textwidth}{!}{
\input{out/img/government_selection_rule_jobfix_groupmean_joint_plot.tex}
}
\par}
\input{out/fignotes/government_selection_rule_jobfix_groupmean_joint_plot.tex}
\end{figure}

\Cref{fig:plot-selection-rule-jobfix-groupmean} reports the alternative jobfix construction where the government residual is added back to the unconditional mean of the worker's \texttt{comparing} group.

\Cref{tab:table-govt-wage-informativeness} tests the informativeness of government wages in comparison to wages in similar private sector jobs.

\begin{figure}[t]
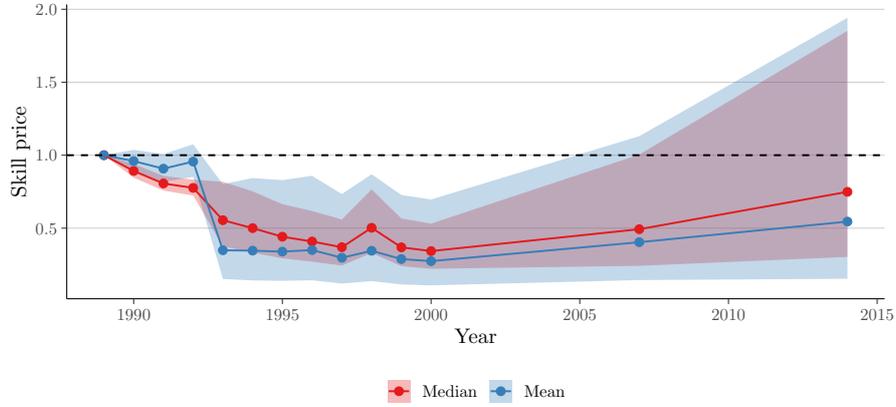

\caption{Comparison of estimation methods for skill price \label{fig:skill-price-comparison}}

{\centering

\resizebox{0.75\textwidth}{!}{
\input{out/img/evolution_skill_prices.tex}
}

\par}
\input{out/fignotes/evolution_skill_prices.tex}
\end{figure}

\begin{figure}[t]
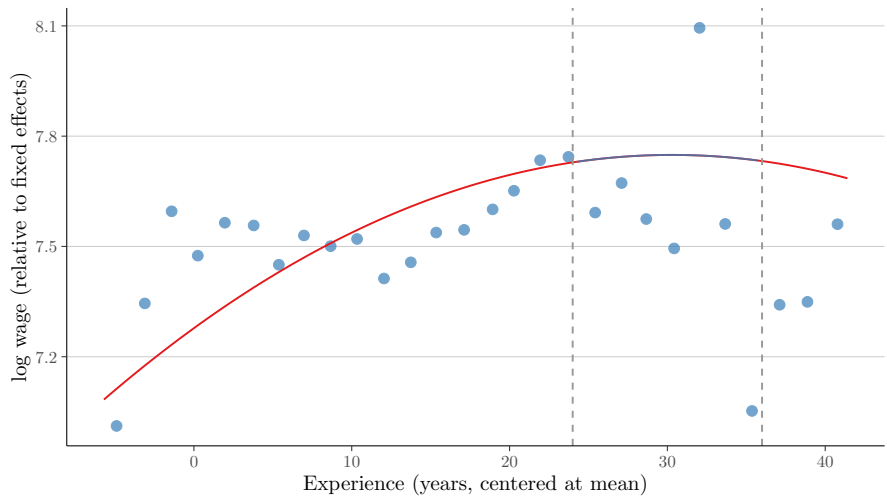

\caption{Comparison of estimated wage profile and data \label{fig:wage-profile}}

{\centering

\resizebox{0.75\textwidth}{!}{
\input{out/img/wage_profile.tex}
}
\par}
\input{out/fignotes/wage_profile.tex}
\end{figure}

\clearpage
\FloatBarrier
\section{Heckman selection correction}\label{sec:heckman-appendix}

This section provides full estimation details for the Heckman selection correction summarised in \Cref{sec:robustness-selection}.

\paragraph{First stage and exclusion test.}
\Cref{tab:heckman-first-stage} presents two columns.
The first column (``Probit'') is the first-stage selection equation: a probit of private-sector employment on all skill-related covariates, experience controls (including Mundlak means), province fixed effects, and the exclusion restriction, that is whether either parent was ever a comparable private-sector worker.
The first-stage Wald statistic of $\HeckmanFirstWald$ ($p \HeckmanFirstP$) confirms that the instrument is strongly relevant.

The second column (``Exclusion test'') tests the exclusion restriction.
The exclusion restriction requires that parental private-sector employment does not directly affect the worker's private-sector wage conditional on individual skill $z_i$, i.e.\ it affects wages only through sector selection (the inverse Mills ratio) once the latter is controlled for.
We follow standard practice for Heckman selection models and check whether the instrument is predictive of wages in the outcome equation after controlling for the inverse Mills ratio and observable skills $f(X_i)$.\footnote{In the standard linear IV model, this test is infeasible: with $y = x\beta + \delta z + \varepsilon$ and a first stage $x = z\gamma + u$, one can only identify the reduced form $y = (\gamma\beta + \delta)z + \ldots$, in which the direct effect $\delta$ cannot be separated from $\gamma\beta$. In the Heckman selection model, by contrast, the non-linearity of the inverse Mills ratio breaks this collinearity and allows $\delta$ to be separately identified from the selection effect \parencite[see][]{heckman1979sample}. To implement the test semi-parametrically, we use the approach of \textcite{robinson1988root}: we partial out skill-related covariates from the outcome, the inverse Mills ratio, and the instrument using GBM, and then test whether the instrument residual enters the second-stage OLS on residuals.}
The instrument is not significant in this augmented equation (Wald statistic $\HeckmanExclWald$, $p = \HeckmanExclP$), consistent with the exclusion restriction.

\begin{table}
\centering
\begin{talltblr}[         %% tabularray outer open
caption={Heckman selection model\label{tab:heckman-first-stage}},
note{}={\textbf{Notes:} \textit{First stage: probit of private-sector employment on all skill covariates, experience controls (Mundlak means and province fixed effects included), and the exclusion restriction. Exclusion test: Robinson GBM partials out skill covariates from all variables; OLS on residuals tests whether the instrument enters the outcome equation conditional on the IMR. Individual-specific characteristics include Big Five personality traits, education, Islamic education, cognitive and language test scores, Mundlak means of experience, and a parent-panel-missingness indicator. Heteroskedasticity-robust standard errors.}},
]                     %% tabularray outer close
{                     %% tabularray inner open
width={1\linewidth},
colspec={X[]X[]X[]},
column{2,3}={}{halign=c,},
column{1}={}{halign=l,},
hline{10}={1,2,3}{solid, black, 0.05em},
colspec={X[2]X[]X[]},
hline{1}={1-3}{solid, black, 0.08em},hline{19}={1-3}{solid, black, 0.08em},
cells={font=\footnotesize},
}                     %% tabularray inner close
\toprule
& Probit & Exclusion test \\ \midrule %% TinyTableHeader
Experience & -0.035 & 0.028 \\
& (0.008) & (0.006) \\
Experience square & 0.001 & -0.000 \\
& (0.000) & (0.000) \\
Parent ever private & 0.292 & 0.058 \\
& (0.062) & (0.057) \\
Inverse Mills ratio &  & 0.539 \\
&  & (0.179) \\
Observations & 62095 & 50040 \\
R2 &  & 0.04 \\
Individual-specific characteristics: X & X & X \\
SEs & Heteroskedasticity-robust & Heteroskedasticity-robust \\
Province dummies & X &  \\
First-stage Wald & 21.836 &  \\
First-stage p & 0.000 &  \\
Exclusion test Wald &  & 1.065 \\
Exclusion test p &  & 0.302 \\
\bottomrule
\end{talltblr}
\end{table}

\paragraph{Outcome equation with selection correction.}
\Cref{tab:table-skill-fe-heckman-robinson} reports the Robinson GBM outcome equation including the inverse Mills ratio as an additional control.
The inverse Mills ratio is positive ($\hat{\gamma} = \HeckmanImrCoef$), indicating that selection on unobservables is present.
The positive sign implies that workers who are more likely to enter the private sector (higher selection probability) also earn higher wages conditional on observables, consistent with positive selection on unobservables into the private sector.
The experience profile is similar to the baseline specification in \Cref{tab:table-skill-fe}.

\begin{table}
\centering
\begin{talltblr}[         %% tabularray outer open
caption={Robinson GBM wage profile},
note{}={\textbf{Notes:} \textit{Sample restricted to private-sector observations. Robinson GBM selection correction: probit first stage gives IMR; cross-validated GBM estimates conditional mean for Y, IMR, experience, experience squared, and Mundlak means; OLS on residuals with province FE gives all coefficients. $\hat{\gamma}_{\text{OLS}} = 0.2852$.}},
label={tab:table-skill-fe-heckman-robinson},
]                     %% tabularray outer close
{                     %% tabularray inner open
width={1\linewidth},
colspec={X[]X[]},
column{1,2}={}{halign=l,},
colspec={X[2]X[]},
hline{1}={1-2}{solid, black, 0.08em},hline{5}={1-2}{solid, black, 0.08em},
}                     %% tabularray inner close
\toprule
Term & Coefficient \\ \midrule %% TinyTableHeader
Inverse Mills ratio & 0.285205 \\
Experience & 0.023509 \\
Experience square & -0.000212 \\
\bottomrule
\end{talltblr}
\end{table}

\paragraph{Skill prediction.}
\Cref{tab:table-correlation-skill-estimates-heckman-robinson} reports the prediction accuracy and cross-method correlations for the skill estimates derived from the Heckman-corrected residuals.
The out-of-fold $R^2$ values and inter-method correlations are closely in line with the baseline specification (\Cref{tab:table-correlation-skill-estimates}), confirming that the skill predictions are not sensitive to the selection correction.
\Cref{tab:gbm-variable-importance-heckman-robinson} reports GBM variable importance for the Heckman specification; the ranking of predictors is essentially unchanged relative to the baseline.

\paragraph{Compression and level effects.}
Regressing Heckman-corrected on baseline skills (both demeaned within sample) through the origin measures how much the selection correction compresses or expands the cross-sectional dispersion of estimated skills: a slope of one indicates no change.
For the GBM estimator on the private/public worker sample, this slope is $\HeckmanGbmSlope$, indicating that the Heckman correction moderately compresses the cross-sectional dispersion of estimated skills.
The weighted mean GBM skill gap between government workers and comparable private-sector workers narrows from $\HeckmanGovtSkillDiffBaseline$ log points under the baseline specification to $\HeckmanGovtSkillDiffHeckman$ under the Heckman correction.
This is consistent with the positive inverse Mills ratio: correcting for positive selection on unobservables into the private sector reduces estimated skill residuals for private-sector workers, moderately narrowing the government-private skill gap.

\begin{table}
\centering
\begin{talltblr}[         %% tabularray outer open
caption={Skill estimate correlations and prediction accuracy by method},
note{}={\textbf{Notes:} \textit{Correlations are Pearson correlations. The in-sample R2 uses fitted values for the model trained on the full sample. The reported out-of-fold R2 is the average out-of-fold R2 across all folds. All R2 are computed using: R2 = 1 - SSE/TSS. Correlations and R2 use survey weights.}},
label={tab:table-correlation-skill-estimates-heckman-robinson},
]                     %% tabularray outer close
{                     %% tabularray inner open
width={1\linewidth},
colspec={X[]X[]X[]X[]X[]X[]X[]X[]},
column{1,2,3,4,5,6,7,8}={}{halign=l,},
colspec={X[2]X[]X[]X[]X[]X[]X[]X[]},
hline{1}={1-8}{solid, black, 0.08em},hline{7}={1-8}{solid, black, 0.08em},
}                     %% tabularray inner close
\toprule
Method & Edu OLS & Basis & LASSO & RF & GBM & In-sample R2 & Out-of-fold R2 \\ \midrule %% TinyTableHeader
Edu OLS & 1.0 & 0.53 & 0.76 & 0.60 & 0.74 & 0.098 & 0.070 \\
Basis & 0.53 & 1.0 & 0.76 & 0.77 & 0.64 & 0.32 & -0.12 \\
LASSO & 0.76 & 0.76 & 1.0 & 0.74 & 0.87 & 0.20 & 0.063 \\
RF & 0.60 & 0.77 & 0.74 & 1.0 & 0.69 & 0.59 & 0.082 \\
GBM & 0.74 & 0.64 & 0.87 & 0.69 & 1.0 & 0.17 & 0.082 \\
\bottomrule
\end{talltblr}
\end{table}

\begin{table}
\centering
\begin{talltblr}[         %% tabularray outer open
caption={GBM variable importance},
note{}={\textbf{Notes:} \textit{Relative influence measure from the GBM model: sum of squared improvements from each split, averaged across trees and normalized to 100.}},
label={tab:gbm-variable-importance-heckman-robinson},
]                     %% tabularray outer close
{                     %% tabularray inner open
width={1\linewidth},
colspec={X[]X[]},
column{1,2}={}{halign=l,},
colspec={X[2]X[]},
hline{1}={1-2}{solid, black, 0.08em},hline{27}={1-2}{solid, black, 0.08em},
}                     %% tabularray inner close
\toprule
Variable & Variable importance \\ \midrule %% TinyTableHeader
Education: Higher education & 100.0 \\
Raven's IQ test score & 54.9 \\
Big Five: Extraversion & 44.1 \\
Reads newspapers (other language) & 43.6 \\
Big Five: Conscientiousness & 41.4 \\
Big Five: Neuroticism & 39.6 \\
Word ability & 30.7 \\
Word recall test score & 29.8 \\
Counting backwards test score & 25.8 \\
Speaks Bahasa Indonesia & 24.5 \\
Big Five: Openness & 19.4 \\
Delayed word recall test score & 10.9 \\
Education: Less than primary & 10.6 \\
Date recall (2/3 correct) & 8.3 \\
Education: Senior secondary & 7.9 \\
Big Five: Agreeableness & 7.6 \\
Writes letters in Indonesian & 6.2 \\
Date recall (3/3 correct) & 3.7 \\
Writes letters in other language & 2.8 \\
Ever attended school & 0.8 \\
Education: Junior secondary & 0.0 \\
Education: Primary & 0.0 \\
Education: Missing & 0.0 \\
Islamic education & 0.0 \\
Speaks Bugis & 0.0 \\
\bottomrule
\end{talltblr}
\end{table}

\clearpage
\FloatBarrier
\section{Further details on extension "multi-dimensional skills \& more flexible experience profile"}
\label{appendix-multidim-skills}

Here, we provide further details on extending the estimation to allow for a factor-structure in the skill estimation of the form:
\begin{equation}
H_{i,e,t} = exp(z_i) \times exp(g_i \times \delta_e)
\end{equation} where \(z_i\) are individual time-fixed skills at labor market entry as before, \(\delta_e\) are arbitrary experience dummies to flexibly capture experience profiles, and \(g_i\) denotes new individual-specific learning capabilities.
Government worker skills are defined as \(z_i = h_{i,0,t}\), assuming that \(\delta_0 = 0\), and the correlation between \(z_i\), \(g_i\) and experience dummies is left unrestricted.
The factor structure allows individual-specific skills and individual-specific wage-experience profiles, but restricts the shape of the latter to a factor structure.
This is a strict generalization of our baseline model and can be an important extension in cases where empirically observed concave wage-experience profiles differ in slope across individuals.
Specifically, it has been documented in specific contexts that wage-experience profiles are steeper for highly educated than for less educated individuals and that the variance of the slope of wage-experience profiles is increasing with observed skills such as education \parencite{primiceriHeterogeneousLifecycleProfiles2009,lagakosLifeCycleWage2018}.
Alternatively, one can allow for multiple dimensions of skills by incorporating multiple factors, which are usually restricted to be orthogonal \parencite{ahnPanelDataModels2013,baiPanelDataModels2009}.

The factor structure can be estimated using the two-stage estimator from \parencite{pesaranEstimationInferenceLarge2006}.
Using Monte Carlo simulations in a previous version of the paper, we checked that the \textcite{pesaranEstimationInferenceLarge2006} estimator performs much better in estimating the factor structure than alternative estimators such as a within estimator and two different Concentrated Maximum Likelihood estimators.
This performance even holds with unbalanced and potentially non-stationary panel data as observed in real-world applications.
We are happy to share these results upon request.
To the best of our knowledge, this is a novel result, because we are not aware of any studies that have looked at the performance of factor model estimators for the individual-level estimates itself, in contrast to treatment effect parameters that are estimated in the presence of individual-level effects.

\textcite{pesaranEstimationInferenceLarge2006} proposes to use cross-sectional averages to estimate experience factors \(\delta_e\).
For this, one can first demean the series to get: \begin{equation}
\widetilde{w_{i,e}}^* \equiv \widetilde{w_{i,e}} - \widetilde{w_{i,\bullet}} = g(z_i)(\delta_e - \delta_\bullet) + (\epsilon_{i,e} - \epsilon_{i,\bullet})
\end{equation}
Using the economic structure of the problem gives \(\delta_0 = 0\), so that identification of \((\delta_e - \delta_\bullet)\) also separately identifies the two terms.
Using the fact that in the assumed data-generating process: \(lim_{n \to \infty} \widetilde{w_{\bullet,e}}^* = \overline{g(z)}(\delta_e - \delta_\bullet)\), we can write:
\begin{equation}
\frac{\widetilde{w_{\bullet,e}}^*}{\widetilde{w_{\bullet,0}}^*} \to 1-\frac{\delta_e}{\delta_\bullet}
\end{equation}

\(\widetilde{w_{\bullet,e}}^*\) gives as many equations as there are experience levels.
However, with the previous restriction of \(\delta_0 = 0\), we lose one restriction which requires to directly use: \(\widetilde{w_{\bullet,e}}^* \approx \overline{g(z)}(\delta_e - \delta_\bullet)\).
Given that we are not directly interested in the estimates of \(\delta\) nor \(g(z_i)\), we can instead choose any non-zero normalization to obtain the same estimates of individual skills \(log(z_i)\).
For any normalization of \(\overline{g(z)}\), we obtain an estimate of \(\delta\).
We can then include these in the following experience-series regression for each individual to estimate \(log(z_i)\) and \(g(z_i)\):
\begin{equation}
\widetilde{w_{i,e}} = log(z_i) + \widehat{\delta_e}g(z_i) + \epsilon_{i,e}
\end{equation}
The individual-level skill estimates are the constant of the regression, giving the estimate \(\widehat{log(z_i)}\).
The derived estimate of private sector skills is generically inconsistent in a panel with fixed $T$ due to the incidental parameters problem.
This is generally true of any estimator that gives individual-specific estimates.
For example, the standard within estimator also gives inconsistent estimates of the individual levels as long as the time dimension is fixed.
Intuitively, the setup only allows us to extract a noisy signal from the even noisier wage data.

\end{document}